\documentclass[longauth]{aa}  

\usepackage{graphicx}
\usepackage{txfonts}
\usepackage[colorlinks=true,linkcolor=blue,citecolor=blue]{hyperref}
\usepackage{float}
\usepackage{subfloat}
\usepackage{subcaption}
\usepackage[flushleft]{threeparttable}
\begin{document} 
\title{SN~2018is: A low-luminosity Type IIP supernova with narrow hydrogen emission lines at early phases}
   \subtitle{}

   \author{R. Dastidar
          \inst{1,2}\fnmsep\thanks{E-mail: rdastidr@gmail.com},
          K. Misra\inst{3},
          S. Valenti\inst{4},
          D. J. Sand\inst{5},
          A. Pastorello\inst{6},
          A. Reguitti\inst{7,6},
          G. Pignata\inst{8},
          S. Benetti\inst{6},
          S. Bose\inst{9}, 
          A. Gangopadhyay\inst{10},
          M. Singh\inst{11},
          L. Tomasella\inst{6},
          J. E. Andrews\inst{12},
          I. Arcavi\inst{13,14},
          C. Ashall\inst{15,16},
          C. Bilinski\inst{5},
          K. A. Bostroem\inst{5},
          D. A. H. Buckley\inst{17, 18,19},
          G. Cannizzaro\inst{20}, 
          L. Chomiuk\inst{21},
          E. Congiu\inst{22},
          S. Dong\inst{23,24,25}, 
          Y. Dong\inst{4},
          N. Elias-Rosa\inst{6},
          M. Fraser\inst{26}, 
          C. Gall\inst{27}, 
          M. Gromadzki\inst{28},
          D. Hiramatsu\inst{29,30},
          G. Hosseinzadeh\inst{31},
          D. A. Howell\inst{32,33},
          E. Y. Hsiao\inst{34},
          C. McCully\inst{32,33},
          N. Smith\inst{5},
          J. Strader\inst{21}
          }

   \institute{Instituto de Astrofísica, Universidad Andres Bello, Fernandez Concha 700, Las Condes, Santiago RM, Chile
         \and
             Millennium Institute of Astrophysics (MAS), Nuncio Monsenor Sòtero Sanz 100, Providencia, Santiago RM, Chile
         \and
             Aryabhatta Research Institute of observational sciencES, Manora Peak, Nainital, 263001, India
         \and
             Department of Physics and Astronomy, University of California, Davis, 1 Shields Avenue, Davis, CA 95616-5270, USA
         \and
            Steward Observatory, University of Arizona, 933 North Cherry Avenue, Tucson, AZ 85721-0065, USA
        \and
            INAF Osservatorio Astronomico di Padova, Vicolo dell’Osservatorio 5, I-35122 Padova, Italy
        \and
            INAF Osservatorio Astronomico di Brera, Via E. Bianchi 46, I-23807 Merate (LC), Italy
        \and
            Instituto de Alta Investigación, Universidad de Tarapacá, Santiago RM, Chile
        \and
            Department of Physics and Astronomy, Aarhus University, Ny Munkegade 120, DK-8000 Aarhus C, Denmark
        \and
            Department of Astronomy, The Oskar Klein Center, Stockholm University, AlbaNova, 106 91 Stockholm, Sweden    
        \and
            Indian Institute of Astrophysics, Koramangala 2nd Block, Bangalore 560034, India
        \and
            Gemini Observatory, 670 North A‘ohoku Place, Hilo, HI 96720-2700, USA  
        \and
            The School of Physics and Astronomy, Tel Aviv University, Tel Aviv 69978, Israel
        \and
            CIFAR Azrieli Global Scholars program, CIFAR, Toronto, ON M5G 1M1, Canada
        \and
            Institute for Astronomy, University of Hawai'i at Manoa, 2680 Woodlawn Dr., Hawai'i, HI 96822, USA 
        \and
            Department of Physics, Virginia Tech, Blacksburg, VA 24061, USA
        \and
            South African Astronomical Observatory, PO Box 9, Observatory 7935, Cape Town, South Africa
        \and
            Department of Astronomy, University of Cape Town, Private Bag X3, Rondebosch 7701, South Africa
        \and
            Department of Physics, University of the Free State, P.O. Box 339, Bloemfontein 9300, South Africa
        \and
            SURF, Science Park 140, 1098 XG Amsterdam, the Netherlands
        \and
            Department of Physics and Astronomy, Michigan State University, 567 Wilson Road, East Lansing, MI 48824, USA 0000-0002-1468-9668
        \and
            European Southern Observatory, Alonso de C´ordova 3107, Casilla 19, Santiago 19001, Chile
        \and
            Department of Astronomy, School of Physics, Peking University, Yiheyuan Rd. 5, Haidian District, Beijing, China, 100871
        \and
            Kavli Institute for Astronomy and Astrophysics, Peking University, Yi He Yuan Road 5, Hai Dian District, Beijing 100871, People’s Republic of China
        \and
            National Astronomical Observatories, Chinese Academy of Science, 20A Datun Road, Chaoyang District, Beijing 100101, China
        \and
            UCD School of Physics, L.M.I. Main Building, Beech Hill Road, Dublin 4, D04 P7W1, Ireland  
        \and
            DARK, Niels Bohr Institute, University of Copenhagen, Jagtvej 128, 2200 Copenhagen, Denmark
        \and
            Astronomical Observatory, University of Warsaw, Al. Ujazdowskie 4, 00-478 Warszawa, Poland
        \and
            Center for Astrophysics, Harvard \& Smithsonian, 60 Garden Street, Cambridge, MA 02138-1516, USA
        \and
            The NSF AI Institute for Artificial Intelligence and Fundamental Interactions, USA
        \and
            Department of Astronomy \& Astrophysics, University of California, San Diego, 9500 Gilman Drive, MC 0424, La Jolla, CA 92093-0424, USA
        \and
            Las Cumbres Observatory, 6740 Cortona Dr, Suite 102, Goleta, CA 93117-5575, USA
        \and
            Department of Physics, University of California, Santa Barbara, CA 93106-9530, USA
        \and
            Department of Physics, Florida State University, 77 Chieftan Way, Tallahassee, FL 32306, USA  
             }

   \date{Received ; accepted }

% \abstract{}{}{}{}{} 
% 5 {} token are mandatory
 
     \abstract
      {We present a comprehensive photometric and spectroscopic study of the Type IIP SN~2018is. The $V$-band luminosity and the expansion velocity at 50 days post-explosion are $-$15.1$\pm$0.2 mag (corrected for A$_V$=1.34 mag) and 1400 km s$^{-1}$, classifying it as a low-luminosity SN II. The recombination phase in the $V$-band is shorter, lasting around 110 days, and exhibits a steeper decline (1.0 mag per 100 days) compared to most other low-luminosity SNe II. Additionally, the optical and near-infrared spectra display hydrogen emission lines that are strikingly narrow, even for this class. The \ion{Fe}{ii} and \ion{Sc}{ii} line velocities are at the lower end of the typical range for low-luminosity SNe II. Semi-analytical modelling of the bolometric light curve suggests an ejecta mass of $\sim$8 M$_\odot$, corresponding to a pre-supernova mass of $\sim$9.5 M$_\odot$, and an explosion energy of $\sim$0.40 $\times$ 10$^{51}$ erg. Hydrodynamical modelling further indicates that the progenitor had a zero-age main sequence mass of 9~M$_\odot$, coupled with a low explosion energy of 0.19 $\times$ 10$^{51}$ erg. The nebular spectrum reveals weak [\ion{O}{i}] $\lambda\lambda$6300,6364 lines, consistent with a moderate-mass progenitor, while features typical of Fe core-collapse events, such as \ion{He}{i}, [\ion{C}{i}], and [\ion{Fe}{i}], are indiscernible. However, the redder colours and low ratio of Ni to Fe abundance do not support an electron-capture scenario either. As a low-luminosity SN II with an atypically steep decline during the photospheric phase and remarkably narrow emission lines, SN~2018is contributes to the diversity observed within this population.}
 
   \titlerunning{SN 2018is: low-luminosity SN}
   \authorrunning{R. Dastidar et al.}
   \maketitle
%
%________________________________________________________________
%%%%%%%%%%%%%%%%%%%%%%%%%%%%%%%%%%%%%%%%%%%%%%%%%%
%%%%%%%%%%%%%%%%% BODY OF PAPER %%%%%%%%%%%%%%%%%%
%
\section{Introduction}
\label{introduction}
Type II Supernovae (SNe II) originate from the core collapse of massive stars ($\gtrsim$ 8 M$_\odot$) that retain a portion of their hydrogen (H) envelope before the explosion. These events are characterised by pronounced H-features in their spectra. Among them, SNe\,IIP are the most common type of core-collapse SNe \citep{Li2011, Graur2017}. These SNe are notable for their relatively constant luminosity phase, commonly referred to as the `plateau', which corresponds to the H-recombination phase and lasts approximately 100 days in the $VRI$ bands. Within the SNe\,IIP category, there is a significant diversity in intrinsic luminosities, with peak absolute magnitudes in the $V$-band ranging from about $-$14 to $-$18 mag. SNe IIP with peak magnitudes $\geq$ $-$15.5\,mag and plateau magnitudes between $-$13.5 to $-$15.5\,mag are classified as low-luminosity SNe\,IIP (LLSNe\,II, \citealt{Pastorello2004, Spiro2014, Muller2020}). This classification contrasts with the average peak absolute magnitude of around $-$16.74\,mag typically observed in normal luminosity SNe IIP ($\sigma$ = 1.01; \citealt{Anderson2014, Galbany2016}).

LLSNe\,II are relatively rare, making up $\sim$5\% of all SNe II \citep{Pastorello2004}. However, this low fraction could be influenced by observational biases, as fainter SNe were more challenging to detect in the early 2000s. With the improved detection capabilities of modern transient surveys, it is increasingly clear that LLSNe II are being detected more frequently than estimated in earlier studies \citep{Spiro2014, Anderson2014}. SNe~1997D \citep{deMello1997, Turatto1998, Benetti2001} and 2005cs \citep{Pastorello2006, Pastorello2009} are prototypical examples of LLSNe\,II, with $V$-band luminosity consistently below M$_V$ = $-$14.65 mag across all observed epochs. Some of the faintest SNe II discovered include SNe~1999br \citep{Zampieri2003, Pastorello2004} and 2010id \citep{Gal-Yam2011}, which exhibited plateau luminosities of M$_V$ = $-$13.76\,mag and M$_r$ = $-$13.85\,mag, respectively. \cite{Pastorello2004} and \cite{Spiro2014} conducted a sample study of LLSNe\,II and characterised their photometric and spectroscopic properties. A distinctive feature of the photospheric spectra of LLSNe~II is the presence of relatively narrow P Cygni profiles, indicative of slow ejecta expansion (a few 1000 km s$^{-1}$) resulting from a low-energy explosion (E$_{\rm exp}$ $\lesssim$ a few times 10$^{50}$ erg). These events also exhibit a lower luminosity during the exponential decay in the nebular phase, indicating the synthesis of a smaller amount of $^{56}$Ni (M$_{\rm Ni}$ $\lesssim$ 10$^{-2}$ M$_\odot$) compared to their normal luminosity counterparts with median M$_{\rm Ni}$ of 0.03-0.04\,M$_\odot$ \citep{Anderson2019, Muller2020, Rodriguez2021}. However, for LLSNe~2016bkv and 2021gmj, with plateau luminosities of $\sim$ $-$14.8 mag and $-$15.4 mag in $V$-band, respectively, the synthesised $^{56}$Ni mass was estimated to be 0.022~M$_\odot$ and 0.014~M$_\odot$ \citep{Hosseinzadeh2018, Nakaoka2018, Murai2024, Meza2024}. The plateau phase of LLSNe~II tends to be longer than that of normal luminosity SNe II, lasting, for instance, 120\,days in SN~2005cs \citep{Pastorello2009} and 140\,days in SN~2018hwm \citep{Reguitti2021}. The extended duration of the plateau is attributed to the interplay between a massive H envelope and a low expansion velocity, as both factors contribute to prolonging this phase through a longer recombination time. There are a handful of SNe IIP that populates the luminosity space between the low and normal luminosity class, typically with plateau luminosities within the narrow range of $-$15.5 to $-$16 mag, e.g., SNe 2008in (M$_V^{\rm 50d}$ = $-$15.5 $\pm$ 0.2 mag, \citealt{Roy2011}), 2009N (M$_V^{\rm 50d}$ = $-$15.7 $\pm$ 0.1 mag, \citealt{Takats2014}), 2009js (M$_V^{\rm 50d}$ = $-$15.9 $\pm$ 0.2 mag, \citealt{Gandhi2013}), 2013am \citep{Zhang2014, Tomasella2018}, 2013K (M$_V^{\rm 50d}$ = $-$15.9 $\pm$ 0.8 mag, \citealt{Tomasella2018}) and 2018aoq (M$_V^{\rm 50d}$ = $-$15.9 $\pm$ 0.2 mag, \citealt{Tsvetkov2019, Tsvetkov2021}). These objects create a continuous distribution of luminosity among Type IIP SNe. The transitional SNe IIP are up to a magnitude brighter than faint ones and the $^{56}$Ni mass yield is comparable to normal luminosity SNe~II. However, the spectra and the low expansion velocities inferred from the spectral lines resemble those observed in LLSNe II. 

There are three LLSNe~II—2003gd, 2005cs, and 2008bk—with progenitor detections in archival pre-SN imaging data (e.g., \citealt{VanDyk2003, Maund2005, Mattila2008}) and confirmed optical disappearance of the progenitors post-explosion \citep{Maund2009, Maund2014}. These observations suggest that red supergiants (RSGs) with initial masses (M$_{\rm ZAMS}$) between 8 and 15 M$_\odot$ are the progenitors of LLSNe II. Hydrodynamical and spectral models (e.g., \citealt{Dessart2013, Jerkstrand2018, Martinez2020}) also support this mass range. Recent hydrodynamic modelling suggests a positive correlation between progenitor mass and explosion energy, such that lower mass progenitors result in less energetic explosions, resulting in a fainter event \citep{Morozova2018, Utrobin2019, Martinez2022}. However, \cite{Zampieri2003} propose massive RSGs as the progenitors of LLSNe~II, with a significant amount of fallback material onto the proto-neutron star, leading to the release of low quantities of $^{56}$Ni. Additionally, the core-collapse of super asymptotic giant branch (SAGB) stars, with masses at the lower end of 8-12\,M$_\odot$, theoretically predicted to lead to electron capture (EC) SNe \citep{Nomoto1984, Nomoto1987}, has also been suggested as a potential origin for LLSNe II \citep{Kitaura2006, Hosseinzadeh2018, Valerin2022}. Thus far, SNe~2016bkv and 2018zd \citep[][but see \citealt{Callis2021}]{Hiramatsu2021, Hosseinzadeh2018} are considered promising candidates for ECSNe, aside from the well-known historical case of SN~1054, the progenitor of the Crab Nebula. While SN~2016bkv was a LLSN II, SN~2018zd had a bright peak of $-$18.40 $\pm$ 0.60 mag and a plateau magnitude of $-$17.79 $\pm$ 0.55, which are much brighter than LLSNe II. Further constraints on progenitor characteristics can be obtained from late-time nebular spectra of SNe II. Theoretical investigations have demonstrated that the forbidden lines in the nebular spectra of SNe II, such as the [\ion{O}{i}] $\lambda\lambda$6300, 6364 doublet, [\ion{Fe}{ii}] $\lambda$7155, and [\ion{Ni}{ii}] $\lambda$7378 lines, can be employed to constrain the progenitor mass and explosion dynamics \citep{Fransson1987, Fransson1989, Woosley2007, Jerkstrand2018}. 

\begin{table}
    \centering
    \caption{Basic information of SN~2018is.}
    \begin{tabular}{ll} 
		\hline\hline
            Host Galaxy & NGC~5054\\
            R.A. & 13$^{\rm h}$16$^{\rm m}$57$^{\rm s}$.35\\
            Dec. & $-$16$^{\rm d}$37$^{\rm m}$04$^{\rm s}$.43\\
            Discovery & JD 2458136.2\\
            Explosion epoch$^{1}$  & JD 2458133.4$\pm$1.1\\
            Redshift$^2$ &  0.005811\\
            Helio. radial velocity$^3$ & 1734 $\pm$ 2 km s$^{-1}$ \\
            (corrected for LG infall onto Virgo) & \\
            E(B$-$V)$_{\rm MW}^{1}$ & 0.0708$\pm$0.0003 mag\\
            E(B$-$V)$_{\rm host}^{1,a}$ (Colour method) & 0.12$\pm$0.06 mag \\
            E(B$-$V)$_{\rm host}^{1,b}$ (\ion{Na}{id}) & 0.36$\pm$0.07 mag \\
            Distance$^{1}$ & 21.3 $\pm$ 1.7 Mpc \\
		\hline
    \end{tabular}
    \newline
	\noindent
$^{1}${This work}, $^{2}${\citet{Pisano2011}, from NED}, $^{3}${From HyperLEDA}, $^{a}${Low reddening estimate}, $^{b}${High reddening estimate}\\
	\noindent
	% $^2$This paper.\\
	% \noindent
	% $^c$This paper.\\
	\label{Tab1:2018is}
\end{table}

%%*************** SN marked figure *****************
\begin{figure}
    \includegraphics[scale=0.45, clip, trim={1.9cm 0.3cm 4.5cm 2.5cm}]{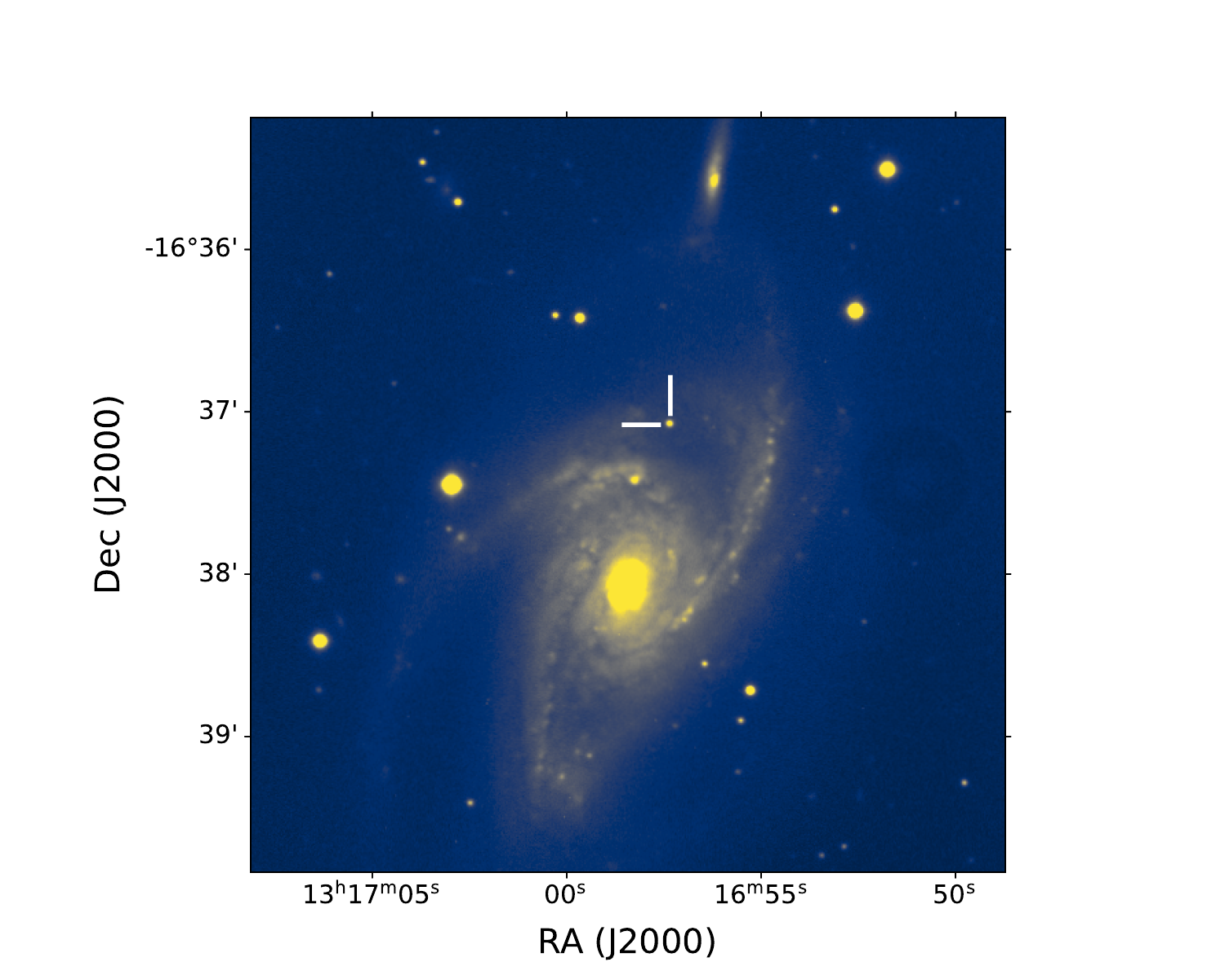}
    \caption{A 300s Sloan-$r$ band image obtained with the 1.82 m Ekar Telescope on 2018 April 19. The location of the SN in the host galaxy NGC 5054 is marked.}
    \label{SN_field}
\end{figure}

In this paper, we present an analysis of the photometric and spectroscopic characteristics of a faint Type II SN, DLT18a (a.k.a. SN~2018is, ATLAS18eca). SN~2018is was discovered during the ongoing Distance Less than 40 Mpc (DLT40, \citealt{Tartaglia2018}) sub-day cadence supernova search, which at that time was largely conducted using the PROMPT5 0.4 m telescope situated at the Cerro Tololo Inter-American Observatory (CTIO, \citealt{Wyatt2018}). The SN was first detected on 2018-01-20.3 UT (JD 2458138.8) at the coordinates R.A.: 13$^{\rm h}$16$^{\rm m}$57$^{\rm s}$.35, Dec.: $-$16$^{\rm d}$37$^{\rm m}$04$^{\rm s}$.43, exhibiting a magnitude of approximately $R \sim 17.9$ mag in the nearby galaxy NGC 5054, which at that time was just coming from behind the Sun. A follow-up confirmation image was obtained on 2018-01-20.6~UT utilising a 0.4 m telescope at the Siding Spring Observatory in New South Wales, Australia, as part of the Las Cumbres Observatory (LCO) telescope network \citep{Brown2013}. A subsequent optical spectrum, acquired on 2018-01-21.3~UT with the Goodman Spectrograph mounted on the Southern Astrophysical Research (SOAR) telescope, displayed a blue continuum, along with prominent lines of \ion{H}{$\alpha$}, \ion{H}{$\beta$} and \ion{He}{i} with approximate velocities of 4000 km s$^{-1}$ \citep{Sand2018} akin to LLSNe II, particularly about 1-2 weeks after the explosion. However, the absolute magnitude of M$_V \sim -13$ mag could also indicate a luminous blue variable (LBV) outburst. Further analysis of the light curve and spectroscopic properties in Section \ref{Sec4:lightcurve} and \ref{Sec5:spectral_features} provided conclusive evidence categorising SN~2018is as a SN II. Throughout this paper, we adopt a flat $\Lambda$CDM cosmological model with parameters H$_0$ = 67.4$\pm$0.5\,km\,s$^{-1}$\,Mpc$^{-1}$, $\Omega_M$ = 0.3, and $\Omega_\Lambda$ = 0.7 \citep{Planck2020}. The basic parameters of the SN and the host galaxy are listed in Table~\ref{Tab1:2018is}.

The paper is organised as follows: Section \ref{DataReduction} describes the instrumental setups and the tools employed for reducing the photometric and spectroscopic data of SN~2018is. In Section \ref{Sec3:Parameters}, estimation of the explosion epoch of SN~2018is, distance to the SN, and reddening of the host galaxy are reported. Section~\ref{Sec4:lightcurve} presents the light curve evolution, and the spectral features are discussed in Section~\ref{Sec5:spectral_features}. The photometric and spectroscopic parameters of SN~2018is are compared with other SNe~II in Section \ref{sec6}. In Section \ref{Sec6:Modelling}, the modelling of the bolometric light curve using both semi-analytical and hydrodynamical methods are discussed. Finally, the nature of the progenitor is discussed in \ref{Sec7:Discussion} and an overall summary of this work is presented in Section \ref{summary}.

\section{Observations and Data Reduction}
\label{DataReduction}

The observing campaign of SN~2018is commenced a few hours after its DLT40 discovery, using instruments equipped with broadband {\it UBVgriz} filters. PROMPT5 unfiltered DLT40 images were reduced as described in \cite{Tartaglia2018}, using the dedicated pipeline and calibrated to the $r$ band. Observations in the {\it UBVgri} bands were conducted through the Global Supernova Project (GSP) using the 0.4m, 1m, and 2m telescopes of the Las Cumbres Observatory (LCO). Pre-processing of the images, including bias correction and flat-fielding, was conducted using the BANZAI pipeline \citep{McCully2018}. Subsequent data reduction was performed with \texttt{lcogtsnpipe} \citep{Valenti2016}, a PyRAF-based photometric reduction pipeline. Since the SN is offset from the host galaxy in a region with a smooth background, image subtraction was not required. The {\it UBV} band data were calibrated to Vega magnitudes \citep{Stetson2000}, using standard fields observed on the same night by the same telescope. The {\it gri} band data were calibrated to AB magnitudes using \citet{Henden2009}. Additional optical photometry in $BVgriz$ filters was obtained with (i) the optical imaging component of the Infrared-Optical imager: Optical (IO:O, \citealt{Barnsley2016}), mounted on the 2m Liverpool Telescope (LT, \citealt{Steele2004}), (ii) the Alhambra Faint Object Spectrograph and Camera (ALFOSC) and Stand-by camera (STANcam) on the 2.56 m Nordic Optical Telescope (NOT, \citealt{Djupvik2010}), and (iii) the Asiago Faint Object Spectrograph and Camera (AFOSC) on the 1.82 m Copernico Telescope. The data from these telescopes were reduced similarly, and PSF photometry was performed using \texttt{DAOPHOT II} \citep{Stetson1987} to compute the instrumental magnitudes of the SN. The compiled photometry for SN~2018is is presented in Table~\ref{phot}. We also included the ATLAS forced photometry data \citep{Tonry2018, Smith2020} in our work which is tabulated in Table~\ref{atlas_phot}.

SN~2018is was also observed with the Ultra-Violet/Optical Telescope (UVOT; \citealt{Roming2005}) onboard the Neil Gehrels {\it Swift} Observatory \citep{Gehrels2004}. The observations were carried out in three UV ({\it uvw2}: $\lambda_c$ = 1928 \AA, {\it uvm2}:  $\lambda_c$ = 2246 \AA, {\it uvw1}: $\lambda_c$ = 2600 \AA) and three optical ({\it u, b, v}) filters at four epochs. UV photometric data were obtained from the {\it Swift} Optical/Ultraviolet Supernova Archive (SOUSA\footnote{https://archive.stsci.edu/prepds/sousa/}; \citealt{Brown2014}). The reduction procedure is outlined in \citet{Brown2009}, which includes the subtraction of the host galaxy count rate. For estimating the magnitudes, revised zero points and time dependent sensitivity were adopted from \citet{Breeveld2011}. The {\it Swift} UVOT magnitudes are listed in Table \ref{uvot}. 

Spectroscopic observations of SN~2018is were carried out from 3.4 to 384 days post-discovery, using (i) the Robert Stobie Spectrograph (RSS, \citealt{Burgh2003}) with the PG300 lines/mm grating on the Southern African Large Telescope (SALT; \citealt{Buckley2006}), which covers 3400 - 9000 $\AA$, at a resolution of $\sim$18 $\AA$ with a 1.5$^{\prime\prime}$ slit, (ii) the Goodman High Throughput Spectrograph (GHTS-R; \citealt{Clemens2004}) on the Southern Astrophysical Research Telescope (SOAR), (iii) the ALFOSC on the NOT, (iv) the Double Beam Spectrograph (DBSP; \citealt{Oke1982}) mounted on the 200-inch Hale telescope at Palomar Observatory (P200), (v) the Boller and Chivens (B\&C) Spectrograph with the 300 lines/mm grating on the University of Arizona’s Bok 2.3 m telescope located at Kitt Peak Observatory, which were reduced in a standard way with IRAF \citep{Tody1986, Tody1993} routines, (vi) the Blue Channel (BC) spectrograph on the 6.5 m MMT, with the 1200 lines/mm grating covering a range of $\sim$5700-7000 $\AA$, (vii) the Optical System for Imaging and low-Intermediate-Resolution Integrated Spectroscopy (OSIRIS, grating ID R1000B) mounted on the Gran Telescopio Canarias (GTC), and (viii) the Low Resolution Imaging Spectrometer (LRIS; \citealt{Oke1995}) on the 10 m Keck-I telescope. The LRIS spectrum was taken using a 1$''$ aperture with the 560 dichroic to split the beam between the 600/4000 grism on the blue side and the 400/8500 grating on the red side. Taken together, the merged spectrum spans $\sim$3200-10200 $\AA$. In addition, a near-infrared low resolution spectrum was obtained with FIRE at the Magellan 6.5\,m telescope (8000–25000 \AA). The log of spectroscopic observations is provided in Table~\ref{spec_observations}. 

Standard procedures were followed for the spectroscopic data, which were reduced within \texttt{IRAF}\footnote{\texttt{IRAF} refers to the Image Reduction and Analysis Facility distributed by the National Optical Astronomy Observatories, which was operated by the Association of Universities for Research in Astronomy, Inc., under a cooperative agreement with the National Science Foundation.}. The \texttt{APALL} task was employed to extract one-dimensional spectra, which were subsequently calibrated in wavelength and flux using arc lamps and spectrophotometric standard star spectra, respectively. These standards were observed at comparable airmasses either on the same night or on adjacent nights. Night sky emission lines in the spectra were used to validate the accuracy of wavelength calibration, and necessary shifts were applied as required. In order to account for absolute flux calibration, the spectra were scaled with respect to the photometric data and further corrected for the redshift of the host galaxy.

\section{Parameters of SN 2018is}
\label{Sec3:Parameters}

\subsection{Explosion epoch and distance}
\label{distance}
The last non-detection date of SN~2018is was recovered from the forced photometry light curve and determined to be 2018 January 12.6 (JD 2458131.1), with a limiting magnitude of 19.85 in the ATLAS $o$-filter. The SN was first detected in the ATLAS $o$-filter at a magnitude of 18.01 $\pm$ 0.06 on 2018 January 17.7 (JD 2458136.2), 2.6 days before the DLT40 detection on 2018-01-20.3 UT (JD 2458138.8). Therefore, using the last non-detection and first detection from ATLAS, the explosion date is estimated to be 2458133.6 $\pm$ 2.6. An alternate method for estimating the explosion epoch is by cross-correlating the SN spectra with a library of spectral templates using the SuperNova IDentification package ({\tt SNID}, \citealt{Blondin2007}) as done in \cite{Gutierrez2017}. We performed spectral matching on the first two spectra of SN~2018is, obtained on 2018 January 21.0 and 21.3 (0.5 and 0.7 day after discovery, respectively). The best match was determined based on the {\tt SNID} `rlap' parameter, which quantifies the quality of fit, with a higher value indicating a better correlation. In addition to the `rlap' parameter, the top three spectral matches provided by \texttt{SNID} were checked visually. We found a good match between the spectra of SN~2018is and SN~2006bc (available in the SN IIP templates in \texttt{SNID} from \citealt{Gutierrez2017}). SN~2006bc is a LLSN II, with absolute plateau magnitude of $-$15.1 mag in $V$-band and with a $\pm$4 day uncertainty in the explosion epoch obtained from a photometric non-detection \citep{Anderson2014}. For both spectra of SN~2018is, the best match was found to be with the 9 ($\pm$ 4) day spectrum of SN~2006bc, with a higher `rlap' value for the second spectrum of SN~2018is. Considering the age of the supernova on 2018 January 21.3 to be 9 ($\pm$ 4) days post-explosion, the explosion epoch is estimated to be 2018 January 12.3 ($\pm$ 4), corresponding to JD 2458130.8.

\begin{table*}
 \centering
  \caption{The derived parameters of SN~2018is for the EPM distance estimate: the angular size ($\theta$), photospheric temperature ($T$), and the interpolated photospheric velocity ($v_{ph}$).}
  \begin{tabular}{@{}lcllllll@{}}
  \hline
  \hline 
  $t^\dagger$ & $\theta_{BV}$ & $T_{BV}$ & $\theta_{BVI}$ & $T_{BVI}$ & $\theta_{VI}$ & $T_{VI}$  & $v_{ph}$ \\  
   (days) & ($10^9$ km Mpc$^{-1}$) & (K) & ($10^9$ km Mpc$^{-1}$) & (K) & ($10^9$ km Mpc$^{-1}$) & (K) & (km s$^{-1}$) \\ 
  \hline
 4.6 & 1.78 (0.09) & 13208 (767)  & 1.25 (0.08) & 16609 (923) & 0.96 (0.07) & 19347 (1019) & 3967 (357)\\
22.4 & 2.84 (0.06) & 8027 (210)  & 2.48 (0.08) & 9695 (256)  & 2.05 (0.16)  & 11766 (741) & 2322 (252)\\
25.7 & 3.16 (0.08) & 6941 (253)  & 2.70 (0.13) & 8866 (403) & 2.03 (0.23)  & 12027 (1084) & 2242 (303)\\
27.0 & 3.16 (0.11) & 6213 (177)  & 3.03 (0.09) & 7438 (188)  & 2.45 (0.33)  & 10077 (1399) & 2060 (291)\\
35.7 & 3.22 (0.03) & 6128 (64)   & 3.18 (0.04) & 7165 (102)  & 2.59 (0.11)  & 9747 (383) & 1944 (261)\\
41.5 & 3.23 (0.02) & 5503 (224)  & 3.22 (0.05) & 6511 (268) & 2.74 (0.37)  & 9164 (1123) & 1763 (272)\\
45.2 & 3.24 (0.02) & 5764 (159)  & 3.21 (0.03) & 6778 (181) & 2.82 (0.23)  & 9092 (610) & 1649 (313)\\
 \hline
\end{tabular}	
 \label{epm}		  
 \begin{tablenotes}
       \item[a]{$^\dagger$ since discovery, $t_0$= JD 2458136.2}
     \end{tablenotes}
\end{table*}

A number of redshift-independent distance estimates are available in the NASA Extragalactic Database (NED) for NGC~5054, the host galaxy of SN~2018is. These estimates span a range from 12.40 Mpc to 27.30 Mpc. The Virgo infall distance to the galaxy NGC 5054 is 25.7 $\pm$ 0.2 Mpc, based on the recessional velocity of the galaxy, v$_{\rm Vir}$ = 1734 $\pm$ 2 km s$^{-1}$, from HyperLeda \citep{Makarov2014}. The distance estimate of NGC 5054 in the Cosmicflows-3 catalog \citep{Tully2016} is 18.2 $\pm$ 2.5 Mpc.

To obtain an independent distance estimate, we apply the Expanding Photosphere Method (EPM) utilising the early photometric and spectroscopic data of SN~2018is. The EPM is a variant of the Baade-Wesselink method to estimate SN distances \citep{Kirshner1974}.  We follow the steps and techniques outlined in \citet{Dastidar2018} to implement EPM in SN~2018is. During the early phases, the SN ejecta is fully ionised, and electron scattering is the primary source of opacity at the photosphere. In this phase, the SN can be approximated as radiating like a diluted blackbody. The EPM compares the linear and angular radius of the homologously expanding optically thick SN ejecta to compute the SN distance. The angular radius of the expanding ejecta at any time $t$ can be approximated as 

\begin{equation}
    \theta = \frac{\rm R}{\rm D} = 
    \sqrt{ \frac{f_\lambda 10^{0.4A_\lambda}} { \zeta_\lambda^2 (T_c)\pi B_\lambda (T_c) } }
    \label{theta}
\end{equation}

\noindent
where $B_\lambda$ is the Planck function at colour temperature $T_c$, $f_\lambda$ is the flux density received at Earth, $A_\lambda$ is the extinction at wavelength $\lambda$, and $\zeta_\lambda (T_c)$ is the colour temperature dependent `dilution factor'. Here, $R = v_{ph}(t-t_0)$, where $(t-t_0)$ is the time since explosion and $v_{ph}$ is the photospheric velocity at the corresponding epoch. Eq.~\ref{theta} can be written in terms of magnitudes obtained from broadband photometry integrated over the filter response function. The convolution of the filter response function was computed by \citet{Hamuy2001}. The dilution factors $\zeta$ can be expressed as a function of $T_c$, as described in \citet{Dessart2005}. We converted the Sloan $r$ and $i$ magnitudes to $I$ magnitudes by using the equations given by \cite{Lupton2005}. The extinction at the central wavelengths of the $BVI$-bands were determined using the high reddening value, A$_V$=1.34 mag (see Sect.~\ref{extinction}).

Employing coefficients from \citet{Dessart2005}, we estimated $\theta$ in three filter combinations, \{$BV$\}, \{$BVI$\}, and \{$VI$\}, and listed the values in Table~\ref{epm}. The photospheric velocity $v_{ph}$ was determined using the \ion{H}{$\beta$} line up to 4.6 days after discovery and the \ion{Sc}{ii} $\lambda$6246 line up to 45.0 days after discovery. We interpolated the velocity to the epochs of photometry using Automated Loess Regression (ALR, \citealt{Rodriguez2019}). 

\begin{figure}
    \centering
	\includegraphics[scale=0.54, clip, trim={0.cm 0cm 0cm 0.3cm}]{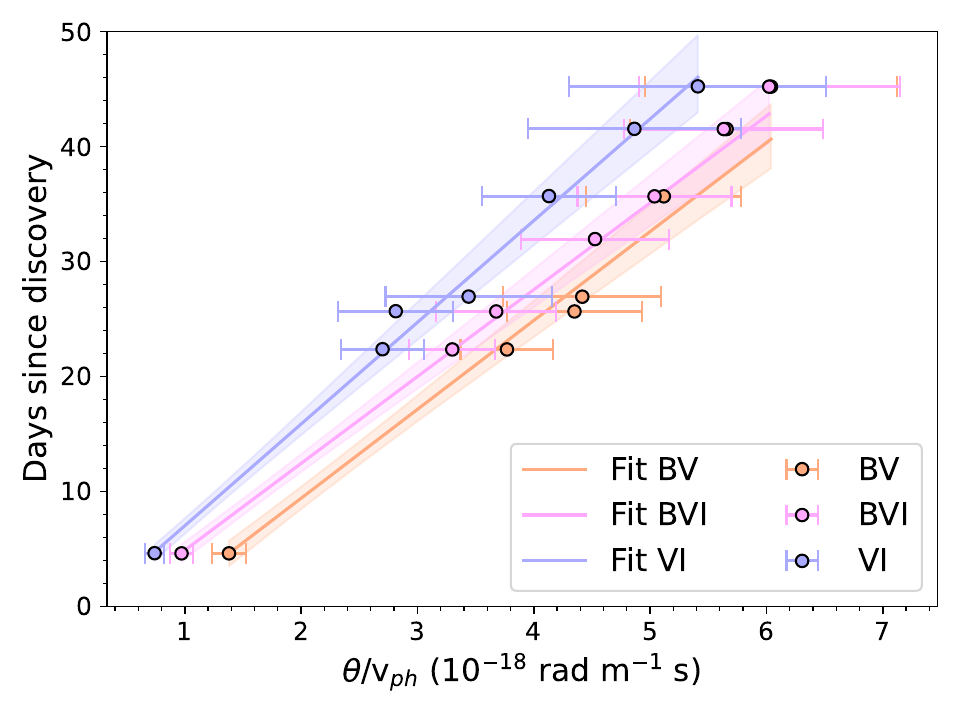}
    \vspace{0.5cm}
    \includegraphics[scale=0.6, clip, trim={0.cm 0cm 0cm 0.5cm}]{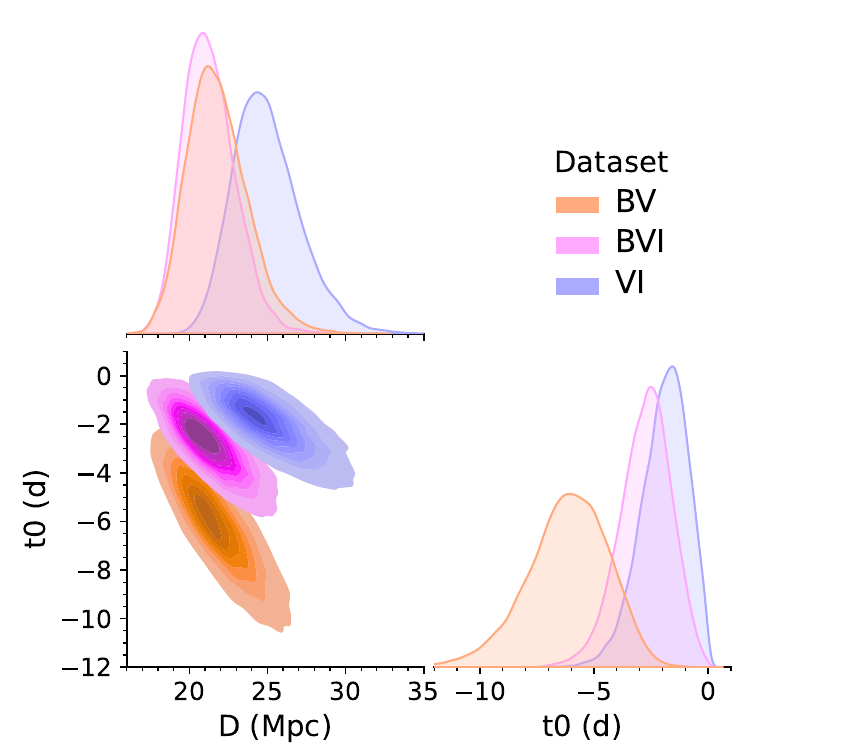}
    \caption{Top panel shows the linear fit to ${\rm t}$ vs $\theta/v_{ph}$ for the three filter combinations \{$BV$\}, \{$BVI$\} and \{$VI$\} to determine the explosion epoch and distance. Bottom panel shows the one and two dimensional projections of the posterior probability distributions of D and t$_0$ for the three filter sets in the corner plot.}

    \label{epm_results}
\end{figure}

A linear fit is performed on $t$ vs $\theta/v_{ph}$ to derive the distance, following the equation:

\begin{equation}
    t = {\rm D}(\theta/v_{ph}) + t_0
	\label{eq_epm}
\end{equation}

\noindent
where the slope of the linear equation gives the distance, and the y-intercept provides the time of explosion $t_0$. We used \texttt{emcee} \citep{Foreman2013} to perform the linear fit, and the best fit to the \{$BV$\}, \{$BVI$\} and \{$VI$\} filter sets, along with their 1 $\sigma$ confidence intervals, are shown in the upper panel of Figure \ref{epm_results}. The bottom panel of this figure shows the joint likelihoods of the parameters t$_0$ and D for the three filter sets along with the marginalised likelihood functions. The distances are estimated as the mean and standard deviation of the marginalised functions, which are 21.8 $\pm$ 1.9 Mpc, 21.3 $\pm$ 1.7 Mpc and 24.5 $\pm$ 2.2 Mpc for \{$BV$\}, \{$BVI$\}, and \{$VI$\} filter sets, respectively. The corresponding intercept values are $-$6.2 $\pm$ 1.8, $-$2.8 $\pm$ 1.1, and $-$1.9 $\pm$ 1.0 days with respect to the discovery date from ATLAS. We will use the distance and t$_0$ obtained from the \{$BVI$\} filter set in the rest of the paper. 
Thus, the estimated explosion epoch from EPM is JD 2458133.4$\pm$1.1 (2018 January 14.9). This value is consistent with those estimated from last non-detection and \texttt{SNID} within the errors, and we will use the EPM estimated explosion epoch throughout this work. The EPM estimated distance (21.3 $\pm$ 1.7 Mpc) is small compared to the Virgo infall distance; however it is in agreement with the Cosmicflows-3 distance within the errors. We will use the EPM derived distance in the rest of this paper.

\subsection{Extinction}
\label{extinction}

\begin{figure}
    \includegraphics[scale=0.45, clip, trim={0.cm 0cm 0cm 1.5cm}]{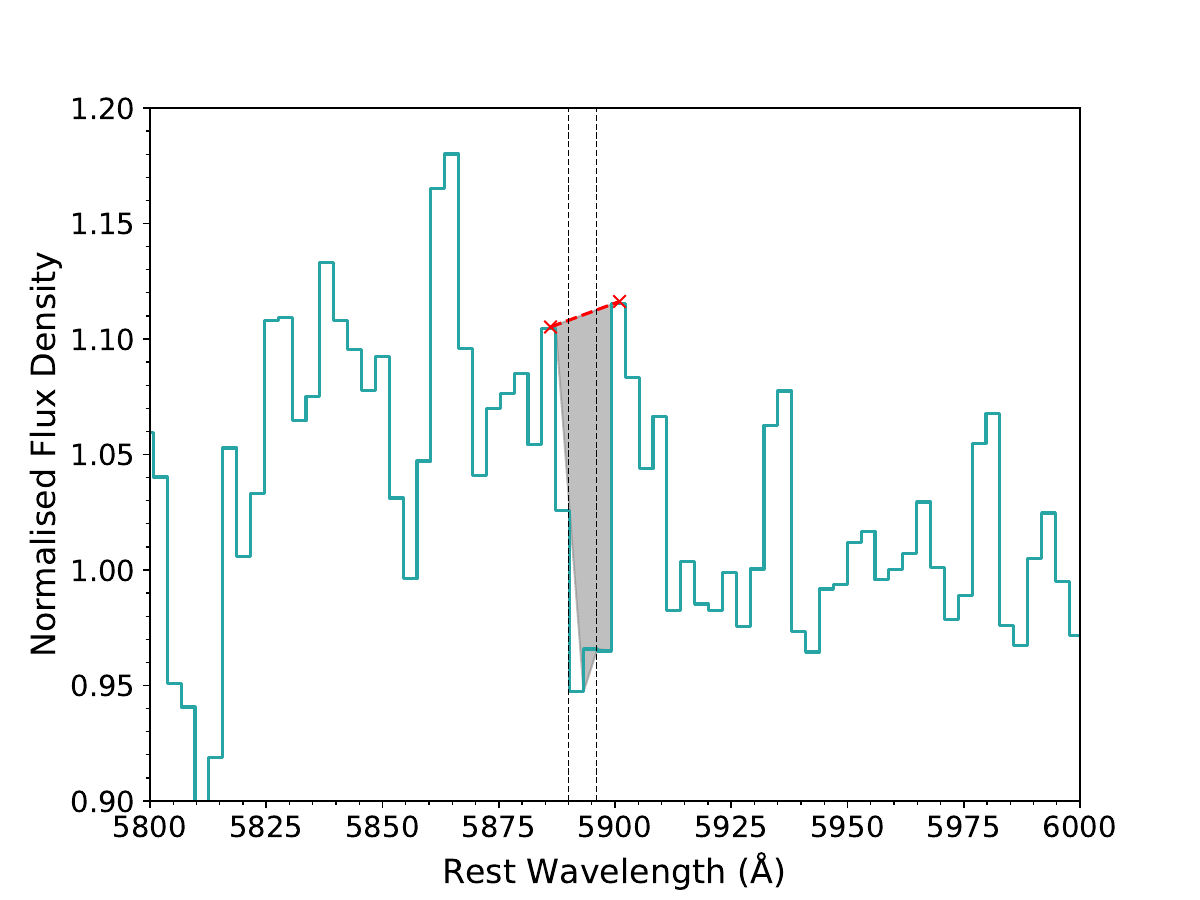}
    \caption{Cut-out of the SALT spectrum showing the blended \ion{Na}{id} lines.}
    \label{fig:extinction}
\end{figure}

The extinction along the line of sight to the SN due to dust in both the Milky Way (MW) and the host galaxy plays a crucial role in studying the intrinsic nature of the event. Based on the infrared dust maps provided by \cite{Schlafly2011}, the Galactic reddening for SN~2018is is E($B-V$)$_{\rm MW}$ = 0.0708$\pm$0.0003 mag.

The empirical relations correlating the \ion{Na}{id} ($\lambda\lambda$ 5890, 5896) equivalent width with the colour excess (e.g. \citealt{Munari1997, Poznanski2012}) have been used for various SNe (e.g., \citealt{Nakaoka2018,Meza2024}). The rest-frame SALT spectrum of SN~2018is, obtained 3.9 days post-explosion, shows a conspicuous feature around 5893~\AA{} (as shown in Fig \ref{fig:extinction}). The equivalent width of this feature is estimated to be 1.26 $\pm$ 0.01 \AA. Using equation 9 from \cite{Poznanski2012}, we estimate E($B-V$)$_{\rm host}$ to be 0.42 $\pm$ 0.08 mag. Given that \cite{Poznanski2012} used the dust maps from \cite{Schlegel1998}, we further multiplied E($B-V$)$_{\rm host}$ by the re-normalisation factor of 0.86, as suggested by \cite{Schlafly2011}, resulting in E($B-V$)$_{\rm host}$ = 0.36 $\pm$ 0.07 mag. We note that the accuracy of using the \ion{Na}{id} line in low-resolution spectra has been challenged \citep{Poznanski2011, Phillips2013}, hence the E($B-V$)$_{\rm host}$ estimated using this method can only be considered as an upper limit.

\begin{figure*}
    \includegraphics[scale=0.9, clip, trim={0.4cm 0.2cm 0.0cm 1.1cm}]{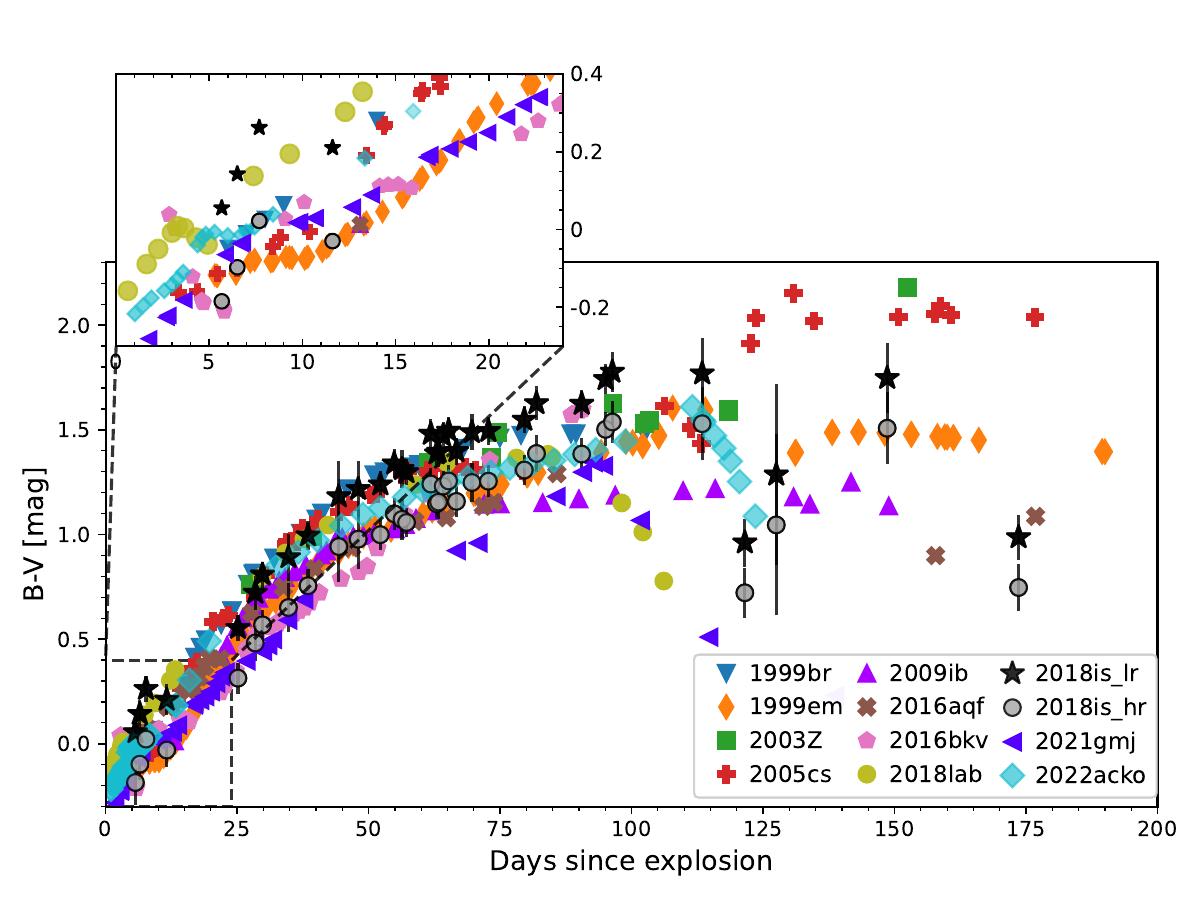}
    \caption{The ($B-V$) colour evolution of SN~2018is, corrected for the high (A$_V$=1.34 mag) and low (A$_V$=0.59 mag) extinction scenarios, are compared with other SNe II.}
    \label{fig:bvcolor_expl}
\end{figure*}

The $V$-band magnitudes at 50 days, considering only Milky Way (MW) extinction as well as MW plus host extinction, are $-$13.94 and $-$15.08 mag, respectively. Furthermore, using \texttt{SNID}, the early spectra of SN~2018is were found to closely match those of the LLSN 2006bc. This, along with the low expansion velocities discussed in Section~\ref{sec6.1}, suggests that SN~2018is belongs to the LLSNe II category. Therefore, as an independent estimate of the host galaxy extinction, we compared the MW-corrected colour evolution of SN~2018is with that of the well-observed LLSN II, SN~2005cs, and calculated the shift required for SN~2018is's colour to align with that of SN~2005cs. The E($B-V$)$_{\rm host}$ estimated using this method is 0.12 $\pm$ 0.06 mag, which is three times lower than the value obtained from the \ion{Na}{id} feature. The resulting colour curve, using both estimates, is shown in Fig.~\ref{fig:bvcolor_expl}.

Given the caveats associated with different methods for calculating host-galaxy extinction, and with no method being definitively preferable, we will consider two values of extinction in this study. Therefore, we adopt a total $E(B-V)_{\rm tot}$ of 0.19$\pm$0.06 as the low-extinction estimate and 0.43$\pm$0.07 mag as the high-extinction estimate, with the corresponding A$_{V}$ values being -- A$_{\rm V,lr}$= 0.59$\pm$0.19 mag and A$_{\rm V,hr}$= 1.34 $\pm$ 0.21 mag, assuming total-to-selective extinction ratio, R$_{\rm V}$=3.1, and using the extinction function from \cite{Gordon2023}.  

\section{Light curve evolution}
\label{Sec4:lightcurve}

\begin{figure}
    \includegraphics[width=\columnwidth, clip, trim ={0cm 0.0cm 0cm 0.cm}]{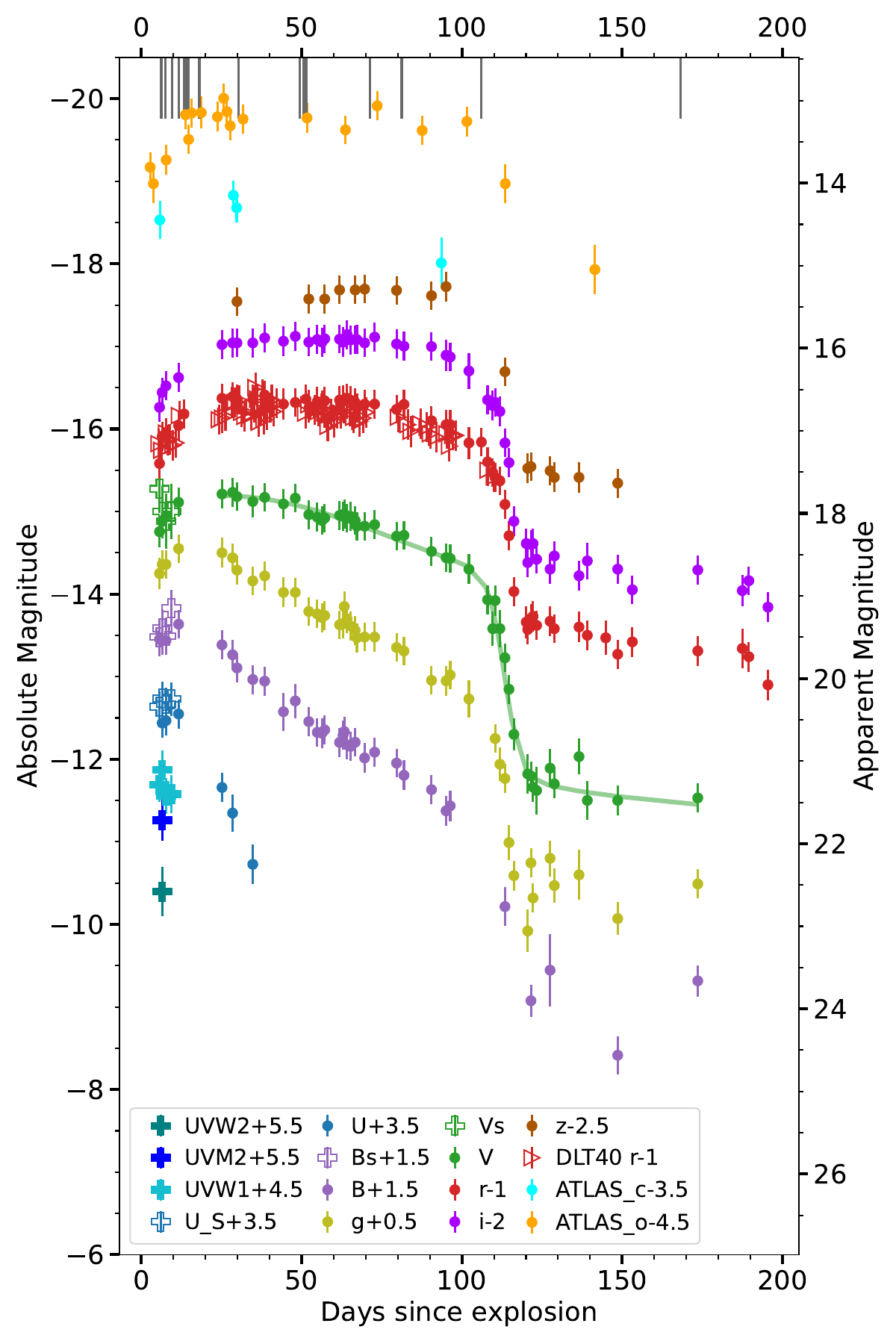}
    \caption{Absolute (corrected for A$_V$ = 1.34 mag) and apparent magnitude UV and optical light curves of SN~2018is, shifted arbitrarily for clarity. Vertical gray lines mark the epochs of spectroscopic observations. Parameterised fit to the $V$-band light curve \citet{Valenti2016} is also shown.}
    \label{fig:18is_LC}
\end{figure}

\begin{figure}
    \includegraphics[scale=0.57, clip, trim={0.3cm 0.cm 0.0cm 0.3cm}]{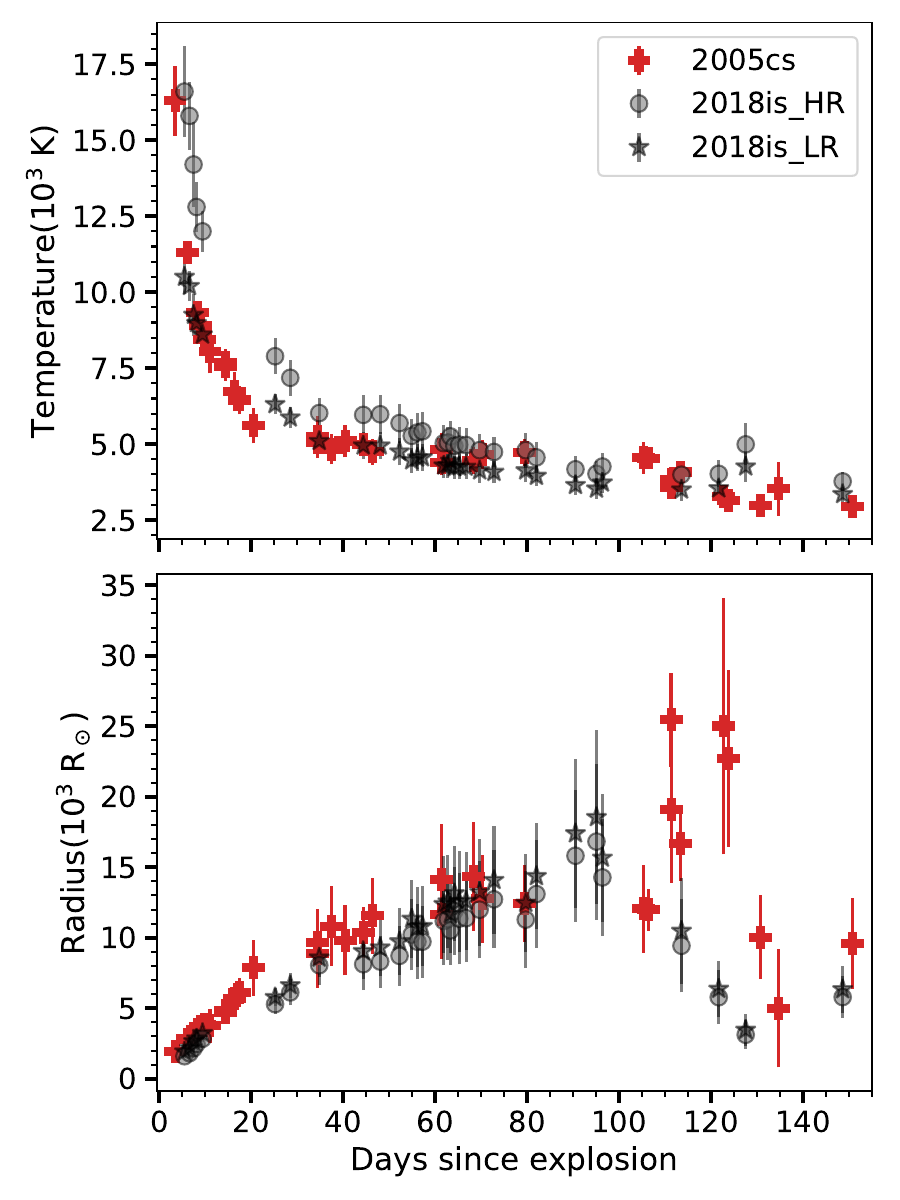}
    \caption{The temperature and radius evolution of SN~2018is obtained by fitting a blackbody to the SED constructed from the observed photometric fluxes.}
    \label{fig:18is_temp_rad}
\end{figure}

The multiband optical and UV absolute and apparent magnitude light curves of SN~2018is are shown in Fig.~\ref{fig:18is_LC}. The high reddening value, which is converted to extinction estimates in the different bands using the dust-extinction law from \cite{Gordon2023}, has been used to derive the absolute magnitudes. Initially, the light curves exhibit an increase in brightness for the first 10 days following the assumed explosion epoch, followed by a decrease in the bluer filters. Meanwhile, in the $r$ and $i$ band,  the luminosity increases until 28 days and remain nearly constant thereafter. The plateau phase, lasting nearly 100 days, is relatively short for typical LLSNe II \citep{Pastorello2009, Spiro2014}. The $V$-band light curve shows a linear decline with a slope of 1.04 $\pm$ 0.03 mag (100 d)$^{-1}$, which is steeper than usual for LLSNe II \citep{Pastorello2009, Spiro2014}. Post-maximum decline rates in the $r$ and $i$ bands are 0.26 $\pm$ 0.02 mag (100 d)$^{-1}$ and 0.02 $\pm$ 0.02 mag (100 d)$^{-1}$, respectively, consistent with other LLSNe~II \citep{Pastorello2009, Spiro2014}.

Using Equation 1 from \cite{Valenti2016}, which is the same as Equation 4 from \cite{Olivares2010} without the Gaussian component, we fit the $V$-band light curve to derive parameters that can be compared to those of other LLSNe~II. From the best fit, over-plotted on the $V$-band light curve in Figure~\ref{fig:18is_LC}, we determined t$_{\rm PT}$, the time from explosion to the transition point between the end of the plateau and the start of the radioactive tail phase, to be 113.9$\pm$ 1.1 d. The parameter w$_0$ gives an estimate of the duration of the post-plateau decline until the onset of the radioactive tail phase to be (6$\times$w$_0$) 15 days. The parameter a$_0$ = 2.4$\pm$0.1\,mag, which quantifies the magnitude drop when the light curve transitions from the photospheric phase to the radioactive tail, is typical for SNe IIP \citep{Olivares2010}, although, under-luminous objects generally show a larger drop of about 3-5 mag (\citealt{Spiro2014,Valenti2016}). 

\begin{table}
\caption{Light curve slopes in different phases.}
\label{Tab:LC_slopes}
\renewcommand{\arraystretch}{1.1}
\centering
\begin{tabular}{llllll}
\hline
Band & t$_{\rm start}$ & t$_{\rm stop}$ & slope      \\
     &     (d)     &  (d)       & (mag/100d) \\
\hline
B    & 24.6 &   94.5 &	2.60 $\pm$ 0.05    \\
     & 120.9 &  173.0 &  0.11 $\pm$ 0.42   \\
\hline
g    & 24.6 &	 95.8 &	 2.10 $\pm$ 0.03 \\
     & 121.6 &  173.0 &	0.26 $\pm$ 0.15 \\
\hline
V    & 24.6 &   94.5 &	1.04 $\pm$ 0.03  \\
     & 119.9 &  173.0 & 0.51 $\pm$ 0.16 \\
\hline
r    & 24.6 &	 89.9 &	 0.26 $\pm$ 0.02  \\
     & 120.9 &  194.9 &	0.77 $\pm$ 0.08  \\
\hline
i    & 24.5 &	 89.9 &	 0.02 $\pm$ 0.02  \\
     & 122.7 &  194.9 &	0.55 $\pm$ 0.06  \\
\hline
\end{tabular}
 \end{table}

The decay rates in the tail phase of the bolometric light curve of LLSNe II are generally smaller than the 0.98 mag (100 d)$^{-1}$ expected from the $^{56}$Co to $^{56}$Fe decay assuming complete $\gamma$-ray trapping \citep{Pastorello2009, Spiro2014}. For SN~2018is, the decline rates in the tail phase for the $B$, $g$, $V$, $r$ and $i$-band light curves are 0.3 $\pm$ 0.4, 0.4 $\pm$ 0.2, 0.7 $\pm$ 0.2, 0.8 $\pm$ 0.1 and 0.6 $\pm$ 0.1 mag (100 d)$^{-1}$, respectively. The light curves of SN~2005cs exhibit similar decline rates between 140-320 d in the $BVRI$ bands, measured at 0.32, 0.46, 0.71, and 0.77 mag (100 d)$^{-1}$, respectively. \citep{Spiro2014} calculated the decline rate of SN~2003Z in the tail phase (>150 d) in $VRI$ bands to be 0.67, 1.05, and 0.58 mag (100 d)$^{-1}$, respectively. It has been proposed that the shallower slope in LLSNe II is due to an additional radiation source generated in the warm inner ejecta \citep{Utrobin2007}.

In order to compare the light curve and spectral properties of SN~2018is with other Type II SNe, we constructed a comparison sample constituting the normal luminosity Type II SN~1999em, and a number of intermediate and LLSNe II, whose distances, reddening and references are listed in Table~\ref{comparison_sample}. 

\subsection{Colour and temperature evolution}
The extinction-corrected colour evolution of SN~2018is is compared to a subset of SNe II from the comparison sample in Fig.~\ref{fig:bvcolor_expl}. In the low-extinction case, SN~2018is exhibits a colour that falls on the red end of the sample. If only the MW extinction correction were applied, the colour of SN~2018is would be 0.12 mag redder than the colour obtained using the low-extinction estimate. Studies by \cite{Pastorello2009} and \cite{Spiro2014} have noted that LLSNe II, when compared to normal SNe II, tend to have redder intrinsic colours. 

In SN~2018is, a rapid increase in the $B-V$ colour by around 0.2 mag is observed in the first 5 days, as shown in the zoom-in plot. This is followed by a slight decrease and then a rapid rise after 10 days. Similar behaviour is noted in the colour evolution of SNe~2018lab and 2022acko \citep{Pearson2023, Bostroem2023}. SNe~2018lab, 2021gmj and 2022acko exhibit a trend towards bluer colours after the end of the recombination phase, around 100 and 110 days, respectively, unlike SNe~2003Z and 2005cs, which evolves towards redder colours. However, due to the large error bars associated with the colour of SN~2018is after the recombination phase, it is difficult to ascertain whether it followed a pattern similar to SN~2005cs or SN~2022acko. When using the higher reddening estimate, SN~2018is is about 0.24 mag bluer than with the low reddening estimate, aligning more closely with the colour evolution of SNe~2016aqf, 1999em, and 2009ib \citep{Muller2020, Takats2015}.

As the SN expands and cools, the photospheric temperature and radius evolve over time. These parameters can be traced by fitting a blackbody model to the spectral energy distribution (SED) at different epochs. The SED is constructed using \texttt{SuperBol} \citep{Nicholl2018}, as described in Section~\ref{Sec6:Modelling}. The top and bottom panels of Figure~\ref{fig:18is_temp_rad} illustrate the temperature and the radius evolution of SN~2018is for both high and low reddening scenarios, compared to that of SN~2005cs. In the high reddening scenario, the temperatures during the first 60 days are higher than those of SN~2005cs. However, the temperature evolution in the low reddening scenario aligns more closely with SN~2005cs. Overall, SN~2018is shows a gradual temperature decline during the first 30 days, followed by a slower decline from around 6000\,K to 4000\,K over the next 100 days. The radius increases rapidly in the first 30 days and then continues to expand slowly, similar to the evolution observed in SN~2005cs. Around day 110, a rapid decline in radius is observed in SN~2018is, whereas SN~2005cs exhibits an increase in radius before declining, although we note that the error bars are large at these times. 

%%%%%%%%%%%%%%%%%%% Section: Absolute Magnitude and Ni mass
\subsection{Absolute magnitude comparison and \texorpdfstring{$^{56}$Ni}{Lg} mass}
\begin{figure}
	\includegraphics[scale=0.37, clip, trim={0.7cm 0.8cm 0cm 2.0cm}]{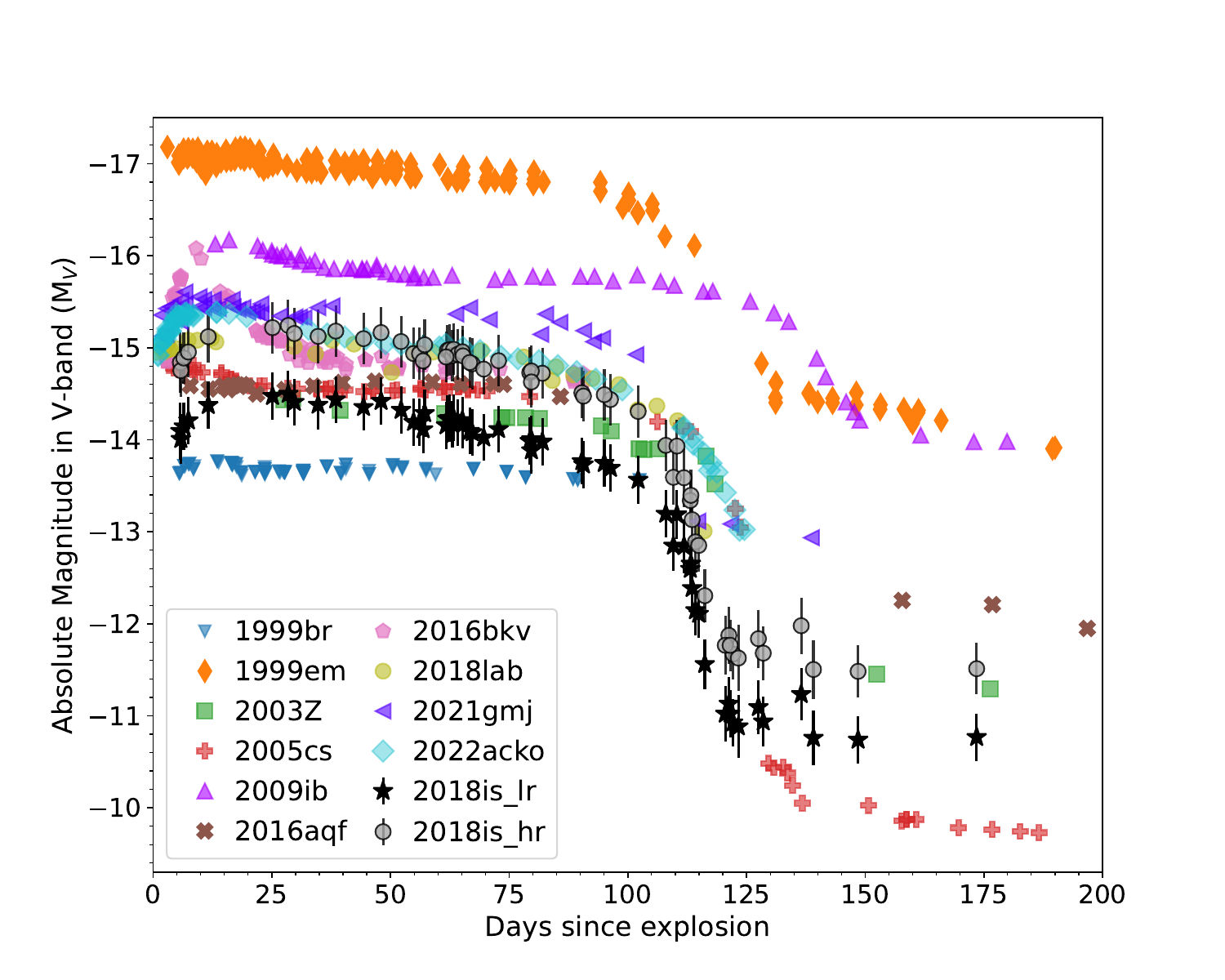}
    \caption{Comparison of absolute $V$-band light curves of SN~2018is with other SNe~II. The magnitudes are corrected for distance and reddening.}
    \label{fig:absmag_compare}
\end{figure}

In Figure~\ref{fig:absmag_compare}, the absolute $V$-band magnitude of SN~2018is is compared to a subset of SNe II from the comparison sample (Table~\ref{comparison_sample}). Typically, LLSNe II exhibit long plateaus (\textgreater 100 days), although also some normal luminosity SNe II (e.g. SN~2015ba, \citealt{Dastidar2018}) have extended plateaus. In H-rich SNe II, the plateau length is primarily influenced by the H envelope mass and the ejecta velocity, with a modest contribution from the energy released by $^{56}$Ni decay \citep{Kasen2009, Morozova2018, Kozyreva2019, Martinez2022}. The low ejecta velocity in LLSNe II, which is typically a factor of a few lower than in normal luminosity Type II SNe, results in higher ejecta densities. This slows the recombination wave, thereby extending the plateau duration. In normal luminosity Type II SNe, while higher $^{56}$Ni yields can extend the plateau length by 10–20\%\citep{Kasen2009}, the higher expansion velocity leads to lower densities and a faster-moving recombination wave, resulting in an overall shorter plateau. 

LLSNe~II events like SNe~2003Z and 2005cs exhibit longer plateaus compared to SN~2018is, with similar plateau luminosities, when the low reddening scenario is considered. In the high reddening scenario, the plateau luminosity of SN~2018is is a closer match to SNe~2018lab and 2022acko. The fall from the plateau in the case of SN~2018is is adequately sampled and has a similar plateau length as SN~2018lab. Moreover, from this figure, it is apparent that SNe II with the similar plateau luminosities can have a range of tail-luminosities ($-$9.5 to $-$12.5), depending on the amount of $^{56}$Ni produced in the explosion. 

For SN~2018is, the synthesised $^{56}$Ni mass is estimated from the tail bolometric luminosity, which is obtained from the tail $V$-band magnitudes using a bolometric correction (BC) from \cite{Hamuy2003}. The tail $V$-band magnitudes at two epochs, 148.6 and 173.6 d, were considered, and the corresponding mean bolometric luminosities are 8.3 $\pm$ 1.4 $\times$ 10$^{39}$ erg s$^{-1}$ and 1.6 $\pm$ 0.3 $\times$ 10$^{40}$ erg s$^{-1}$ for the low and high reddening scenarios, respectively. This results in a mean $^{56}$Ni mass of 0.0026 $\pm$ 0.0004 M$_\odot$ and 0.0051 $\pm$ 0.0009 M$_\odot$, for the low and high reddening scenarios, respectively. In addition, using BCs in the $Vri$ bands from \cite{Rodriguez2019}, determined using a larger sample of SNe II, the $^{56}$Ni masses are 0.0027 $\pm$ 0.0009, 0.0030 $\pm$ 0.0008, 0.0030 $\pm$ 0.0007 M$_\odot$, respectively, for the low reddening scenario and 0.0052 $\pm$ 0.0017, 0.0053 $\pm$ 0.0015, 0.0046 $\pm$ 0.0011 M$_\odot$, respectively, for the high reddening scenario, closely matching the earlier value. The weighted average of $^{56}$Ni mass using BCs from \cite{Rodriguez2019} are 0.0029 $\pm$ 0.0004 and 0.0049 $\pm$ 0.0008 M$_\odot$. Both values align well with those of other LLSNe~II.

\section{Key spectral features} \label{Sec5:spectral_features}
%%%*************** Fig: Spectral Sequence ******************
\begin{figure}
    \includegraphics[scale=0.34, clip, trim={1.6cm 1.0cm 0.5cm 2.9cm}]{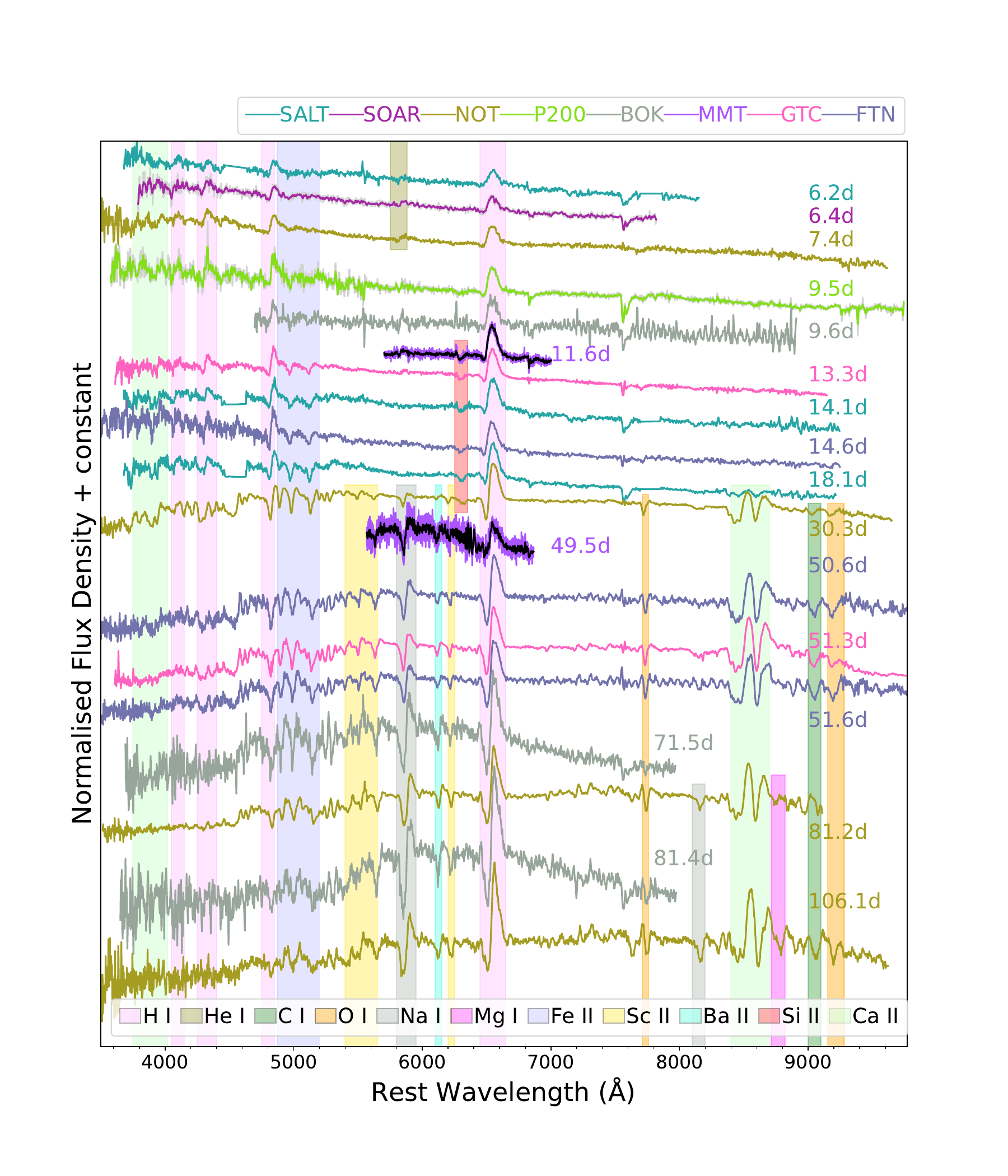}
    \caption{The spectral evolution of SN~2018is from 6.2 to 106.1\,day is shown and the prominent features are marked.}
    \label{fig:spec_seq1}
\end{figure}

\begin{figure*}
    \includegraphics[width=\textwidth, clip, trim={2.2cm 0.65cm 2.1cm 2.4cm}]{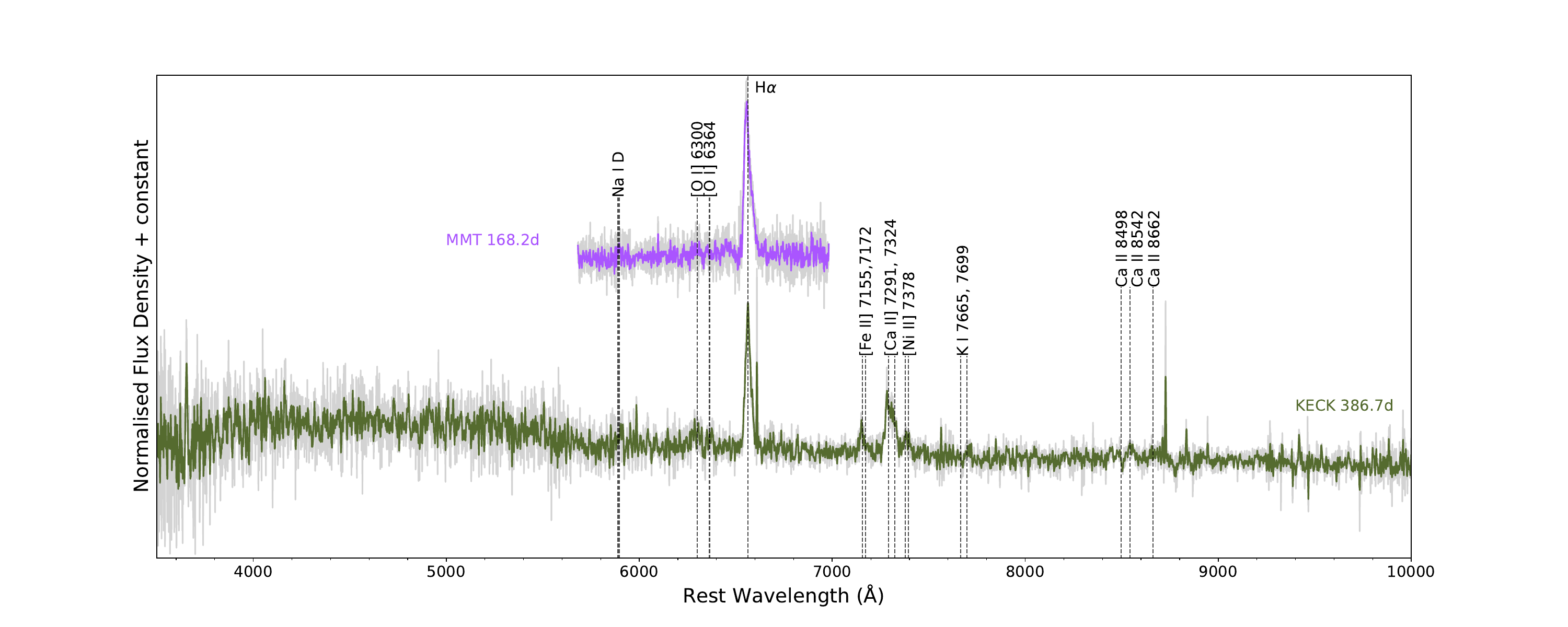}
    \caption{The nebular phase spectra at 168.2 and 386.7 d of SN~2018is are shown and the prominent features are marked.}
    \label{fig:spec_seq3}
\end{figure*}

The first spectrum of SN~2018is was obtained 6.2 days post-explosion. The spectral evolution from 6.2--106.1 days of SN~2018is is shown in Figure~\ref{fig:spec_seq1}. Until 7.4 d, the spectra of SN~2018is show a blue continuum with weak and shallow P~Cygni profiles of the Balmer H lines. A weak \ion{He}{i} $\lambda$5876 feature is visible until 7.4\,days, after which it disappears. The \ion{Fe}{ii} $\lambda$5018 line emerges as early as 9.5\,days. An absorption feature at $\lambda$6300, which we identify as \ion{Si}{ii} $\lambda$6355, is discernible between 11.6 and 30.3 days, disappearing afterwards. This feature was also detected in the early spectra of SN~2005cs \citep{Pastorello2006}. We attribute this feature to \ion{Si}{ii}, given its velocity similarity with other metal lines (around 3000 km s$^{-1}$ at 11.6 days and 1300 km s$^{-1}$ at 30.3 days).

During the photospheric phase (30.3 days onwards), the spectral evolution slows down. The H Balmer absorption strengthens due to recombination, and the decreasing photospheric temperature leads to increased metal line formation, escalating the line blanketing. Strengthening of the NIR \ion{Ca}{ii} triplet at around $\lambda$8500, \ion{Na}{id}, \ion{Sc}{ii} $\lambda\lambda$5527, 6246, \ion{Ba}{ii} $\lambda$6142 and \ion{O}{i} $\lambda$7774 lines are also observed. Two absorption features appear redwards of 9000~\AA\,from the 30.3 day spectrum, which we identify as \ion{C}{i} $\lambda$9100 and \ion{O}{i} $\lambda$9263. The \ion{H}{$\alpha$} absorption feature becomes structured starting from 71.5 days, a characteristic observed during the late plateau phase of LLSNe II. This is primarily attributed to the formation of \ion{Ba}{ii} $\lambda$6497 at lower temperatures \citep{Lisakov2017}. In contrast, standard luminosity SNe II, with their higher expansion velocities, display broader and more blended absorption features. The more complex \ion{H}{$\alpha$} profile observed in LLSNe II is a consequence of the slower expansion velocities, which prevent the line blending that occurs at higher velocities. In addition to the slower expansion rates influencing the \ion{H}{$\alpha$} profile, the lower production of $^{56}$Ni in LLSNe II results in less efficient heating of the ejecta, making the \ion{Ba}{ii} line more prominent compared to normal SNe II. Additionally, the \ion{Mg}{i} feature is discernible in the 81.2 and 106.1 day spectra, appearing redward of the \ion{Ca}{ii} NIR triplet.

Two nebular phase spectra of SN~2018is, at 168.2 days and 386.7 days, are shown in Figure~\ref{fig:spec_seq3}. In the 386.7 day spectrum, besides prominent \ion{H}{$\alpha$} emission, the strongest feature is the $\lambda$7300 doublet emission, associated with [\ion{Ca}{ii}] lines $\lambda\lambda$7291,7324. An unblended emission blueward of [\ion{Ca}{ii}] lines is identified as [\ion{Fe}{ii}] $\lambda$7155, observed in the late time spectra of several SNe~II (e.g. 2016aqf, 2016bkv). The [\ion{O}{i}] $\lambda\lambda$6300, 6364 doublet can be identified in this spectrum, while the \ion{Ca}{ii} NIR triplet is inconspicuous. A weak [\ion{Ni}{ii}] $\lambda$7378, produced by stable $^{58}$Ni, a nuclear burning ash, is identifiable. The presence of [\ion{Ni}{ii}] is a unique signature of neutron excess in the innermost Fe-rich layer, hence is an crucial tracer of explosive burning conditions \citep{Wanajo2009}. This feature has been analysed and discussed in more detail in Section \ref{sec7.3}. 

\subsection{Features in the NIR spectrum}
\begin{figure*}
    \includegraphics[width=\textwidth, clip, trim={2.2cm 0.65cm 2.1cm 2.4cm}]{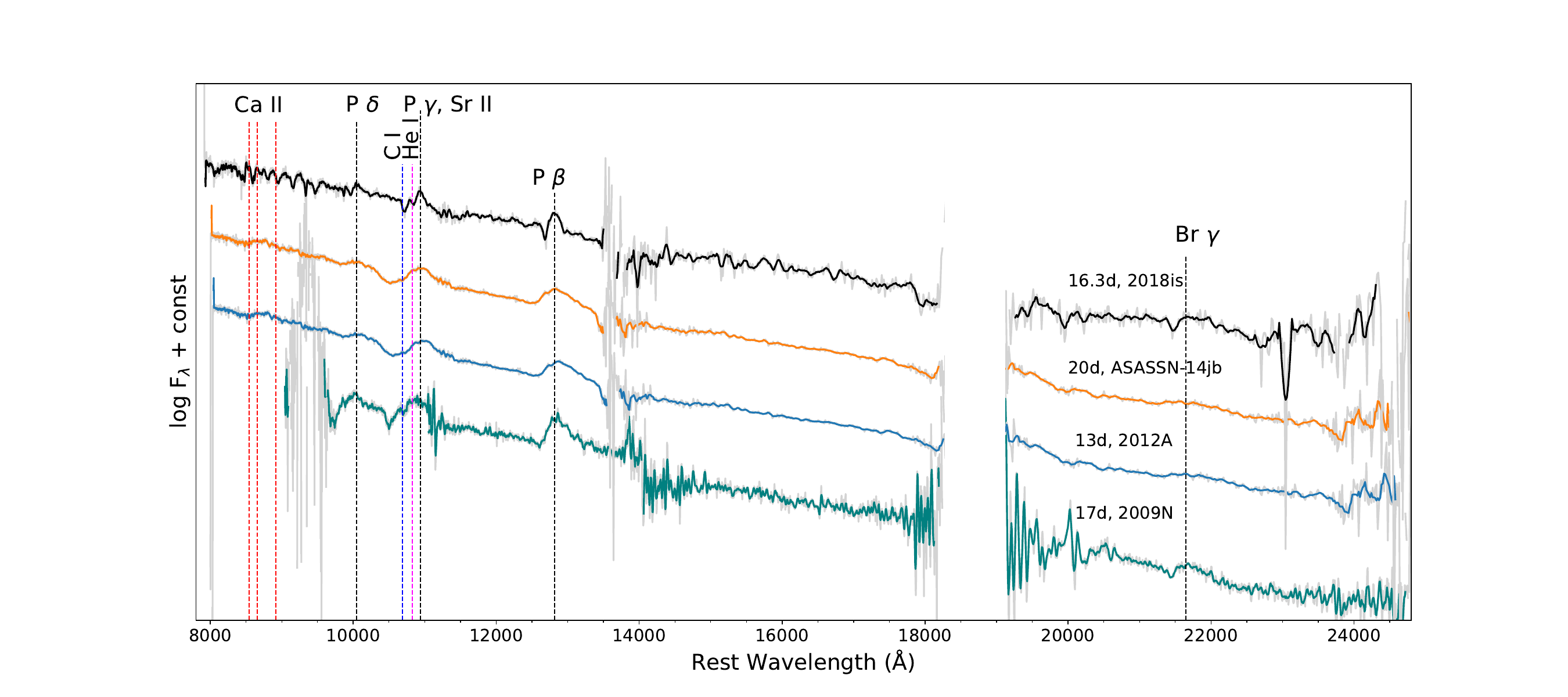}
    \caption{The +16.3 d NIR spectrum of SN~2018is is compared to intermediate luminosity Type II SNe~2009N, 2012A as well as a LLSN ASASSN-14jb.}
    \label{fig:nir_spec}
\end{figure*}

The 16.3 day NIR spectrum of SN~2018is is shown in Fig.~\ref{fig:nir_spec}, alongside those of SN~2009N \citep{Takats2014} and two other Type II SNe from \cite{Davis2019} for comparison. The spectrum of SN~2018is displays the \ion{Ca}{ii} NIR triplet, partial blends of \ion{P}{$\gamma$} $\lambda$10940 and \ion{Sr}{ii} $\lambda$10915 and \ion{P}{$\beta$} $\lambda$12820, and \ion{Br}{$\gamma$} $\lambda$21650. The location of \ion{P}{$\alpha$} $\lambda$18750 is contaminated by strong telluric absorption, making it indiscernible. 

In normal luminosity Type II SNe, such as SN~2012A \citep{Tomasella2013}, \ion{P}{$\gamma$} and \ion{He}{i} $\lambda$10830 lines are strongly blended which gives rise to a broad absorption feature. However, LLSNe II, with their narrow features, exhibit partially unblended \ion{P}{$\gamma$}, \ion{Sr}{ii}, and \ion{He}{i} lines, as seen in the spectra of SNe~2009N and 2018is. \ion{C}{i} $\lambda$10691 also contributes to the broad absorption which is clearly absent in SN~2018is. In normal Type II SNe with a relatively high $^{56}$Ni yield, the absorption in this region is primarily dominated by \ion{He}{i}, which is produced via non-thermal excitation when $^{56}$Ni is located close to the He region in the ejecta \citep{Graham1988}. However, in LLSNe II, where the $^{56}$Ni yield is minimal, \ion{He}{i} is not expected to significantly contribute to this feature. Instead, \ion{P}{$\gamma$}, \ion{Sr}{ii} and \ion{C}{i} dominate the absorption \citep{Pastorello2009}. 

The \ion{P}{$\beta$} line, visible in all the SNe, exhibits a symmetric P~Cygni profile, which is dominated by its emission feature with very little absorption in SNe~2012A, ASASSN-14jb, while in SN~2018is a narrow absorption component is visible. The Brackett series hydrogen line, \ion{Br}{$\gamma$} is prominent in absorption in both SNe~2009N and 2018is. 

Overall, in the NIR spectra, the emission features of the comparison SNe are broad at early times, even in the case of ASASSN-14jb, which is a LLSN (M$_V^{\rm peak}$ = $-$14.93 mag). In contrast, the features of SN~2018is are remarkably narrower, which is also evident in the optical spectra comparison discussed in Section~\ref{sec6.1}.

\begin{figure*}
    \includegraphics[width=0.79\columnwidth, clip, trim = {0.2cm 0.2cm 0cm 2.4cm}]{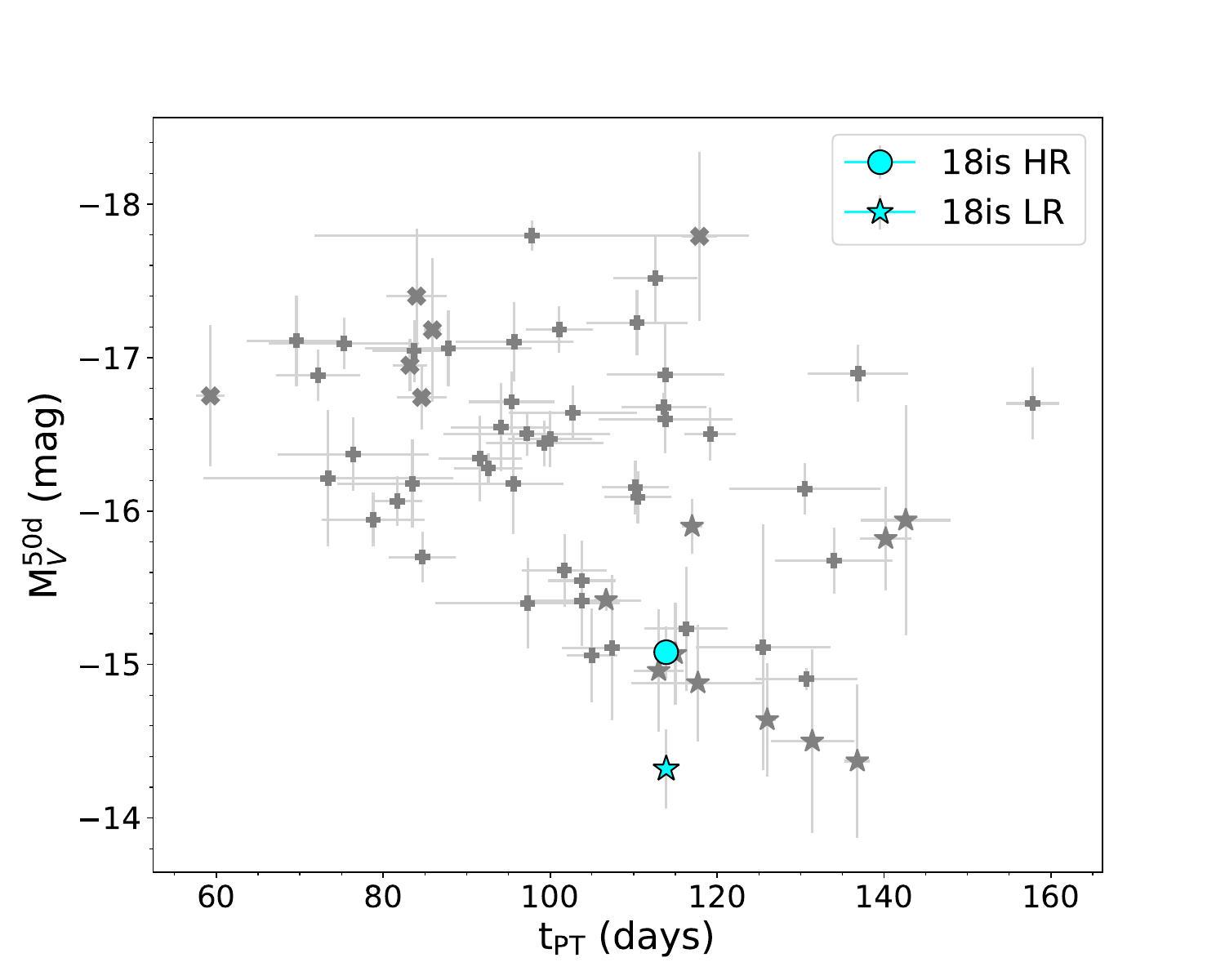}
    \hspace*{-0.83cm}
    \includegraphics[width=0.75\columnwidth, clip, trim = {1.5cm 0.2cm 0cm 2.4cm}]{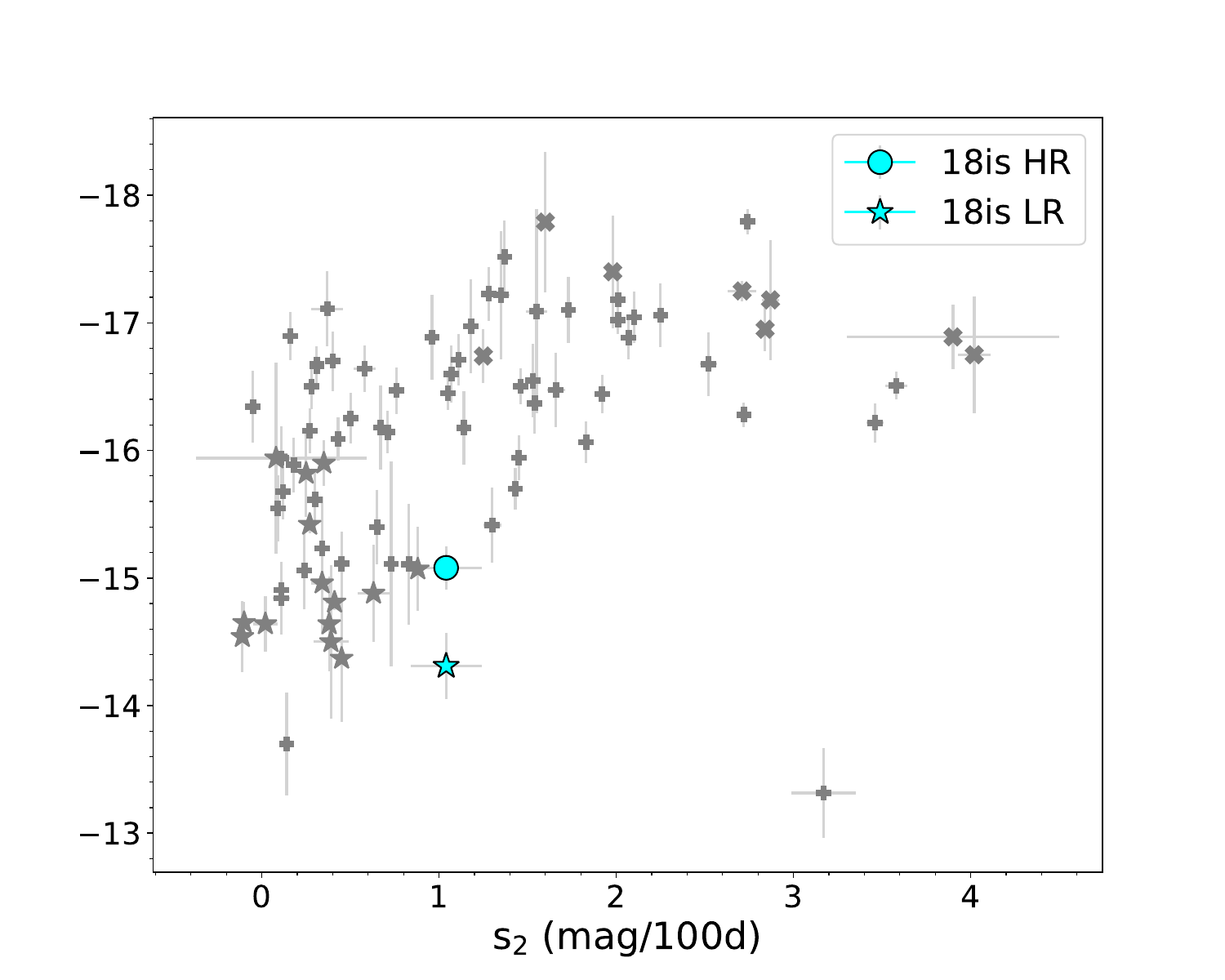}
    \hspace*{-0.83cm}
    \includegraphics[width=0.75\columnwidth, clip, trim = {1.5cm 0.2cm 0cm 2.4cm}]{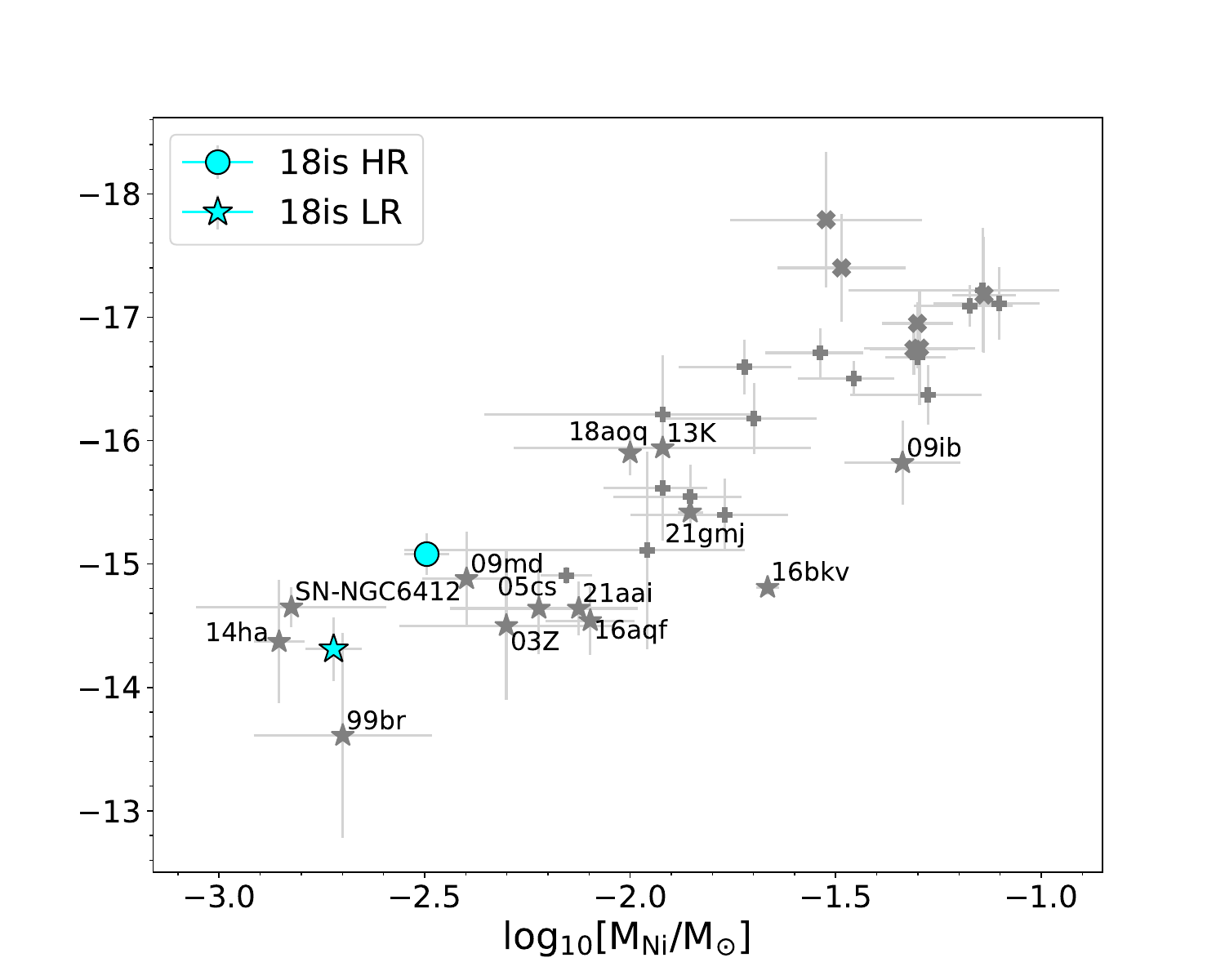}
    \caption{The position of SN~2018is on the absolute magnitude at 50 day (based on the low reddening (LR) and high reddening (HR) scenarios) vs. t$_{\rm PT}$,  $V$-band slope (s$_2$) and $^{56}$Ni mass plot are shown alongside other SNe II. The sample from \citet{Anderson2014} and \citet{Valenti2016} are marked with a `+', from \citet{Dastidar2024} with an `x', and from this work with a `$\star$'.}
    \label{fig:Mv50_s50_tpt}
\end{figure*}

\section{Juxtaposition with other type II SNe}
\label{sec6}
\subsection{Comparison based on photometric parameters}
We compare a few photometric parameters of SN~2018is with those of a sample of SNe II from the literature \citep{Anderson2014, Valenti2016, Dastidar2024}. In Fig.~\ref{fig:Mv50_s50_tpt}, the absolute magnitude at 50 day (M$_V^{\rm 50d}$) is plotted against t$_{\rm PT}$, slope during the plateau phase (s$_2$), and log(M$_{\rm Ni}$). A general trend is observed where higher luminosity SNe~II tend to have a shorter and steeper plateau, as well as a larger $^{56}$Ni mass yield, consistent with findings from earlier studies (e.g. \citealt{Anderson2014,Valenti2016}). While the absolute magnitude in the high-reddening scenario for SN~2018is fits this trend, in the low-reddening scenario, SN~2018is appears slightly offset from the general trend in the M$_V^{\rm 50d}$ vs. t$_{\rm PT}$ and s$_2$ plots.  

\begin{figure}
    \includegraphics[width=1.2\columnwidth, clip, trim = {2.2cm 0.7cm 0cm 2.4cm}]{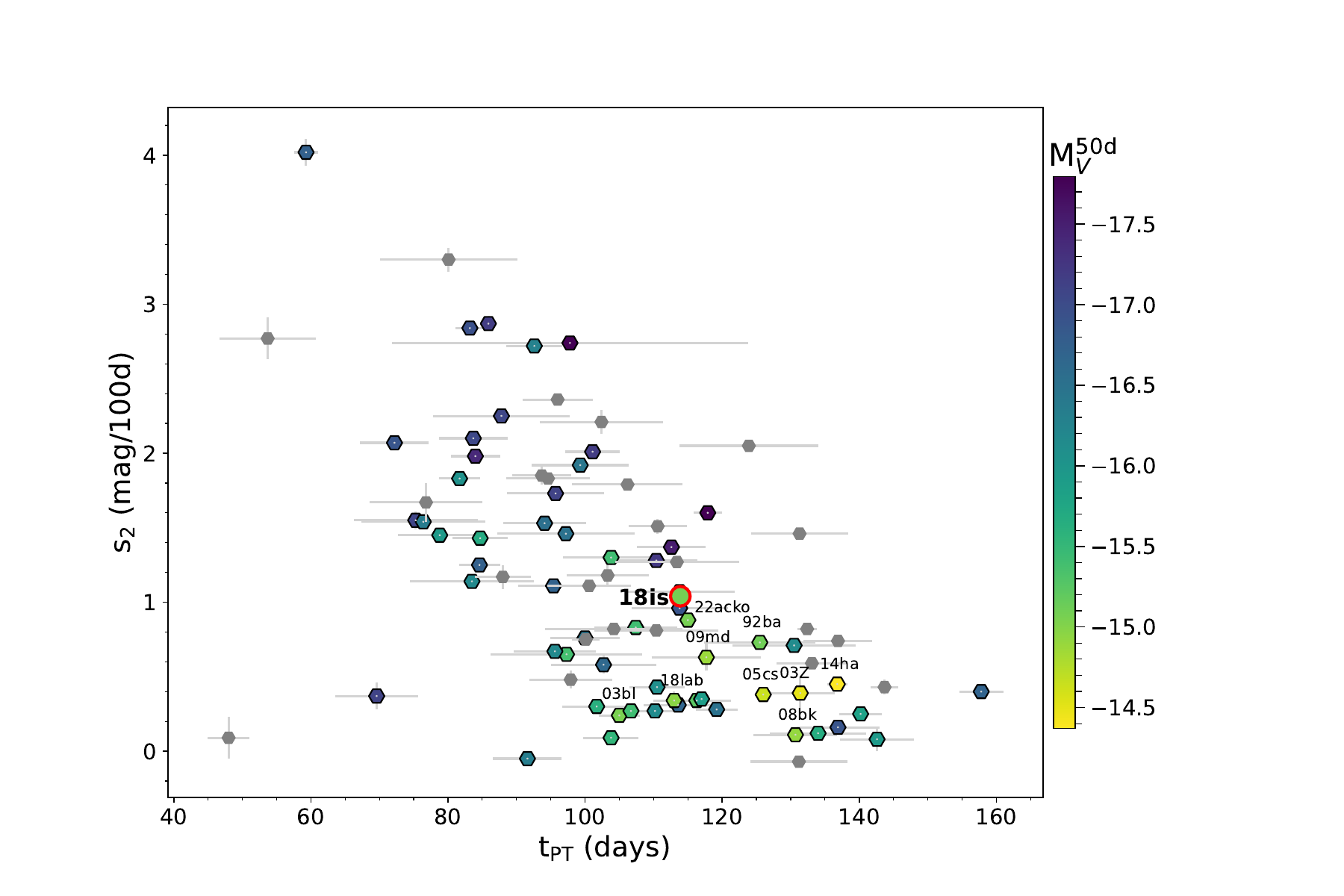}
    \caption{Position of SN~2018is on the $V$-band slope (s$_2$) vs. t$_{\rm PT}$ plot, alongside other SNe II. The points are colour-coded with M$_V^{\rm 50d}$ values. SNe for which M$_V^{\rm 50d}$ is not available are shown in gray. SN~2018is is colour-coded with M$_V^{\rm 50d}$ based on the high reddening scenario.}
    \label{fig:s50_tpt}
\end{figure}

In Figure~\ref{fig:s50_tpt}, we have plotted t$_{\rm PT}$ vs s$_2$ for the SNe II sample from the literature alongside SN~2018is. These parameters are known to exhibit a negative correlation, where SNe II with longer plateaus tend to have shallower decline rates. The figure is colour-coded by the $V$-band absolute magnitude at 50 day. SNe for which M$_V^{\rm 50d}$ is not available are shown in gray. While SN~2018is follows the overall trend, its decline rate is higher than that of other LLSNe II. It also has a shorter t$_{\rm PT}$ similar to SNe~2018lab and 2022acko\footnote{The t$_{\rm PT}$ of SNe~2018lab and 2022acko are approximate values obtained from \cite{Pearson2023} and \cite{Bostroem2023} as estimations using fitting were not possible in these cases due to absence of tail phase $V$-band data.}. 

\begin{figure}
	\includegraphics[width=1.1\columnwidth, clip, trim = {0cm 0cm 0cm 0cm}]{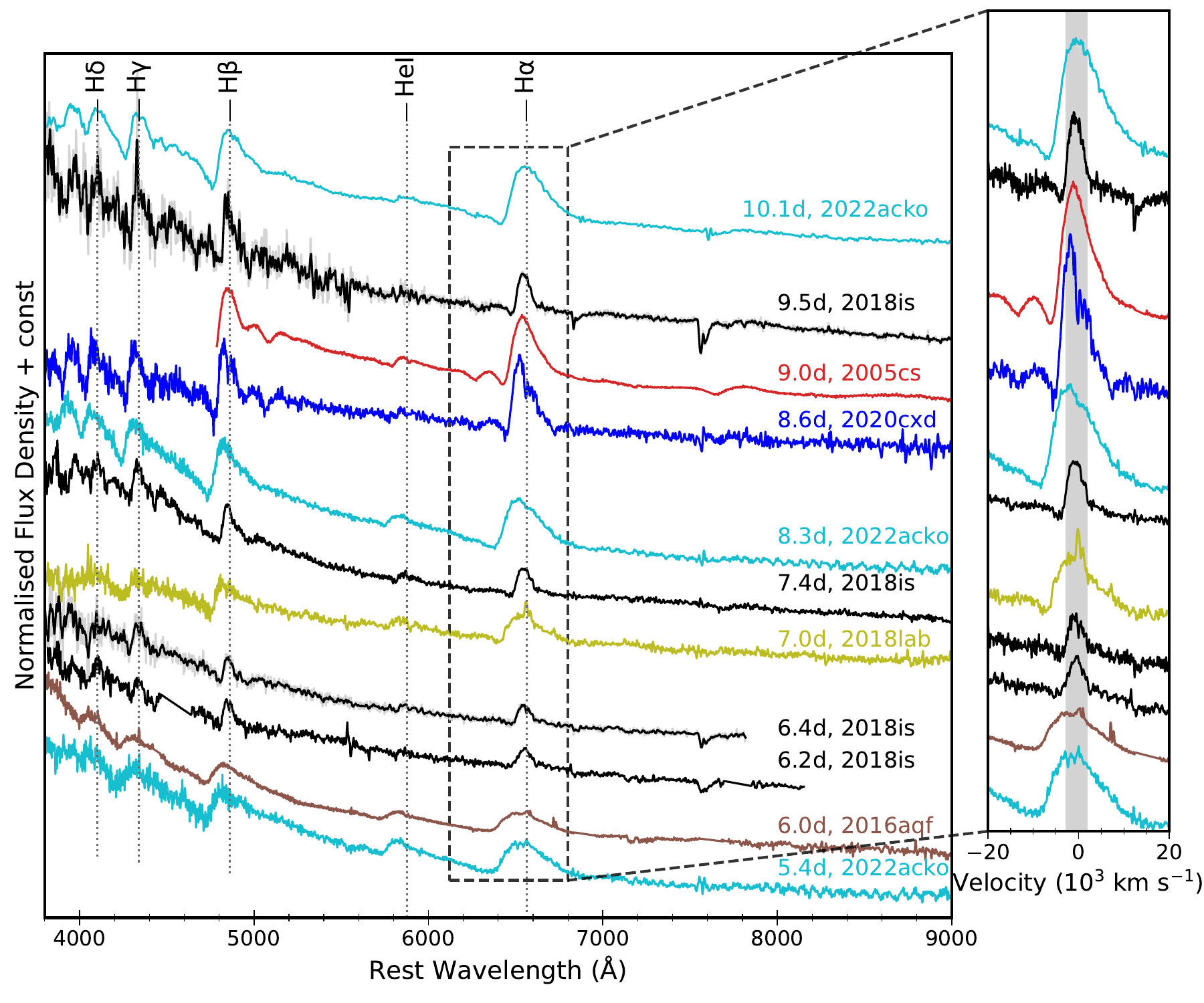}
    \caption{Early spectra of SN2018is are compared to spectra of other LLSNe II at similar epochs. The right panel shows the \ion{H}{$\alpha$} region in velocity space, with the shaded area representing the FWHM of the \ion{H}{$\alpha$} line for SN~2018is.}
    \label{fig:spec_compare_early}
\end{figure}

\begin{figure}
    \centering
    \includegraphics[scale=0.44, clip, trim={1.cm 5.3cm 1.5cm 4.9cm}]{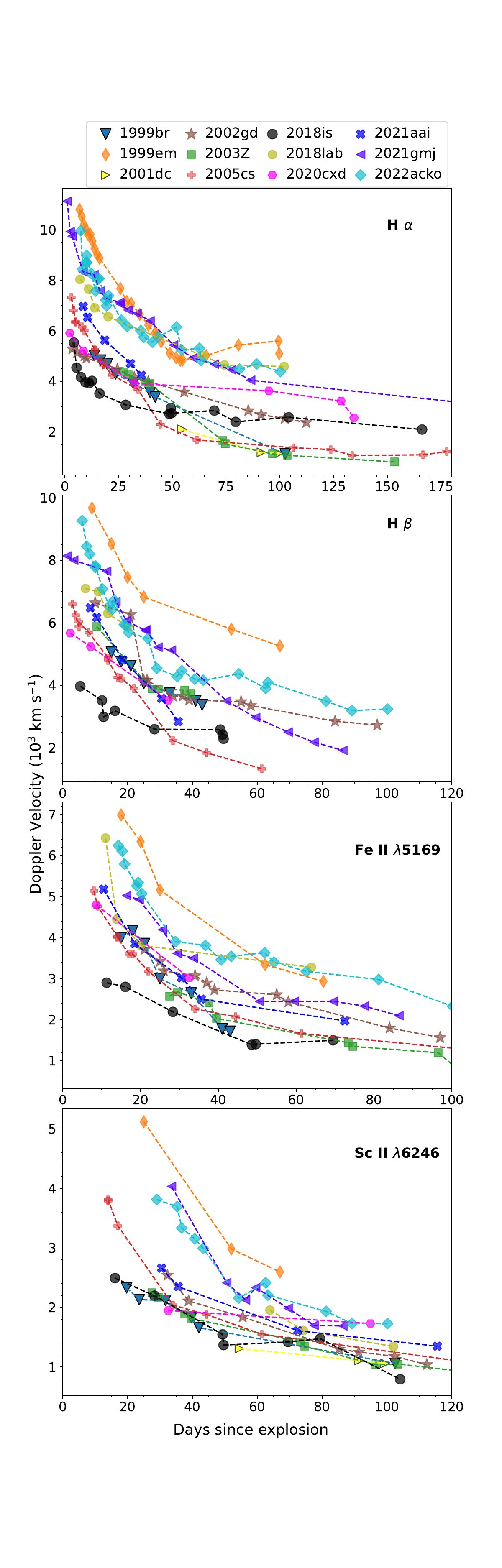}
    \caption{The velocity evolution of the H Balmer and metal lines are compared with a sample of low and standard luminosity SNe~IIP.}
    \label{fig:exp_vel_compare}
\end{figure}

\subsection{Comparison based on spectroscopic features}
\label{sec6.1}
Figure~\ref{fig:spec_compare_early} shows the comparison of the early spectra of SN~2018is with similar epoch spectra of other LLSNe~II. All the spectra exhibit a blue continuum with superimposed P Cygni profiles of H Balmer lines, featuring weak absorption components. The emission components of the H Balmer lines in SN~2018is are narrower than those in the spectra of the comparison sample, indicating a lower expansion velocity of the ejecta compared to the others.

The velocities of various elements in the ejecta, such as \ion{H}{$\alpha$}, \ion{H}{$\beta$}, \ion{Fe}{ii} $\lambda$5169, and \ion{Sc}{ii} $\lambda$6246 lines, were estimated from the position of their absorption minima and are compared to a subset of SNe from the comparison sample as shown in Figure~\ref{fig:exp_vel_compare}. Before 30\,days, the H Balmer line velocities of SN~2018is are the lowest among the comparison SNe. After 30\,days, the H Balmer line velocities settle at around 3000 km s$^{-1}$ with little evolution thereafter. In contrast, the velocities of the H Balmer lines in SNe~1999br, 2003Z and 2005cs, which were higher in the early phases, drop below those of SN~2018is at later phases.

Meanwhile, SNe~2002gd, 2020cxd \citep{Valerin2022}, 2021gmj and 2022acko also exhibit a flattening in velocity evolution after an initial rapid decline, similar to SN~2018is. This flattening could occur if the inner layers of the SN ejecta are relatively H-poor, causing the H Balmer absorption to originate from the H-rich outer (and therefore higher-velocity) ejecta layers even during the later phases \citep{Faran2014}. This scenario is typical when the pre-SN star has a low H-envelope mass, as proposed for fast-declining SNe II. However, in the case of LLSNe II, the formation of \ion{Ba}{ii} $\lambda$6497 is the probable cause for the flattening, which complicates the estimation of the true absorption minimum of \ion{H}{$\alpha$}. Compared to the sample of LLSNe~II, the expansion velocity of \ion{Fe}{ii} $\lambda$5169 in SN~2018is is consistently lower at all epochs. Similarly, the \ion{Sc}{ii} $\lambda$6246 velocities are on the lower end of the comparison sample. The photospheric velocity, inferred from the \ion{Sc}{ii} $\lambda$6246 minimum, rapidly decreases from $\sim$2500 km s$^{-1}$ at about two weeks to less than 800 km s$^{-1}$ at $\sim$100 days. 

\begin{figure}
    \includegraphics[scale=0.41, clip, trim={1.5cm 2.5cm 1.6cm 3.45cm}]{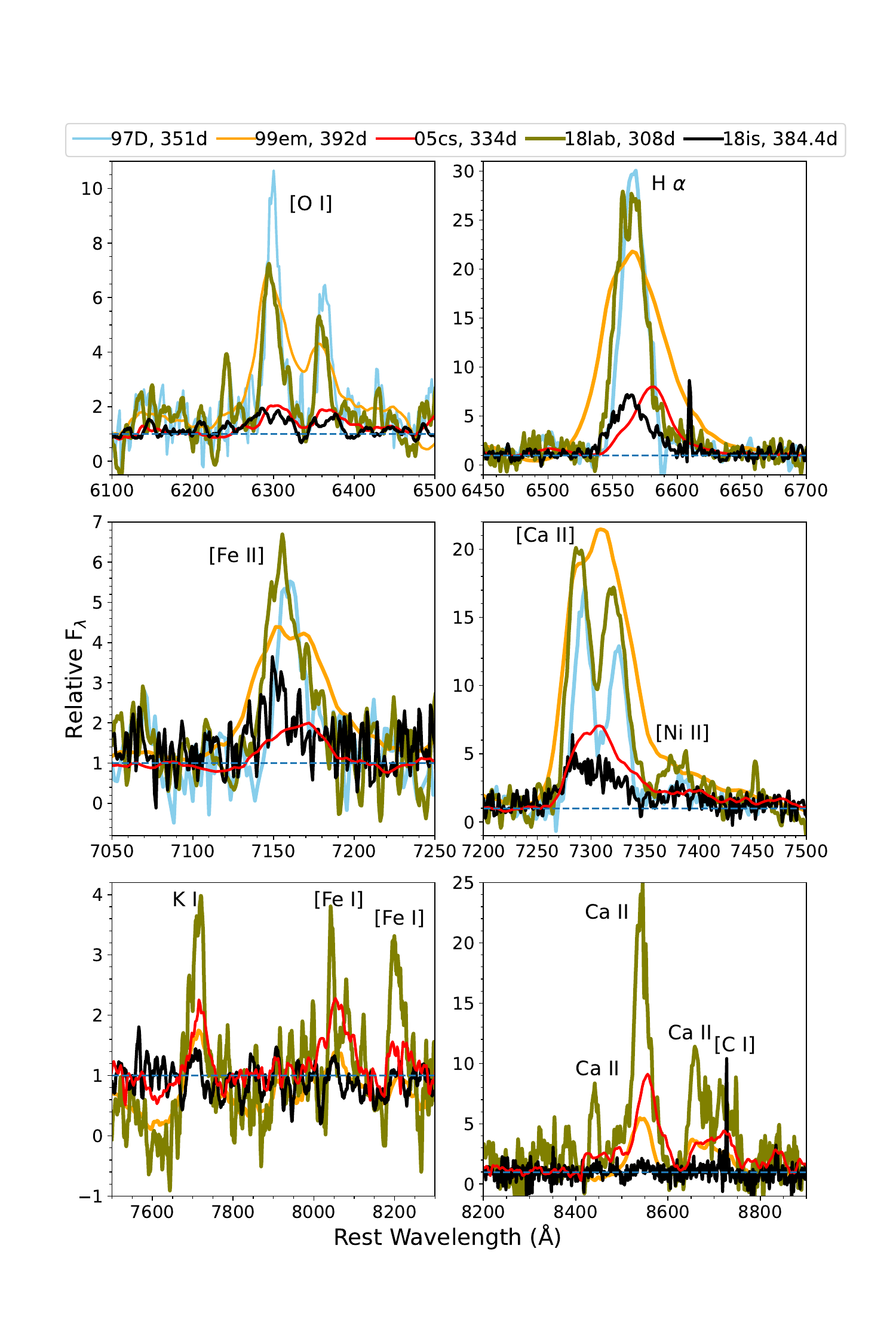}
    \caption{Comparison of the nebular spectra of SN~2018is with other LLSNe~II at similar epochs. The panels show the key spectral features [O I] $\lambda\lambda$6300, 6364, \ion{H}{$\alpha$}, [\ion{Fe}{ii}], [\ion{Ca}{ii}], [\ion{Ni}{ii}], [\ion{Fe}{i}], and \ion{Ca}{ii} lines. }
    \label{fig:OI_Ha}
\end{figure}

Cut-outs of the main features in the nebular phase spectrum of SN~2018is, along with a subset of the comparison sample, are shown in Figure~\ref{fig:OI_Ha}. Most of the features in the nebular spectrum of SN~2018is are weaker compared to other LLSNe~II. [\ion{O}{i}] $\lambda\lambda$6300, 6364, and \ion{H}{$\alpha$} are similar in strength to those in SN~2005cs, while SNe~1997D and 2018lab exhibit much stronger features. However, features such as [\ion{Fe}{i}], [\ion{C}{i}] and \ion{Ca}{ii} NIR triplet, which are present in the spectrum of SN~2005cs, are not discernible in the spectrum of SN~2018is, which could be due to low signal-to-noise ratio in the NIR. 

\begin{figure}
    \includegraphics[scale=0.55, clip, trim = {0cm 0.8cm 0cm 1.3cm}]{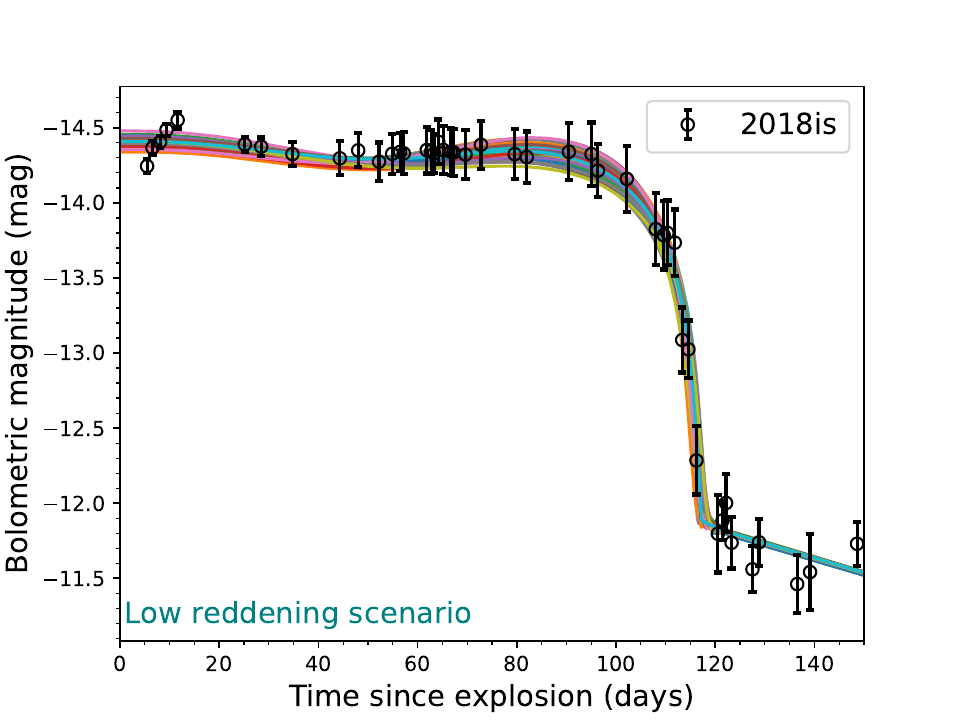}
    \includegraphics[scale=0.55, clip, trim = {0cm 0.cm 0cm 1.3cm}]{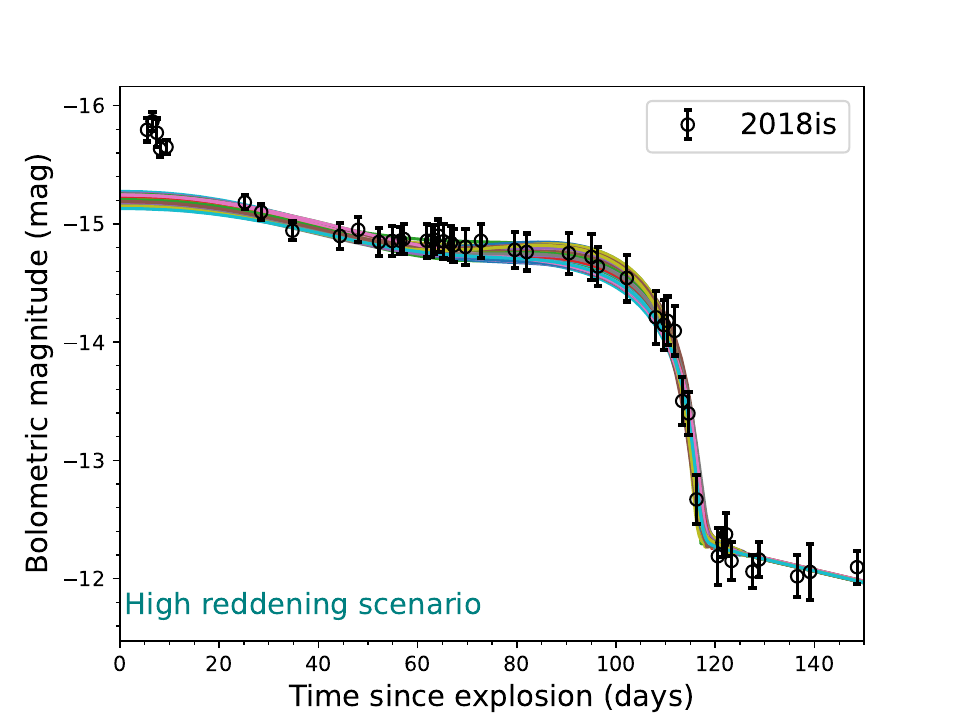}
    \caption{The bolometric magnitude evolution of SN~2018is considering both the low extinction and high extinction estimates are shown in the top and bottom panels, respectively, and the corresponding 50 best-fitted light curves are over-plotted.}
    \label{fig:Nagy}
\end{figure}

\section{Bolometric Light Curve Modelling} \label{Sec6:Modelling}
\subsection{Semi-analytical modelling}
In order to derive estimates of the explosion and progenitor parameters from the bolometric light curve, we employed the semi-analytical modelling of \cite{Nagy2014}. The bolometric light curve of SN\,2018is is constructed using \texttt{SuperBol} \citep{Nicholl2018}, by using $uvw1$ and $UBgVri$ magnitudes up to 9.4 days, followed by $UBgVri$ magnitudes up to 34.8 days and thereafter $BgVri$ magnitudes. This was done to ensure that the UV contribution is taken care of in the early part of the light curve. The de-reddened magnitudes supplied to the routine are interpolated to a common set of epochs and converted to fluxes. The fluxes are used to construct the SED at all the epochs. The routine then fits a blackbody function to the SED, extrapolating to the UV and IR regimes to estimate the true bolometric luminosity. 

We use the Markov Chain Monte Carlo (MCMC) version of the semi-analytic light curve code of \cite{Nagy2014}, developed in \cite{Jager2020}, for fitting the output model to the observed light curves. The semi-analytic code of \cite{Nagy2014} is based on the original model of \cite{Arnett1989}. The model assumes a homologously expanding, spherically symmetric SN ejecta and uses the diffusion approximation for the radiation transport. However, the simple diffusion-recombination model that assumes constant opacity in the ejecta limits the accuracy of the derived physical parameters. For example, the constant opacity approximation results in a negative correlation between Thomson opacity ($\kappa$) and ejecta mass (M$_{\rm ej}$), which has a significant effect on the estimated M$_{\rm ej}$. The correlation between the various explosion parameters makes it important to explore the parameter space with the MCMC approach. \cite{Jager2020} applied this approach to obtain the best estimates and uncertainties for the core parameters.

In our case, the parameters are the initial radius ($\mathrm{R_0}$), the ejected mass ($\mathrm{M_{ej}}$), and the energies (total explosion energy: $\mathrm{E_0}$ = $\mathrm{E_{kin}}$ + $\mathrm{E_{th}}$, kinetic: $\mathrm{E_{kin}}$, thermal: $\mathrm{E_{th}}$), ejecta velocity ($\mathrm{v_{exp}}$) and $^{56}$Ni mass. We searched for the best-fit parameters in the parameter space: $\mathrm{R_0}$: (2--10) $\times$ 10$^{13}$ cm, $\mathrm{M_{ej}}$: 4--20 M$_\odot$, $\mathrm{E_{kin}}$: 0.1--2 foe (1 foe = 10$^{51}$ erg), $\mathrm{E_{th}}$: 0.001--1 foe, $\mathrm{\kappa}$: 0.05--0.4 cm$^2$ g$^{-1}$. The recombination temperature is fixed to 5500 K. The initial expansion velocity is set to 2500 km s$^{-1}$ with an uncertainty of 250 km s$^{-1}$, which is basically the velocity obtained from the minima of the \ion{Sc}{ii} $\lambda$6246 absorption profile at 18 day as the starting day of the fitting is set to 20 day after the explosion. The parameter estimates obtained from the MCMC routine for both the low and high reddening scenarios are listed in Table~\ref{Nagy}. 

The reported parameter values are the mean of the joint posterior, corresponding to the best-fitting solution. The uncertainty limits are derived from taking the 95th percentile of the marginalised posterior probability density function and subtracting the 50th percentile for the upper error. The lower error is estimated by subtracting the 5th percentile from the 50th percentile. The top and bottom panel of Figure~\ref{fig:Nagy} shows the observed bolometric magnitude evolution of SN~2018is corrected for low and high extinction estimate, respectively, and the corresponding best fifty model light curves from the 3$\times$10$^5$ iterations in the MCMC, while the posterior distribution of the parameters and their correlations are shown in Figures~\ref{fig:corner_plot1} and \ref{fig:corner_plot2}. The parameter estimates and their uncertainties are also provided in Table~\ref{Nagy}. There are known parameter correlations \citep{Arnett1989, Nagy2014} between R$_0$ and E$_{\rm th}$, ejected mass and opacity; kinetic energy and opacity, which can also be seen in the corner plot. Since the parameter pairs: E$_{\rm th}$ - R$_0$ are significantly correlated, separate determination of the quantities in the pair is not possible. Rather, the product of the pair should be used as an independent parameter. Assuming a remnant neutron star mass of 1.5 - 2 M$_\odot$, the lower limit of the progenitor mass would be $\sim$ 8-10 M$_\odot$, which indicates that SN~2018is is most likely arising from the collapse of a progenitor close to the lower mass limit for core collapse.

\begin{table}
\centering
\caption{The best-fit core parameters for the true bolometric light curve of SN\,2018is using \citet{Nagy2014} and \citet{Jager2020} for the low reddening (LR) and high reddening (HR) scenarios.\label{Nagy}}
\setlength{\tabcolsep}{2.5pt} % Default value: 6pt
\renewcommand{\arraystretch}{1.3} % Default value: 1
\begin{tabular}{lccl}
\hline\hline
Parameters               & LR                      & HR                     &   Remarks                \\
\hline
R$_0$ (R$_\odot$)        & 637$^{+363}_{-111}$     & 690$^{+296}_{-149}$    & Initial radius of ejecta\\
M$_{\rm ej}$ (M$_\odot$) & 6.0$^{+1.5}_{-1.0}$     & 8.2$^{+1.8}_{-1.4}$    & Ejecta mass\\
E$_{\rm kin}$ (foe)      & 0.27$^{+0.06}_{-0.04}$  & 0.36$^{+0.08}_{-0.06}$ & Initial kinetic energy \\
E$_{\rm th}$ (foe)       & 0.03$^{+0.01}_{-0.01}$ & 0.04$^{+0.01}_{-0.01}$ & Initial thermal energy \\
%T$_{rec} (K)$       & 5500$^{+250}_{-100}$   &  & Recombination temperature\\
$\kappa$ (cm$^2$/g)      & 0.26$^{+0.10}_{-0.08}$  & 0.14$^{+0.06}_{-0.15}$ & Thomson scattering opacity\\
M$_{\rm Ni}$ (10$^{-3}$ M$_\odot$)    & 3.17$^{+0.03}_{-0.01}$ & 4.71$^{+0.03}_{-0.02}$ & $^{56}$Ni mass\\
\hline
\end{tabular}
\end{table}

\begin{table}
\centering
\caption{Pre-supernova structure summary and CSM parameters from \texttt{SNEC} models.} \label{comp_model}
\renewcommand{\arraystretch}{1.1}
\setlength{\tabcolsep}{5pt}
%\footnotesize
\begin{tabular}{cccccccc}
\hline
Mass   & R$_\star$  & M$_{\rm H}$  & M$_{\rm He}$ & R$_{\rm CSM}$ & A$_{\rm V,tot}$      & M$_{\rm CSM}$ & E$_{\rm exp}$\\
(M$_\odot$)  & (R$_\odot$)  & (M$_\odot$)  & (M$_\odot$) & (R$_\odot$) & (mag)      & (M$_\odot$) & (foe)\\
\hline
9                 &   418                  &  5.8 & 1.6   &    600   & 1.18 & $>$0.17 & 0.19\\ %+0.4
\hline
\end{tabular}
\end{table}

\subsection{1D hydrodynamical modelling} \label{sec:SNEC}
We use the open-source 1D radiation hydrodynamics code, Supernova Explosion Code (\texttt{SNEC}, \citealt{Morozova2015}), for multi-band light curve modelling to infer the progenitor parameters and explosion properties of SN~2018is. \texttt{SNEC} is a local thermodynamic equilibrium (LTE) code that employs grey opacities without spectral calculations. The code takes progenitor model, explosion energy, $^{56}$Ni mass, and $^{56}$Ni mixing as inputs and simulates a range of outputs, including multi-band and bolometric light curves, photospheric velocity, and temperature evolution. 

For the progenitor, we adopt a set of non-rotating solar metallicity RSG models from \cite{Sukhbold2016}, which are computed using the stellar evolution code \texttt{KEPLER} \citep{Weaver1978}. The lowest mass limit of the progenitor models for Fe-core collapse, as generated in \cite{Sukhbold2016}, is 9 M$_\odot$. The length of the plateau phase in the simulated light curves was observed to increase with higher ZAMS mass, while it decreased with an increase in explosion energy. We generated a grid of \texttt{SNEC} models encompassing ZAMS masses between 9 and 11\,M$_\odot$ in steps of 0.5\,M$_\odot$, explosion energy within 0.1 - 0.5\,foe in steps of 0.01 foe, while maintaining a constant $^{56}$Ni mass of 0.0049\,M$_\odot$. The $^{56}$Ni mixing parameter, representing the mass-coordinate until which $^{56}$Ni is mixed outwards, is fixed to 2\,M$_\odot$. 

\begin{figure}
	\includegraphics[scale=0.45, clip, trim={0.4cm 0.cm .0cm 0cm}]{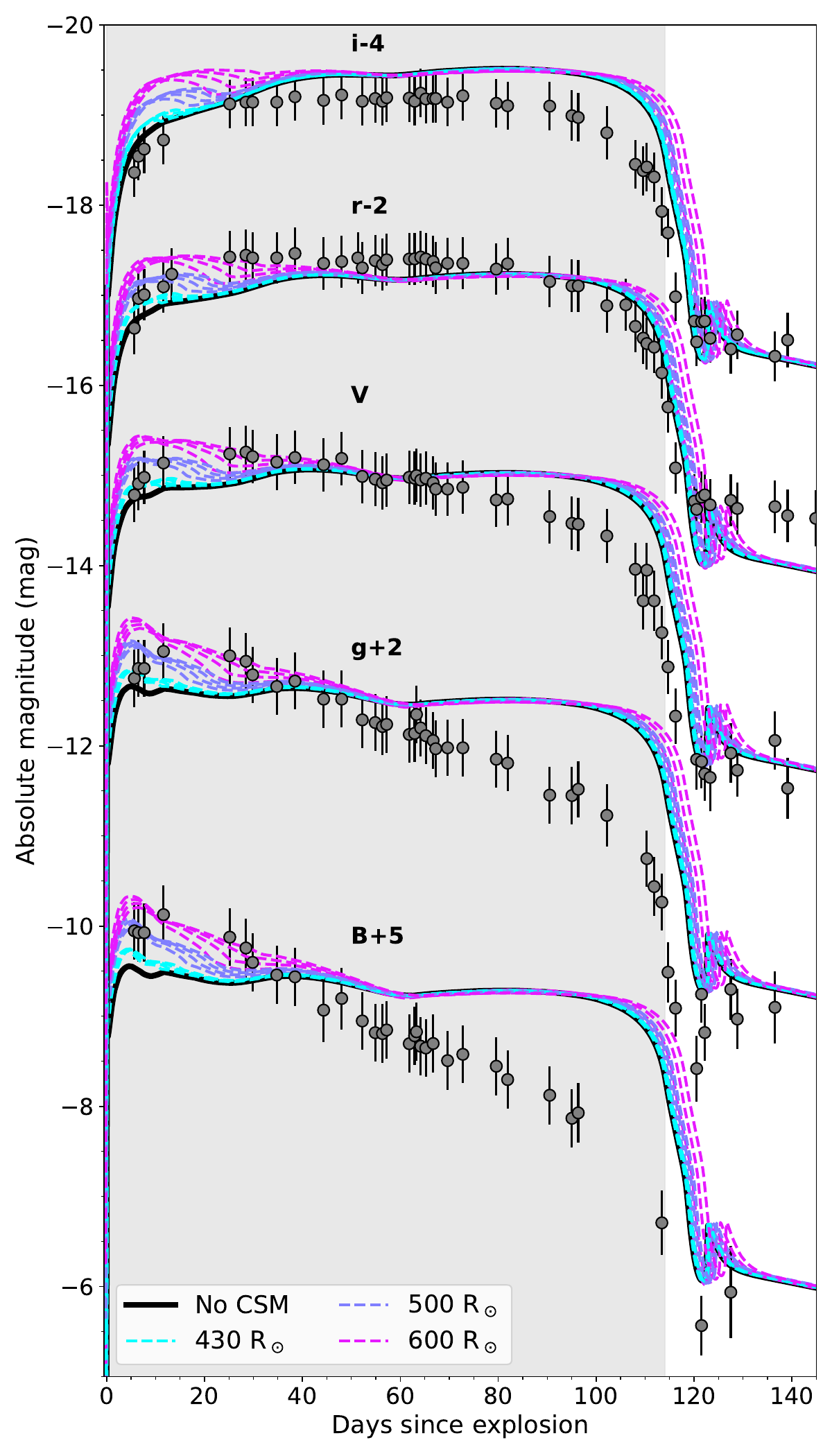}
    \caption{Multi-band model light curves for scenarios with `No CSM' and CSM extents of 430, 500 and 600 R$_\odot$, with K$_{\rm CSM}$ values of 2, 3, 4, and 5$\times$10$^{18}$ g cm$^{-1}$, are shown alongside the observed light curves. The shaded region marks the area below t$_{PT}$, after which non-LTE conditions dominate, making SNEC models invalid.}
    \label{fig:snec_model}
\end{figure}

The best-fitting progenitor model is determined by finding the minimal $\chi^2$, computed as
\begin{equation}
\chi^2 = \sum\limits_{\lambda} \frac{1}{N_\lambda}\sum\limits_{t=30d}^{t<t_{\rm PT}} \left(\frac{m^\mathrm{obs}_\lambda(t) - m^\mathrm{model}_\lambda(t)}{\Delta m^\mathrm{obs}_\lambda(t)}\right)^2 + \left(\frac{v^\mathrm{obs}(t) - v^\mathrm{model}(t)}{\Delta v^\mathrm{obs}(t)}\right)^2,
    \label{eq:chi2}
\end{equation}
where $m^\mathrm{obs}_\lambda$(t) and $\Delta m^\mathrm{obs}_\lambda$(t) are the magnitudes and their corresponding errors at time t, $m^\mathrm{model}_\lambda$(t) are the model magnitudes at time t, $\lambda$ denotes the {$Vri$} filters, $N_\lambda$ is the total number of observed data points in filter $\lambda$, and $v^\mathrm{obs}$(t) and $\Delta v^\mathrm{obs}$(t) are the velocities and their errors at time t. We compute $m^\mathrm{model}_\lambda$ by varying $A_V$ between the Galactic extinction value (0.22 mag) and the high extinction value considered in this work (1.34 mag). For the fitting, we also allow the distance to vary between 19.6 Mpc and 23 Mpc based on the mean value and the error in the distance (21.3$\pm$1.7 Mpc). We did not include the bluer band light curves, such as the $UBg$ bands, for estimating the minimum chi-square, as these light curves are more susceptible to being affected by line blanketing at later phases. We identified the best-fit solution as corresponding to a 9.0\,M$_\odot$ ZAMS star, with a pre-SN mass of 8.75\,M$_\odot$, and a pre-SN stellar radius of 418\,R$_\odot$, and an explosion energy of 0.19 foe. 

To improve the fit to the early light curve ($<$30 days), we added a wind-like circumstellar medium (CSM) profile to the best-fit progenitor model, as in \cite{Dastidar2024} and \cite{Reguitti2024}. The density profile follows ${\rho} = {\rm K}_{\rm CSM}r^{-2}$, where K$_{\rm CSM}$ is the mass-loading parameter. We generated models with K$_{\rm CSM}$ values from 2 to 5$\times$10$^{18}$ g cm$^{-1}$ and CSM extents (R$_{\rm CSM}$) of 430, 500, and 600 R$_\odot$.

\begin{figure}
	\includegraphics[scale=0.45, clip, trim={0.cm 0.cm 0.cm 0.cm}]{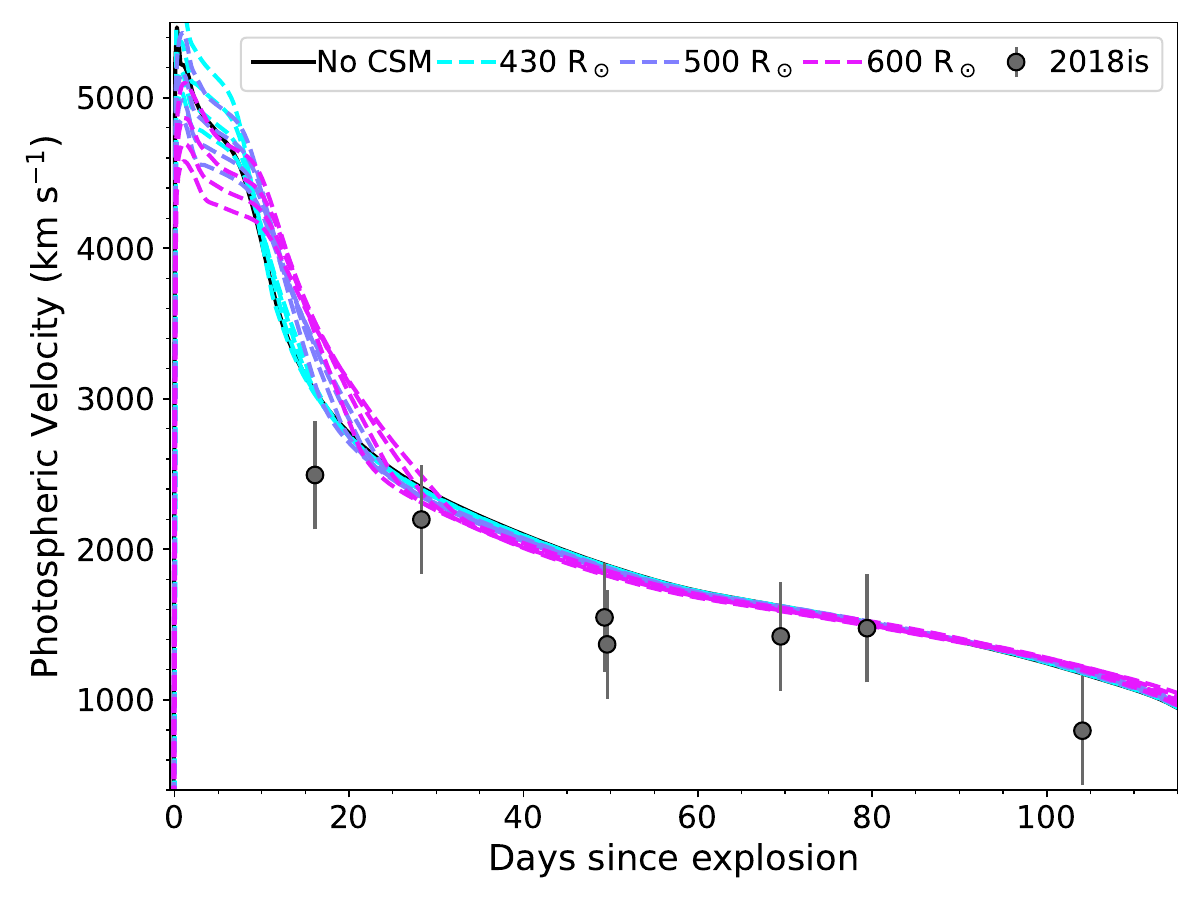}
    \caption{Model photospheric velocities for scenarios with `No CSM' and CSM extents of 430, 500 and 600 R$_\odot$, and K$_{\rm CSM}$ values of 3, 4, 5$\times$10$^{18}$ g cm$^{-1}$ are shown, alongside the line velocity of \ion{Sc}{ii} $\lambda$6246 for SN~2018is.}
    \label{fig:snec_vel}
\end{figure}

\begin{figure}
	\includegraphics[scale=0.5, clip, trim={0.3cm 0.cm 0.5cm 1.5cm}]{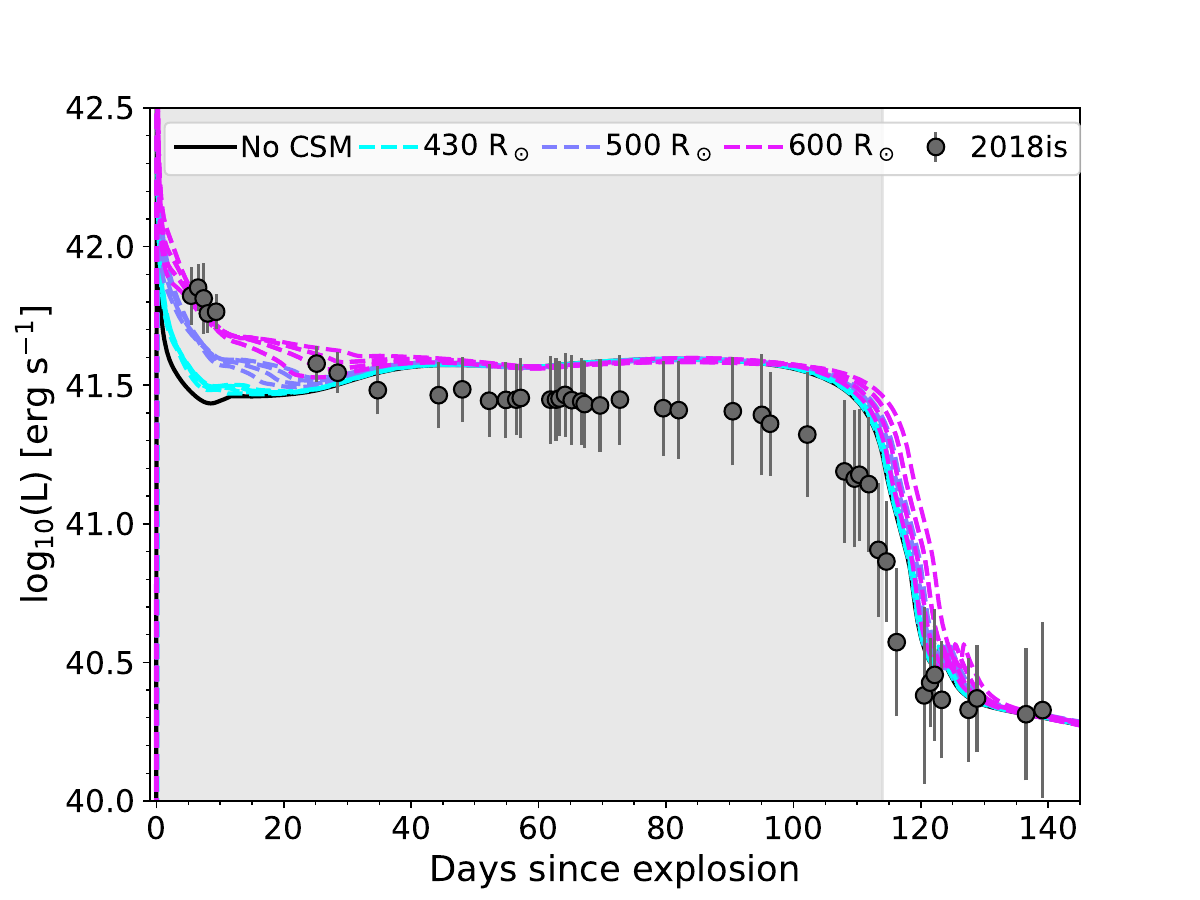}
    \caption{Model bolometric luminosities for scenarios with `No CSM' and CSM extents of 430, 500 and 600 R$_\odot$, and K$_{\rm CSM}$ values of 3, 4, 5$\times$10$^{18}$ g cm$^{-1}$ are shown, alongside the observed bolometric luminosity of SN~2018is. The shaded region marks the area below t$_{PT}$, after which non-LTE conditions dominate, making SNEC models invalid.}
    \label{fig:snec_lum}
\end{figure}

Fig.~\ref{fig:snec_model} shows the absolute model magnitudes alongside the observed light curves. The distance and extinction correction (A$_V$) applied to the observed light curves are 23 Mpc and 1.18 mag, respectively. Thus, these models suggest a total E($B-V$) value of 0.38 mag, close to the high reddening estimate. Fig.~\ref{fig:snec_vel} compares model photospheric velocities with the \ion{Sc}{ii} line velocities, while the model bolometric luminosity alongside the observed bolometric luminosity for SN~2018is is shown in Fig.~\ref{fig:snec_lum}. The early bolometric luminosity is better reproduced with K$_{\rm CSM}$ values of (2-5)$\times$10$^{18}$ g cm$^{-1}$ and R$_{\rm CSM}$ = 600 R$_\odot$, which we will consider as the best-match models. The progenitor, explosion and CSM parameters for the best-match model is tabulated in Table~\ref{comp_model}.

The CSM mass is estimated to be 0.17-0.43 M$_\odot$ using
$$ {\rm M}_{\rm{csm}} = 4 \pi {\rm K}_{\rm{CSM}} ({\rm R}_{\rm{start}} - {\rm R}_{\rm{CSM}}), $$
where ${\rm R}_{\rm{start}}$ = 403 R$_\odot$.

We note that these models overestimated the plateau length by $\sim$3 days, while the model photospheric velocity mostly corresponds to the upper limit of the observed \ion{Sc}{ii} line velocities. The explosion energy, which influences both luminosity and expansion velocity, also plays a crucial role in determining the plateau length, with higher energies generally leading to shorter plateaus. Therefore, the plateau length of SN~2018is could theoretically be modelled with an increased explosion energy, though this would result in velocities exceeding those observed. \cite{Kozyreva2022} showed that plateau length might depend on the viewing angle in asymmetric explosions, offering a possible explanation for the plateau-length discrepancy in SN~2018is.

Overall, light curve modelling indicates that SN~2018is is consistent with a low-energy explosion (10$^{50}$ erg) of a low-mass star (9 M$_\odot$), with the shorter observed plateau potentially due to an asymmetric explosion.

\section{Discussion} 
\label{Sec7:Discussion}
\subsection{Nebular constraints on the progenitor mass}
\label{sec7.3}
\begin{figure*}
	\includegraphics[scale=0.35, clip, trim={0.cm, 0.4cm, 0.cm, 0cm}]{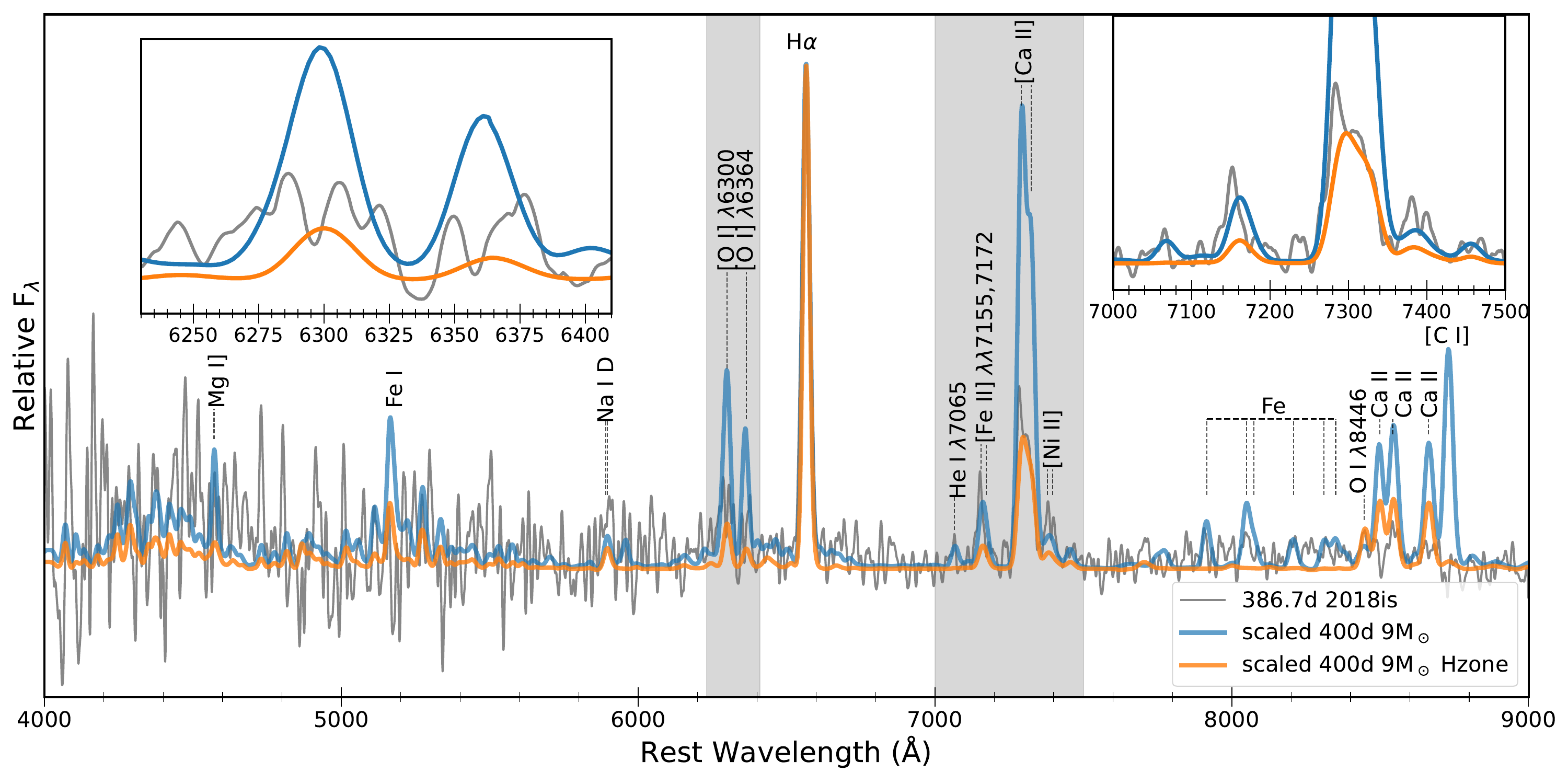}
    \caption{The 386.7 day nebular spectrum of SN~2018is compared to the 9 M$_\odot$ model and hydrogen-zone model of \citet{Jerkstrand2018}. Narrow emission lines (probably) from the host galaxy in the observed spectrum have been removed.}
    \label{fig:Jerkstrand}
\end{figure*}

\subsubsection{Comparison with spectral models}
The ZAMS mass of the progenitor can be constrained by comparing the nebular spectra to the nebular spectral models of \cite{Jerkstrand2018}. In their study, they observed that the luminosity of the [\ion{O}{i}] line is primarily influenced by the progenitor's ZAMS mass \citep{Jerkstrand2012, Jerkstrand2014}. Progenitors with higher masses tend to display more pronounced [\ion{O}{i}] features in their nebular spectra. 

To enable a consistent comparison, model spectrum has been scaled to match the $^{56}$Ni mass of 0.0049 M$_\odot$, distance and epoch of the SN~2018is nebular spectrum. Additionally, owing to uncertainties in extinction, all spectra have been normalised to their respective \ion{H}{$\alpha$} peak flux. 
Figure~\ref{fig:Jerkstrand} shows the comparison of the 390 d nebular spectrum of SN~2018is with the 400 d nebular spectral model for a 9 M$_\odot$ RSG progenitor. The 9 M$_\odot$ progenitor model of \cite{Sukhbold2016}, evolved with \texttt{KEPLER}, was used in \cite{Jerkstrand2018}. Additionally, we plot the 9 M$_\odot$ \textquoteleft pure hydrogen-zone\textquoteright~model in Figure~\ref{fig:Jerkstrand}. In this model, the progenitor is composed entirely of material from the hydrogen envelope. It is anticipated that this model would resemble the nebular spectra of ECSNe, characterised by faint features such as \ion{Mg}{i}], \ion{Fe}{i}, [\ion{O}{i}], \ion{He}{i}, [\ion{C}{i}] $\lambda$8727, and \ion{O}{i} $\lambda$7774, along with a prominent \ion{O}{i} $\lambda$8446 line. The two insets in Figure~\ref{fig:Jerkstrand} zoom in on the [\ion{O}{i}] and [\ion{Ca}{ii}] regions. The [\ion{O}{i}] emission of SN~2018is lies between the two models, while [\ion{Ca}{ii}] is better matched by the H-zone model.

Although, \ion{Fe}{i} lines between 7900 - 8400 \AA~ and [\ion{C}{i}] $\lambda$8727, which are present in the 9 M$_\odot$ RSG progenitor model, are absent in SN~2018is, features such as [\ion{Fe}{ii}], [\ion{Ni}{ii}], and (weak) \ion{He}{i} are discernible. The \ion{O}{i} $\lambda$8446 line, which is an important diagnostic in the nebular spectra of ECSNe and is present in the H-zone model, is also absent in the observed spectrum. The presence of He, Fe, and Ni lines in the observed spectrum may imply the presence of He shell in the ejecta. The 9 M$_\odot$ model also predicts strong [\ion{O}{i}] $\lambda\lambda$6300, 6364 doublet and \ion{Mg}{i}], both of which arise from the O zone. In SN~2018is, while the [\ion{O}{i}] $\lambda\lambda$6300, 6364 doublet is discernible, \ion{Mg}{i}] cannot be identified due to the low SNR of the nebular spectrum. From this comparison, it is evident that the progenitor mass of SN~2018is was 9 M$_\odot$ or lower. The possible presence of a He-shell suggests that it likely underwent Fe-core collapse before exploding as a SN.

\subsubsection{Constraints from forbidden lines}
\begin{figure}
	\includegraphics[scale=0.18, clip, trim={0.cm 0cm 0cm 0.cm}]{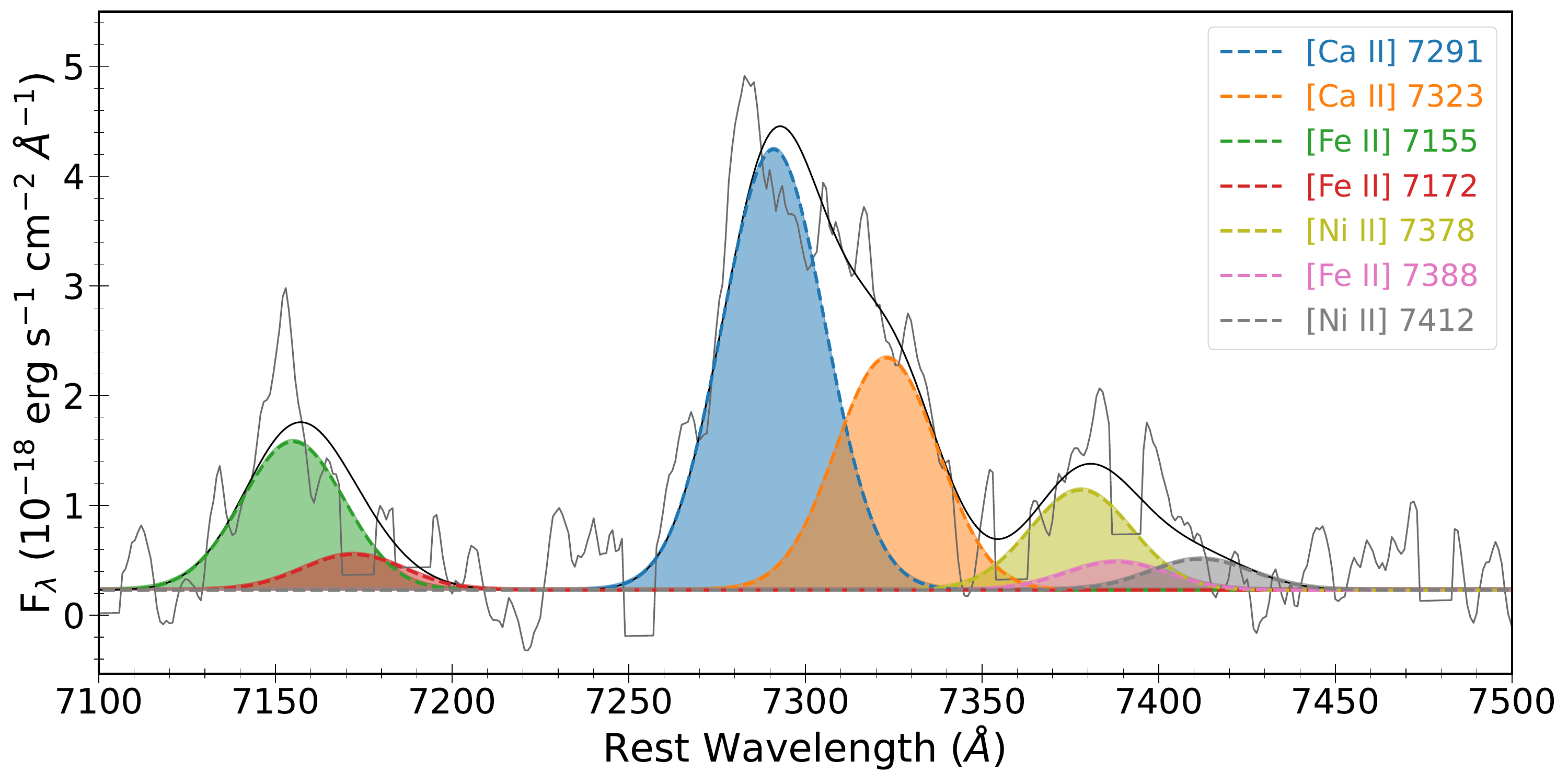}
    \caption{Spectrum cut-out of the 386.7 day spectrum showing the Gaussian fits to determine the [\ion{Ni}{ii}] $\lambda$7378/[\ion{Fe}{ii}] $\lambda$7155 in SN~2018is.}
    \label{fig:Ni_Fe_ratio}
\end{figure}

The nebular spectrum at 386.7 d shows several forbidden transitions, which can aid in constraining the stable Ni to Fe abundance ratio. Theoretical model predicts that the ZAMS mass of the progenitor decreases with increasing Ni/Fe abundance ratio, and hence, this ratio can be used to put constraints on the progenitor mass. The 386.7 d spectrum is calibrated to extrapolated $gVri$-band photometry. Using Gaussian components, we fit the seven dominant line transitions in the 7100-7500 \AA{} region, which are [\ion{Ca}{ii}] $\lambda\lambda$7291, 7323, [\ion{Fe}{ii}] $\lambda$7155, [\ion{Fe}{ii}] $\lambda$7172, [\ion{Fe}{ii}] $\lambda$7388, [\ion{Fe}{ii}] $\lambda$7453, [\ion{Ni}{ii}] $\lambda$7378, and [\ion{Ni}{ii}] $\lambda$7412. The relative luminosities of lines from a given element are taken from \cite{Jerkstrand2015}. So, the iron lines are constrained by L$_{7453}$ = 0.31 L$_{7155}$, L$_{7172}$ = 0.24 L$_{7155}$, L$_{7388}$ = 0.19 L$_{7155}$, and the nickel lines are constrained by L$_{7412}$ = 0.31 L$_{7378}$. A single line width for all lines, the full width at half-maximum (FWHM) velocity V, has been used. The free parameters are then L$_{7291, 7323}$, L$_{7155}$, L$_{7378}$, and V. As shown in Figure~\ref{fig:Ni_Fe_ratio}, a good fit is obtained for L$_{7291, 7323}$ = 1.2 $\times$ 10$^{37}$ erg s$^{-1}$, L$_{7155}$ = (2.7 $\pm$ 0.4) $\times$ 10$^{36}$ erg s$^{-1}$, L$_{7378}$ = (1.8$\pm$0.3) $\times$ 10$^{36}$ erg s$^{-1}$, and  V = 1400 km s$^{-1}$ . From this we determine a ratio L$_{7378}$ /L$_{7155}$ = 0.67. The iron-zone temperature is estimated to lie in the range 2550 - 2650 K, using the ratio of L$_{7155}$ and M$_{^{56}Ni}$ (in the high reddening scenario) to be (8.9$\pm$1.7)$\times$ 10$^{38}$ erg s$^{-1}$ M$_\odot$$^{-1}$. Using the above temperature, the Ni to Fe abundance ratio is found to be 0.04, which is 0.7 times the solar value (0.056, \citealt{Lodders2003}). However, as noted by \cite{Jerkstrand2015}, primordial Fe and Ni may contaminate the observed ratio, potentially leading to an underestimation by a factor of three. Consequently, the estimated Ni/Fe ratio derived from the 386.7 d spectrum should be regarded as a lower limit and could be as high as 0.12. Theoretical studies indicate that CCSNe originating from moderate-mass progenitors (9–11 M$_\odot$) exhibit higher Ni/Fe abundance ratios \citep{Woosley1995}. For example, the models of \cite{Woosley2007} predict a Ni/Fe ratio of 0.04 for a 15 M$_\odot$ ZAMS mass star. Based on these predictions, the upper limit of the progenitor's ZAMS mass for SN~2018is can be constrained to 15 M$_\odot$. Finally, it is important to note that recent theoretical models present conflicting perspectives on the dependence of the Ni/Fe abundance ratio on the progenitor's ZAMS mass. While 1D explosion nucleosynthesis models by \cite{Sukhbold2016} do not show a significant Ni/Fe enhancement for moderate-mass progenitors, 3D models of Fe-core CCSNe \citep{Burrows2024, Wang2024} suggest that initial velocity perturbations can significantly affect the Ni/Fe ratio, yielding values from sub-solar to as high as 50 times the solar value. Thus, using the Ni/Fe abundance ratio to constrain the progenitor’s ZAMS mass remains highly uncertain.

We estimate the [\ion{O}{i}] $\lambda\lambda$6300, 6364 luminosity by Gaussian fit to be 2.9 $\times$ 10$^{36}$ erg s$^{-1}$, assuming a ratio [\ion{O}{i}] $\lambda$6300/ [\ion{O}{i}] $\lambda$6364 = 3. The luminosity ratio of [\ion{Ca}{ii}]/[\ion{O}{i}] is considered a good diagnostic for the He core mass, and consequently the progenitor mass \citep{Fransson1987, Fransson1989}, with higher ratios corresponding to lower ZAMS masses. For SN~2005cs, this ratio was estimated to be $\sim$4.2 $\pm$ 0.6 \citep{Pastorello2009}. The [\ion{Ca}{ii}]/[\ion{O}{i}] luminosity ratio is 4.1 for SN~2018is, which indicates a low-mass progenitor for SN~2018is. However, the line ratio is only a rough diagnostic of the core mass due to the contribution to the emission from primordial O in the hydrogen zone \citep{Jerkstrand2012, Maguire2012} and the effects of mixing, which can complicate the determination of relative abundances based on line strengths, as noted by \cite{Fransson1989}.

\subsection{Investigating the Electron-Capture Nature of SN 2018is}

The progenitor mass of SN~2018is, as estimated from the analysis of nebular spectra and from semi-analytical and hydrodynamical modelling of the bolometric light curve, suggests that the ZAMS progenitor mass is below 9 M$_\odot$. The explosion energy inferred from hydrodynamical modelling is also low, below 0.20 foe, along with a low mass of synthesised $^{56}$Ni. These factors raise the possibility that SN~2018is could be an ECSN, resulting from the core-collapse of an oxygen-magnesium-neon (OMgNe) core in a SAGB star. 

In SAGB stars, the helium-rich layer is almost destroyed during the second dredge up. As a result, ECSNe from single-star progenitors are not expected to have O-rich or He-rich shells. Additionally, progenitor evolution models predict that the H/He layer becomes diluted during the SAGB stage. Consequently, features associated with He burning, such as, \ion{Fe}{i}, \ion{He}{i} $\lambda$7065, and [\ion{C}{i}] $\lambda$8727, are expected to be absent in such cases. 

In contrast, layers of Si, O and He would not be entirely burnt to Fe group elements in Fe-core collapse SN and may provide additional lines of S, Ca, O, C and He to spectra. The presence of He-core material, as in Fe-core SNe, is characterised by signatures of \ion{He}{i} $\lambda$7065, [\ion{C}{i}] $\lambda$8727, [\ion{C}{i}] $\lambda$9850, \ion{O}{i} $\lambda$7774, \ion{Fe}{i} $\lambda$5950 and \ion{Fe}{i} lines between 7900–8500 $\AA$, with the \ion{Fe}{i} and \ion{C}{i} lines being particularly prominent. The absence of these lines would be the distinctive marker for ECSNe. 

Certain properties of SN~2018is, such as its low $^{56}$Ni mass, a low-mass progenitor (below 9 M$_\odot$), and the absence of \ion{Fe}{i}, \ion{He}{i} $\lambda$7065 and [\ion{C}{i}] $\lambda$8727 lines in the nebular spectrum, are consistent with an ECSN. However, the lack of the \ion{O}{i} $\lambda$8446 line, which originates from the H-zone, contradicts this scenario. Additionally, the Ni/Fe abundance ratio upper limit for SN~2018is is 0.12, which is significantly lower than the expected range of 1-2 for ECSNe \citep{Wanajo2009}. Therefore, the nebular spectrum of SN~2018is does not align clearly with either the ECSN or Fe-core SNe scenario. 

Further evidence against SN~2018is being an ECSN comes from light curve modelling of ECSNe, which predicts an explosion as bright as typical SNe~IIP \citep{Tominaga2013,Moriya2014}, with a plateau luminosity around L $\sim$ 10$^{42}$ erg s$^{-1}$, a duration of 60–100 days, and a subsequent luminosity drop of approximately 4 magnitudes. Additionally, ejecta velocities during the plateau phase are expected to exceed 2000 km s$^{-1}$. These predictions do not match the observed properties of SN~2018is. 

Recently, \cite{Sato2024} proposed a new diagnostic to differentiate between ECSNe and Fe-core SNe, based on the $B-V$ and $g-r$ colour indices during the plateau phase. Their models suggest that ECSNe are bluer during the plateau phase compared to Fe-core SNe. For both reddening scenarios, the $B-V$ colour of SN~2018is at half plateau duration (t$_{PT}$/2) exceeds 1 mag, which does not satisfy the criteria outlined in equation C1 required for ECSNe. Therefore, the ECSN scenario for SN~2018is can be clearly dismissed.

\section{Summary}
\label{summary}
We present an analysis of SN~2018is, a low-luminosity Type IIP supernova, based on comprehensive optical photometry and spectroscopy. Through a concerted community effort, we were able to achieve good cadence photometry in the photospheric phase, during the transition to the radioactive tail, and in the radioactive tail phase, which is rare for a LLSN II. SN~2018is exhibits a $V$-band light curve decline rate of 1.04 mag (100d)$^{-1}$, with a plateau phase lasting approximately 110 days, which is shorter and steeper than most other low-luminosity SNe IIP. The optical and near-infrared spectra display hydrogen emission lines that are strikingly narrow, even for this class. The velocity derived from \ion{Fe}{ii} $\lambda$5169 and \ion{Sc}{ii} $\lambda$6246 lines are notably low compared to other SNe in this category. 

The nebular spectrum lacks lines such as \ion{He}{i} $\lambda$7065 and [\ion{C}{i}] $\lambda$8727, which are expected in low-mass progenitors exploding but absent in SN~2018is. According to models, such as those proposed by \cite{Dessart2013a,Jerkstrand2018}, these lines are expected in SNe II with low-mass progenitors, as more massive stars tend to have extended oxygen shells that shield the He shell from gamma-ray deposition. Nevertheless, some LLSNe II, like SN~2005cs, also lack these lines in their spectra. In accordance with the discussion in \cite{Jerkstrand2018}, these observations are more consistent with ECSN rather than Fe-CCSN, as ECSN typically lack lines produced in the He layer. However, the substantially low Ni/Fe abundance ratio, redder colours, and absence of \ion{O}{i} $\lambda$8446 are strong evidences against EC nature of SN~2018is.

Semi-analytical modelling of the bolometric light curve suggests an ejecta mass of approximately 8 M$_\odot$, implying a pre-supernova mass of about 9.5 M$_\odot$, and an explosion energy of 0.40 foe (considering the high reddening scenario). Hydrodynamical modelling further supports a progenitor with a ZAMS mass of 9 M$_\odot$ and a low explosion energy of 0.19 $\times$ 10$^{51}$ erg. Additionally, the models suggest the presence of a dense CSM with a mass of at least 0.17 M$_\odot$ close to the progenitor (within 200 R$_\odot$) which is necessary to reproduce the early light curve. The shorter plateau duration observed in SN~2018is could be explained by explosion asymmetries in low-mass progenitors, as suggested by existing models in the literature. In summary, the rapid $V$-band decline, relatively shorter plateau, and remarkably narrow emission lines make SN~2018is stand out among the population of low-luminosity Type II SNe.

\section*{Acknowledgements}
\begin{small}
We thank the anonymous referee for their insightful comments and suggestions, which helped to improve the paper. R.D. acknowledges funds by ANID grant FONDECYT Post-doctorado No 3220449. K.M. acknowledges the support from the Department of Science and Technology (DST), Govt. of India and Indo-US Science and Technology Forum (IUSSTF) for the WISTEMM fellowship (October 2018 to March 2019) and the Department of Physics, UC Davis, where a significant part of this work was carried out. K.M. acknowledges the support from the BRICS grant DST/ICD/BRICS/Call-5/CoNMuTraMO/2023 (G) funded by the DST, India. This work makes use of observations from the Las Cumbres Observatory network. The LCO team is supported by NSF grants AST-1911225 and AST-1911151. Research by S.V. is supported by NSF grant AST-2008108. Time-domain research by the University of Arizona team and D.J.S. is supported by National Science Foundation (NSF) grants 2108032, 2308181, 2407566, and 2432036 and the Heising-Simons Foundation under grant \#2020-1864. A.P., A. R., N. E. R., L. T., S. B. acknowledge support from the PRIN-INAF 2022, ``Shedding light on the nature of gap transients: from the observations to the models”. G.P. acknowledges support from the National Agency for Research and Development (ANID) through the Millennium Science Initiative Program – ICN12\_009. J.S. acknowledges support from the Packard Foundation. A. R. acknowledges financial support from the GRAWITA Large Program Grant (PI P. D’Avanzo). L.C.\ is grateful for support from NSF programs AST-2107070 and AST-2205628. This research uses data obtained through the Telescope Access Program (TAP), which has been funded by the TAP member institutes.

Observations reported here were in part obtained at the MMT Observatory, a joint facility of the University of Arizona and the Smithsonian Institution. This research is based on observations made with the Nordic Optical Telescope, operated by the Nordic Optical Telescope Scientific Association at the Observatorio del Roque de los Muchachos, La Palma, Spain, of the Instituto de Astrofisica de Canarias; the Gran Telescopio Canarias, installed at the Spanish Observatorio del Roque de los Muchachos of the Instituto de Astrofísica de Canarias, in the island of La Palma; the Liverpool Telescope operated on the island of La Palma by Liverpool John Moores University at the Spanish Observatorio del Roque de los Muchachos of the Instituto de Astrofísica de Canarias with financial support from the UK Science and Technology Facilities Council; the 1.82 m Copernico Telescope of INAF-Osservatorio Astronomico di Padova at Mt. Ekar. Based on observations obtained at the Southern Astrophysical Research (SOAR) telescope, which is a joint project of the Minist\'{e}rio da Ci\^{e}ncia, Tecnologia e Inova\c{c}\~{o}es (MCTI/LNA) do Brasil, the US National Science Foundation’s NOIRLab, the University of North Carolina at Chapel Hill (UNC), and Michigan State University (MSU). We acknowledge Weizmann Interactive Supernova data REPository http://wiserep.weizmann.ac.il (WISeREP, \citealt{Yaron2012}). This research has made use of the NASA/IPAC Extragalactic Database (NED) which is operated by the Jet Propulsion Laboratory, California Institute of Technology, under contract with the National Aeronautics and Space Administration. We acknowledge the usage of the HyperLeda database (http://leda.univ-lyon1.fr). 
\end{small}
%%%%%%%%%%%%%%%%%%%%%%%%%%%%%%%%%%%%%%%%%%%%%%%%%%



\bibliographystyle{aa}
\bibliography{aanda} % if your bibtex file is called example.bib

%%%%%%%%%%%%%%%%%%%%%%%%%%%%%%%%%%%%%%%%%%%%%%%%%%

%%%%%%%%%%%%%%%%% APPENDICES %%%%%%%%%%%%%%%%%%%%%

\appendix

\section{Some extra material}

%Log of optical imaging of SN2020aze through DLT telescope
\begin{table}
 \begin{center}
 \caption{DLT40 \texttt{clear} band photometry of SN~2018is calibrated to the $r$-band.}
 \label{dlt_phot}
 \scalebox{0.95}{
 \begin{tabular}{cccc}
 \hline
UT Date & JD  & Phase$^\dagger$ & $r$ \\
(yyyy-mm-dd) &     & (days)          & (mag) \\
\hline
2018-01-20.3 & 2458138.8 &   5.4 & 17.98 $\pm$ 0.06\\ 
2018-01-21.2 & 2458139.7 &	 6.3 & 18.06 $\pm$ 0.08 \\
2018-01-22.7 & 2458141.2 &   7.8 & 17.85 $\pm$ 0.06 \\
2018-01-23.2 & 2458141.7 &   8.3 & 17.94 $\pm$ 0.07 \\
2018-01-23.7 & 2458142.2 &   8.8 & 17.93 $\pm$ 0.06 \\
2018-01-24.7 & 2458143.2 &   9.8 & 18.00 $\pm$ 0.06 \\
2018-01-25.7 & 2458144.2 &  10.8 & 17.97 $\pm$ 0.06 \\
2018-01-26.7 & 2458145.2 &  11.8 & 17.64 $\pm$ 0.05 \\
2018-02-08.2 & 2458157.7 &  24.3 & 17.69 $\pm$ 0.06 \\
2018-02-09.2 & 2458158.7 &  25.3 & 17.67 $\pm$ 0.05 \\
2018-02-10.2 & 2458159.7 &  26.3 & 17.65 $\pm$ 0.04 \\
2018-02-12.2 & 2458161.6 &  28.2 & 17.48 $\pm$ 0.07 \\
2018-02-13.2 & 2458162.6 &  29.2 & 17.48 $\pm$ 0.05 \\
2018-02-14.3 & 2458163.8 &  30.4 & 17.45 $\pm$ 0.04 \\
2018-02-15.2 & 2458164.6 &  31.2 & 17.60 $\pm$ 0.05 \\
2018-02-16.1 & 2458165.6 &  32.2 & 17.65 $\pm$ 0.05 \\
2018-02-18.1 & 2458167.6 &  34.2 & 17.63 $\pm$ 0.05 \\
2018-02-19.1 & 2458168.6 &  35.2 & 17.54 $\pm$ 0.06 \\
2018-02-19.6 & 2458169.1 &  35.7 & 17.30 $\pm$ 0.05 \\ 
2018-02-20.1 & 2458169.6 &  36.2 & 17.44 $\pm$ 0.05 \\  
2018-02-20.6 & 2458170.1 &  36.7 & 17.72 $\pm$ 0.05 \\
2018-02-21.1 & 2458170.6 &  37.2 & 17.58 $\pm$ 0.04 \\
2018-02-21.6 & 2458171.1 &  37.7 & 17.38 $\pm$ 0.05 \\ 
2018-02-22.1 & 2458171.6 &  38.2 & 17.68 $\pm$ 0.04 \\ 
2018-02-23.1 & 2458172.6 &  39.2 & 17.63 $\pm$ 0.06 \\ 
2018-02-24.1 & 2458173.6 &  40.2 & 17.54 $\pm$ 0.05 \\ 
2018-02-24.6 & 2458174.1 &  40.7 & 17.45 $\pm$ 0.06 \\ 
2018-02-25.1 & 2458174.6 &  41.2 & 17.59 $\pm$ 0.06 \\ 
2018-02-26.1 & 2458175.6 &  42.2 & 17.49 $\pm$ 0.06 \\
2018-03-07.1 & 2458184.6 &  51.2 & 17.61 $\pm$ 0.08 \\
2018-03-08.1 & 2458185.6 &  52.2 & 17.51 $\pm$ 0.05 \\
2018-03-10.1 & 2458187.6 &  54.2 & 17.57 $\pm$ 0.04 \\
2018-03-11.1 & 2458188.6 &  55.2 & 17.59 $\pm$ 0.05 \\
2018-03-12.1 & 2458189.6 &  56.2 & 17.55 $\pm$ 0.05 \\
2018-03-13.1 & 2458190.6 &  57.2 & 17.61 $\pm$ 0.05 \\
2018-03-14.1 & 2458191.6 &  58.2 & 17.77 $\pm$ 0.06 \\
2018-03-15.1 & 2458192.6 &  59.2 & 17.62 $\pm$ 0.05 \\
2018-03-16.1 & 2458193.6 &  60.2 & 17.71 $\pm$ 0.08 \\
2018-03-17.1 & 2458194.6 &  61.2 & 17.62 $\pm$ 0.05 \\
2018-03-18.1 & 2458195.6 &  62.2 & 17.61 $\pm$ 0.06 \\
2018-03-19.1 & 2458196.6 &  63.2 & 17.55 $\pm$ 0.06 \\
2018-03-20.1 & 2458197.6 &  64.2 & 17.50 $\pm$ 0.06 \\
2018-03-21.1 & 2458198.6 &  65.2 & 17.59 $\pm$ 0.05 \\
2018-03-22.1 & 2458199.6 &  66.2 & 17.66 $\pm$ 0.08 \\
2018-03-24.0 & 2458201.5 &  68.1 & 17.71 $\pm$ 0.07 \\
2018-03-25.0 & 2458202.5 &  69.1 & 17.67 $\pm$ 0.06 \\
2018-03-26.0 & 2458203.5 &  70.1 & 17.60 $\pm$ 0.07 \\
2018-04-05.0 & 2458213.5 &  80.1 & 17.66 $\pm$ 0.06 \\
2018-04-08.0 & 2458216.5 &  83.1 & 17.75 $\pm$ 0.07 \\
2018-04-09.0 & 2458217.5 &  84.1 & 17.83 $\pm$ 0.08 \\
2018-04-12.0 & 2458220.5 &  87.1 & 17.77 $\pm$ 0.07 \\
2018-04-14.0 & 2458222.5 &  89.1 & 17.78 $\pm$ 0.06 \\
2018-04-15.0 & 2458223.5 &  90.1 & 17.86 $\pm$ 0.06 \\
2018-04-17.0 & 2458225.5 &  82.1 & 17.90 $\pm$ 0.07 \\
2018-04-20.0 & 2458228.7 &  95.3 & 17.86 $\pm$ 0.04 \\
2018-04-21.0 & 2458229.5 &  96.1 & 18.01 $\pm$ 0.05 \\
2018-04-22.0 & 2458230.5 &  97.1 & 17.88 $\pm$ 0.06 \\
2018-04-23.0 & 2458231.5 &  98.1 & 17.88 $\pm$ 0.07 \\
2018-05-03.0 & 2458241.5 & 108.1 & 18.30 $\pm$ 0.08 \\
2018-05-05.0 & 2458243.5 & 110.1 & 18.38 $\pm$ 0.07 \\
 \hline
 \end{tabular}}
 \end{center}
 $^\dagger$Phase with respect to the explosion epoch (JD = 2458133.4).
\end{table}
\begin{table*}
\caption{Optical photometry of SN~2018is.}
\label{phot}
\centering
\smallskip
\scriptsize
\begin{tabular}{c c c c c c c c c c c}
\hline
UT Date     & JD       & Phase$^a$ & {\it U} & {\it B} & {\it V} & {\it g} & {\it r} & {\it i} & {\it z} &  Tel \\ 
(yyyy-mm-dd)& 2458000+ & (d)       & (mag)   & (mag)   & (mag)   & (mag)   & (mag)   & (mag)   & (mag) &   \\
\hline
2018-01-20.6 & 139.2 &  5.8 & --         & 18.44±0.03 & 18.22±0.07 & 18.52±0.07 & 18.22±0.06 & 18.25±0.03 & --       & 0m4\\
2018-01-21.4 & 139.9 &  6.5 & 17.73±0.03 & 18.46±0.01 & 18.09±0.03 & 18.41±0.02 & 17.89±0.03 & 18.07±0.01 & --       & 1m0\\
2018-01-22.3 & 140.8 &  7.4 & 17.70±0.06 & 18.46±0.01 & 18.02±0.02 & 18.41±0.02 & 17.87±0.02 & 17.99±0.02 & --       & 1m0\\
2018-01-26.5 & 145.0 &  11.6 & 17.62±0.02 & 18.26±0.03 & 17.86±0.03 & 18.22±0.02 & 17.76±0.05 & 17.89±0.04 & --       & 1m0\\
2018-01-28.2 & 146.7 & 13.3 & --         & --         & --         & --         & 17.62±0.01 & --                   & --       & GTC\\
2018-02-09.1 & 158.6 & 25.2 & 18.51±0.01 & 18.51±0.02 & 17.76±0.01 & 18.27±0.05 & 17.43±0.02 & 17.49±0.02 & --       & 1m0\\
2018-02-12.4 & 161.9 & 28.5 & 18.82±0.15 & 18.63±0.05 & 17.74±0.02 & 18.33±0.01 & 17.41±0.03 & 17.47±0.03 & --       & 1m0\\
2018-02-13.7 & 163.2 & 29.8 & --         & 18.77±0.01 & 17.82±0.01 & 18.44±0.01 & 17.47±0.01 & 17.50±0.01 & 17.25±0.01 & NOT-ALFOSC\\
2018-02-18.7 & 168.2 & 34.8 & 19.44±0.16 & 18.93±0.03 & 17.85±0.09 & 18.61±0.03 & 17.44±0.02 & 17.47±0.03 & --         & 1m0\\
2018-02-22.4 & 171.9 & 38.5 & --         & 18.95±0.03 & 17.80±0.02 & 18.55±0.03 & 17.39±0.01 & 17.41±0.01 & --         & 1m0\\
2018-02-28.2 & 177.7 & 44.3 & --         & 19.32±0.15 & 17.88±0.04 & 18.75±0.05 & 17.50±0.07 & 17.45±0.04 & --         & 1m0\\
2018-03-03.9 & 181.4 & 48.0 & --         & 19.19±0.12 & 17.81±0.02 & 18.75±0.04 & 17.48±0.03 & 17.39±0.04 & --         & 1m0\\
2018-03-07.2 & 184.7 & 51.3 & --         & --         & --         & --         & 17.44±0.01 & --                   & --         & GTC\\
2018-03-08.1 & 185.6 & 52.2 & --         & 19.31±0.02 & 17.91±0.01 & 18.75±0.01 & 17.64±0.01 & 17.49±0.01 & 17.22±0.01 & NOT-STANCam\\
2018-03-10.8 & 188.3 & 54.9 & --         & 19.57±0.01 & 18.04±0.03 & 19.01±0.01 & 17.47±0.03 & 17.43±0.03 & --         & 1m0\\
2018-03-12.0 & 189.8 & 56.4 & --         & 19.58±0.06 & 18.08±0.03 & 19.05±0.03 & 17.51±0.02 & 17.46±0.03 & --         & 1m0\\
2018-03-12.8 & 190.3 & 56.8 & --         & --         & 18.12±0.04 & 19.07±0.04 & 17.53±0.04 & --                   & --         & 1m0 \\
2018-03-13.1 & 190.6 & 57.1 & --         & 19.41±0.01 & 17.95±0.01 & 18.80±0.01 & 17.55±0.01 & 17.45±0.01 & 17.22±0.01 & NOT-STANCam\\
2018-03-17.6 & 195.1 & 61.6 & --         & 19.56±0.04 & 18.08±0.01 & 18.87±0.01 & 17.45±0.01 & 17.43±0.01 & 17.11±0.01 & LT\\
2018-03-17.9 & 195.4 & 61.6 & --         & 19.63±0.06 & 18.01±0.03 & 19.18±0.05 & 17.47±0.03 & 17.40±0.01 & --         & 1m0\\
2018-03-18.3 & 196.3 & 62.9 & --         & 19.60±0.02 & 18.03±0.01 & 19.13±0.02 & 17.45±0.02 & 17.46±0.04 & --         & 1m0\\
2018-03-19.2 & 196.7 & 63.3 & --         & 19.56±0.03 & 18.00±0.02 & 18.92±0.07 & --         &         --         & --         & 1m0\\
2018-03-20.1 & 197.6 & 64.2 & --         & 19.72±0.02 & 18.05±0.01 & 19.07±0.04 & 17.43±0.02 & 17.37±0.03 & --         & 1m0\\
2018-03-21.2 & 198.6 & 65.2 & --         & 19.74±0.02 & 18.03±0.018 & 19.16±0.01 & 17.45±0.01 & 17.43±0.02 & --         & 1m0\\
2018-03-22.6 & 200.1 & 66.7 & --         & 19.53±0.02 & 18.14±0.01 & 18.94±0.01 & 17.48±0.01 & 17.41±0.01 & 17.11±0.01 & LT\\
2018-03-23.1 & 200.6 & 67.2 & --         & --         & 18.15±0.02 & 19.30±0.04 & 17.55±0.02 & 17.43±0.04 & --         & 1m0\\
2018-03-25.6 & 203.0 & 69.6 & --         & 19.72±0.06 & 18.21±0.01 & 19.02±0.01 & 17.50±0.01 & 17.45±0.01 & 17.10±0.01 & LT\\
2018-03-28.7 & 206.2 & 72.8 & --         & 19.81±0.02 & 18.13±0.02 & 19.29±0.06 & 17.50±0.03 & 17.40±0.04 & --         & 1m0\\
2018-04-04.3 & 212.8 & 79.2 & --         & --         & 18.25±0.04 & 19.38±0.02 & 17.63±0.03 & 17.50±0.02 & --         & 1m0 \\
2018-04-04.6 & 213.1 & 79.5 & --         & 19.91±0.02 & 18.23±0.01 & 19.26±0.02 & 17.57±0.01 & 17.51±0.02 & 17.09±0.01 & NOT-ALFOSC \\
2018-04-06.9 & 215.3 & 81.9 & --         & 20.09±0.05 & 18.26±0.02 & 19.46±0.01 & 17.53±0.01 & 17.51±0.01 & --         & 1m0\\
2018-04-15.2 & 223.6 & 90.2 & --         & 20.64±0.10 & 18.47±0.01 & 19.91±0.01 & 17.72±0.03 & 17.61±0.02 & --         & 1m0 \\
2018-04-15.6 & 224.0 & 90.6 & --         & 20.07±0.03 & 18.50±0.01 & 19.45±0.01 & 17.70±0.01 & 17.47±0.01 & 17.18±0.01 & LT \\
2018-04-20.0 & 228.5 & 95.1 & --         & 20.68±0.04 & 18.49±0.01 & 19.51±0.05 & 17.29±0.06 & 17.44±0.08 & 17.07±0.04 & 1.82m-AFOSC\\
2018-04-21.3 & 229.8 & 96.4 & --         & 20.46±0.07 & 18.54±0.02 & 19.75±0.01 & 17.75±0.03 & 17.64±0.02 & --         & 1m0\\
2018-04-27.1 & 235.6 & 102.2 & --         & --         & 18.67±0.05 & 20.04±0.15 & 17.97±0.09 & 17.81±0.13 & --         & 1m0\\
2018-05-01.0 & 239.5 & 106.1 & --        & --         & --         & --         & 17.99±0.02 & --                   & --         & GTC\\
2018-05-02.9 & 241.4 & 108.0 & --        & --         & 19.04±0.02 & --         & 18.20±0.03 & 18.16±0.03 & --         & 1m0\\
2018-05-04.5 & 243.0 & 109.6 & --        & --         & 19.39±0.11 & --         & 18.33±0.02 & 18.23±0.05 & --         & 1m0\\
2018-05-05.3 & 243.8 & 110.4 & --        & --         & 19.05±0.06 & 20.52±0.02 & 18.39±0.01 & 18.19±0.02 & --         & 1m0\\
2018-05-06.8 & 245.2 & 111.8 & --        & --         & 19.39±0.15 & 20.83±0.10 & 18.43±0.02 & 18.30±0.05 & --         & 1m0\\
2018-05-08.2 & 246.6 & 113.1 & --        & --         & 19.59±0.04 & 21.00±0.03 & 18.58±0.02 & 18.37±0.02 & --         & 1m0\\
2018-05-08.5 & 247.0 & 113.4 & --        & 21.52±0.16 & 19.85±0.02 & 20.76±0.06 & 18.83±0.02 & 18.74±0.01 & 18.10±0.01 & LT\\
2018-05-09.6 & 248.1 & 114.7 & --        & --         & 20.12±0.04 & 21.78±0.12 & 19.10±0.03 & 18.92±0.02 & --        & 1m0\\
2018-05-11.1 & 249.6 & 116.2 & --        & --         & 20.67±0.09 & 22.18±0.05 & 19.77±0.02 & 19.63±0.04 & --         & 1m0\\
2018-05-15.0 & 253.5 & 120.1 & --        & --         & --         & --         & 20.14±0.09 & 19.90±0.04 & --         & 1m0\\
2018-05-15.4 & 253.9 & 120.5 & --        & --         & 21.21±0.16 & 22.58±0.19 & 20.23±0.05 & 20.11±0.04 & 19.27±0.06 & LT\\
2018-05-16.2 & 254.7 & 121.2 & --        & --         & 21.10±0.14 & 22.36±0.11 & 19.98±0.02 & 19.81±0.07 & --         & 1m0\\
2018-05-16.6 & 255.0 & 121.6 & --        & 22.80±0.09 & 21.21±0.04 & 21.96±0.03 & 20.24±0.02 & 19.94±0.01 & 19.25±0.02 & NOT-ALFOSC\\
2018-05-17.1 & 255.6 & 122.2 & --        & --         & 21.31±0.02 & 22.45±0.01 & 20.07±0.06 & 19.90±0.04 & --         & 1m0\\
2018-05-18.2 & 256.7 & 123.3 & --        & --         & 21.35±0.23 & --         & 20.18±0.01 & 20.09±0.02 & --         & 1m0\\
2018-05-22.4 & 260.9 & 127.5 & --        & 22.29±0.40 & 21.14±0.15 & 21.70±0.13 & 20.13±0.05 & 20.19±0.05 & 19.30±0.05 & LT\\
2018-05-23.7 & 262.2 & 128.8 & --        & --         & 21.30±0.01 & 22.26±0.12 & 20.25±0.03 & 20.08±0.03 & 19.38±0.05 & NOT-ALFOSC\\
2018-05-31.4 & 269.9 & 136.5 & --        & --         & 21.00±0.13 & 21.90±0.25 & 20.20±0.06 & 20.27±0.06 & 19.38±0.06 & LT\\
2018-06-03.0 & 272.5 & 139.1 & --        & --         & 21.47±0.16 & --         & 20.30±0.06 & 20.11±0.14 & --         & 1m0\\
2018-06-08.8 & 278.3 & 144.9 & --        & --         & --         & --         & 20.33±0.12 & --                   & --         & 1m0\\
2018-06-12.5 & 282.0 & 148.6 & --        & 23.46±0.15 & 21.50±0.05 & 22.66±0.10 & 20.56±0.03 & 20.24±0.03 & 19.45±0.03 & NOT-ALFOSC\\
2018-06-17.1 & 286.6 & 153.2 & --        & --         & --         & --         & 20.38±0.05 & 20.46±0.02 & --         & 1m0\\
2018-07-07.5 & 307.0 & 173.6 & --        & 22.56±0.08 & 21.47±0.04 & 22.24±0.04 & 20.52±0.04 & 20.25±0.01 & 19.69±0.02 & NOT-ALFOSC\\
2018-07-21.4 & 320.9 & 187.5 & --        & --         & --         & --         & 20.46±0.16 & 20.47±0.08 & --         & 2m0\\
2018-07-23.4 & 322.9 & 189.5 & --        & --         & --         & --         & 20.56±0.05 & 20.35±0.01 & --         & 2m0\\
2018-07-29.4 & 328.9 & 195.5 & --        & --         & --         & --         & 20.90±0.07 & 20.67±0.04 & --         & 1m0\\
\hline
\end{tabular}
\newline
$^a$since explosion epoch $t_0$ = 2458133.4 (JD) (2018-01-14.9)
\label{photometry}      
\end{table*}

\begin{table*}
\caption{{\it Swift} UVOT photometry of SN~2018is.}
\setlength{\tabcolsep}{3pt}
\renewcommand{\arraystretch}{1.1}
\footnotesize
\centering
\smallskip
\begin{tabular}{c c c c c c c c c}
\hline 
	Date & JD & Phase$^a$ & {\it uvw2} & {\it uvm2} & {\it uvw1} & {\it uvu} & {\it uvb} &  {\it uvv} \\ 
	(yyyy-mm-dd) & 2458000+ & (d) & (mag) & (mag) & (mag) & (mag) & (mag) & (mag)\\
\hline
2018-01-20.4 & 2458138.9 & 5.5 &     --         &   --            & 18.34$\pm$0.12 & 17.64$\pm$0.10 & 18.40$\pm$0.15 & 17.69$\pm$0.24 \\
2018-01-21.5 & 2458140.0 & 6.6 & 19.40$\pm$0.24 &  18.94$\pm$0.18 & 18.16$\pm$0.16 & 17.54$\pm$0.11 & 18.30$\pm$0.15 & 18.00$\pm$0.27 \\
2018-01-23.0 & 2458141.5 & 8.1 &     --         &   --            & 18.48$\pm$0.13 & 17.61$\pm$0.10 & 18.39$\pm$0.14 & 18.08$\pm$0.29 \\
2018-01-24.3 & 2458142.8 & 9.4 &     --         &   --            & 18.46$\pm$0.15 & 17.54$\pm$0.10 & 18.05$\pm$0.13 & 18.00$\pm$0.29 \\
\hline
\end{tabular}
\newline
$^a$since explosion epoch $t_0$ = 2458133.4 JD (2018-01-14.9)
\label{uvot}      
\end{table*}
\begin{table}
 \begin{center}
 \caption{ATLAS $o$ and $c$ band forced photometry of SN~2018is.}
 \label{atlas_phot}
 \scalebox{0.95}{
 \begin{tabular}{ccccc}
 \hline
UT Date & JD  & Phase$^\dagger$ & $o$         & $c$\\
(yyyy-mm-dd) &     & (days)          & (mag)  & (mag)\\
\hline
2018-01-17.7 &  2458136.2 & 2.8   & 18.01 $\pm$ 0.06 & --\\
2018-01-18.7 &  2458137.2 & 3.8   & 18.21 $\pm$ 0.16 & --\\
2018-01-20.7 &  2458139.2 & 5.8   & -- & 18.03 $\pm$ 0.15 \\       
2018-01-22.7 &  2458141.2 & 7.8   & 17.92 $\pm$ 0.04 & --\\ 
2018-01-28.7 &  2458147.2 & 13.8  & 17.38 $\pm$ 0.04 & --\\ 
2018-01-29.6 &  2458148.1 & 14.7  & 17.67 $\pm$ 0.04 & --\\ 
2018-01-30.6 &  2458149.1 & 15.7  & 17.36 $\pm$ 0.04 & --\\ 
2018-02-02.6 &  2458152.1 & 18.7  & 17.35 $\pm$ 0.09 & --\\ 
2018-02-07.6 &  2458157.1 & 23.7  & 17.40 $\pm$ 0.05 & --\\ 
2018-02-09.6 &  2458159.1 & 25.7  & 17.18 $\pm$ 0.02 & --\\ 
2018-02-10.6 &  2458160.1 & 26.7  & 17.34 $\pm$ 0.02 & --\\ 
2018-02-11.6 &  2458161.1 & 27.7  & 17.51 $\pm$ 0.03 & --\\ 
2018-02-12.6 & 2458162.1 & 28.7 & -- & 17.73 $\pm$ 0.03 \\  
2018-02-13.6 & 2458163.1 & 29.7 & -- & 17.88 $\pm$ 0.03 \\  
2018-02-15.6 &  2458165.1 & 31.7  & 17.43 $\pm$ 0.04 & --\\ 
2018-03-07.6 &  2458185.1 & 51.7  & 17.41 $\pm$ 0.04 & --\\ 
2018-03-19.6 &  2458197.1 & 63.7  & 17.56 $\pm$ 0.03 & --\\ 
2018-03-29.5 &  2458207.0 & 73.6  & 17.27 $\pm$ 0.03 & --\\ 
2018-04-12.5 &  2458221.0 & 87.6  & 17.56 $\pm$ 0.03 & --\\ 
2018-04-18.5 & 2458227.0 & 93.6 & -- & 18.55 $\pm$ 0.26 \\ 
2018-04-26.4 &  2458234.9 & 101.5  & 17.46 $\pm$ 0.05 & --\\
2018-05-08.4 &  2458246.9 & 113.5 & 18.21 $\pm$ 0.16 & --\\ 
2018-06-05.4 &  2458274.9 & 141.5 & 19.25 $\pm$ 0.24 & --\\ 
 \hline
 \end{tabular}}
 \end{center}
 $^\dagger$Phase with respect to the explosion epoch (JD = 2458133.4).
\end{table}
\begin{table}
\caption{Log of spectroscopic observations of SN~2018is.}
\setlength{\tabcolsep}{3pt}
\renewcommand{\arraystretch}{1.1}
\scriptsize
\centering
\smallskip
\begin{tabular}{c c c c c c c}
\hline \hline
UT Date       & JD         &Phase$^a$ & Instrument/ & Resolution  & Exposure \\ 	
(UT)          & (2458000+) & (Days)    &   Telescope & ($\lambda$/$\Delta\lambda$) & Time (s) \\
\hline
 2018-01-21.0 & 139.6 & 6.2	 & RSS/SALT      & 1000  &   1500 \\
 2018-01-21.3 & 139.8 & 6.4  & GHTS-R/SOAR   &  850  &    900 \\
 2018-01-22.2 & 140.8 & 7.4  & ALFOSC/NOT    &  360  &   2400 \\
 2018-01-24.4 & 142.9 & 9.5  & DBSP/P200     & 492(b)/893(r) &  3$\times$1000 (r)/3$\times$1000 (b) \\
 2018-01-24.5 & 143.0 & 9.6  & B$\&$C/Bok    & 1880  &    600 \\
 2018-01-26.4 & 145.0 & 11.6  & BCH/MMT       & 4500  &   1200 \\
 2018-01-28.2 & 146.7 & 13.3  & OSIRIS/GTC    & 1018 (R1000B)  &    600 \\
 2018-01-29.0 & 147.5 & 14.1  & RSS/SALT      &  1000 &   1500 \\
 2018-01-29.5 & 148.0 & 14.6  & FLOYDS/FTN    &   380 &   3600 \\
 2018-01-31.2 & 149.7 & 16.3  & FIRE/Magellan &   400 &    126 \\
 2018-02-02.0 & 151.5 & 18.1  & RSS/SALT      &  1000 &   1500 \\
 2018-02-14.2 & 163.7 & 30.3  & ALFOSC/NOT    &   360 &   3600 \\
 2018-03-05.4 & 182.9 & 49.5  & BCH/MMT       &  4500 &    600 \\
 2018-03-06.5 & 184.0 & 50.6  & FLOYDS/FTN    &   380 &   3600 \\
 2018-03-07.2 & 184.7 & 51.3  & OSIRIS/GTC    &  1018 (R1000B) &    600 \\
 2018-03-07.5 & 185.0 & 51.6  & FLOYDS/FTN    &   380 &   3600 \\
 2018-03-27.4 & 204.9 & 71.5  & B$\&$C/Bok    &  1880 &   1200 \\
 2018-04-06.1 & 214.6 & 81.2  & ALFOSC/NOT    &   360 &   3600 \\
 2018-04-06.3 & 214.8 & 81.4  & B$\&$C/Bok    &  1880 &   1200 \\
 2018-05-01.0 & 239.5 & 106.1 & ALFOSC/NOT    &   360 &   2400 \\
 2018-07-02.1 & 301.6 & 168.2 & BCH/MMT       &  4500 &   1200 \\
 2019-02-05.6 & 520.1 & 386.7 & LRIS/Keck I    & 1200-2200 &   4$\times$1200 \\ 
\hline                                   
\end{tabular}
\newline
$^a$ since explosion epoch $t_0$ = 2458133.4 (JD) (2018-01-14.9)
\label{spec_observations}      
\end{table}
\begin{table*}
\centering
\caption{SN II comparison sample.}
\label{comparison_sample}
\renewcommand{\arraystretch}{1.1}
\setlength{\tabcolsep}{3pt}
\footnotesize
\begin{tabular}{llllllll}
\hline
SN       & Host   & Explosion  & Redshift & Distance  & E(B$-$V)$_{\rm host}$   & E(B$-$V)$_{\rm MW}$    & References \\
         & galaxy & Epoch (MJD) &          &  modulus$^\dagger$ (mag)  & (mag)                   &   (mag) &         \\
\hline
1997D       & NGC 1536 & 50361.0 $\pm$ 15.0 & 0.00406 & 33.89 $\pm$ 0.05 & 0.090 $\pm$ 0.111    & 0.0168 $\pm$ 0.0012 & 1,2\\
1999br      & NGC 4900 & 56496.8 $\pm$ 0.2  & 0.00321 & 31.10 $\pm$ 0.80 & 0.044 $\pm$ 0.040    & 0.0204 $\pm$ 0.0002 & 6  \\
1999em      & NGC 1637 & 51476.5 $\pm$ 5.0  & 0.00239 & 30.37 $\pm$ 0.07 & 0.058 $\pm$ 0.019    & 0.0346 $\pm$ 0.0003 & 3, 4, 5\\
2002gd      & NGC 3537 & 52551.5 $\pm$ 2.0  & 0.00892 & 33.01 $\pm$ 0.35 & --                   & 0.0575 $\pm$ 0.0006 & 12 \\
2003Z       & NGC 2742 & 52665.1 $\pm$ 2.4  & 0.00431 & 31.80 $\pm$ 0.60 & --                   & 0.0334 $\pm$ 0.0005 & 11, 12 \\
2005cs      & M51a     & 53548.4 $\pm$ 0.3  & 0.00145 & 29.26 $\pm$ 0.33 & 0.0192               & 0.0308 $\pm$ 0.0015 & 7,8,11\\
2008in      & M61      & 54825.1 $\pm$ 0.8  & 0.005224 & 30.60 $\pm$ 0.20 & 0.076 $\pm$ 0.104 & 0.0193 $\pm$  0.0001 & 10\\
2009N       & NGC 4487 & 54847.6 $\pm$ 1.2  & 0.0035  & 31.67 $\pm$ 0.11 & 0.113 $\pm$ 0.019 & 0.019 $\pm$ 0.001 & 13\\
2009ib      & NGC 1559 & 55041.3 $\pm$ 3.1  & 0.00435 & 31.48 $\pm$ 0.33 &  0.131 $\pm$ 0.025   & 0.0257 $\pm$ 0.0003 & 14\\
2009md      & NGC 3389 & 55162.0 $\pm$ 8.0  & 0.00440 & 31.64 $\pm$ 0.21 & 0.1$^{+0.1}_{-0.05}$ & 0.0234 $\pm$ 0.0002 & 9\\
2013K       & ESO 009-G10 & 56302.0 $\pm$ 5.0  & 0.00807 & 32.80 $\pm$ 0.40 & 0.15$\pm$0.20     &  0.1218 $\pm$ 0.0026 &  16 \\
ASASSN-14ha & NGC 1566    & 56910.0 $\pm$ 1.5  & 0.00500 & 31.27 $\pm$ 0.49 & --                &  0.0079 $\pm$ 0.0002 &  15 \\
2016aqf     & NGC 2101    & 57442.6 $\pm$ 0.3  & 0.00389 & 30.16 $\pm$ 0.27 & --                & 0.0472 $\pm$ 0.0009 & 22\\
2016bkv     & NGC 3184    & 57467.5 $\pm$ 1.2  & 0.00198 & 30.40 $\pm$ 0.18 & --                & 0.0144 $\pm$ 0.0001 & 17, 18 \\
SN-NGC6412  & NGC 6412    & 57210.0 $\pm$ 2.0  & 0.004380 & 31.26$\pm$0.16   & --                &  0.0348 $\pm$ 0.0005 & 21\\
2018aoq     & NGC 4151    & 58207.5 $\pm$ 1.0  & 0.00332 & 31.51 $\pm$ 0.17 & 0.0163 & 0.0237 $\pm$ 0.0011 & 19, 20, 23\\
2018hwm     & IC 2327     & 58425.0 $\pm$ 1.5  & 0.00895 & 33.75 $\pm$ 0.19 & --                & 0.0224 $\pm$ 0.0009 & 24 \\
2018lab     & IC 2163     & 58480.9 $\pm$ 1.0  & 0.0089 & 32.75 $\pm$ 0.40 & 0.15               & 0.0748 $\pm$ 0.0006 & 27 \\
2020cxd     & NGC 6395    & 58897.0 $\pm$ 1.5  & 0.003883 & 31.60 $\pm$ 0.20 & -- & 0.0355 $\pm$ 0.0097 & 25\\
2021aai     & NGC 2268    & 59223.4 $\pm$ 1.0  & 0.007428 & 32.47 $\pm$ 0.20 & 0.20 $\pm$ 0.03 & 0.0548 $\pm$ 0.0010 & 25\\
2021gmj     & NGC 3310    & 59292.8 $\pm$ 0.5  & 0.00331 & 31.25 $\pm$ 0.06 & 0.03 $\pm$ 0.01 & 0.0192 $\pm$ 0.0005 & 28, 29 \\
2022acko    & NGC 1300    & 59918.2 $\pm$ 0.5  & 0.00526 & 31.39 $\pm$ 0.33 & 0.03 $\pm$ 0.01    & 0.0261 $\pm$  0.0003 & 26 \\
\hline
\end{tabular}
\newline
{\footnotesize
$^\dagger$ if redshift dependent, rescaled to H$_0$=67.5 km/s/Mpc.\\
References: [1] \cite{deMello1997}, [2] \cite{Turatto1998}, [3] \cite{Hamuy2001}, [4] \cite{Leonard2002}, [5] \cite{Elmhamdi2003}, [6] \cite{Pastorello2004}, [7] \cite{Tsvetkov2006}, [8] \cite{Pastorello2009}, [9] \cite{Fraser2011}, [10] \cite{Roy2011}, [11] \cite{Faran2014}, [12] \cite{Spiro2014}, [13] \cite{Takats2014}, [14] \cite{Takats2015}, [15] \cite{Valenti2016}, [16] \cite{Tomasella2018}, [17] \cite{Nakaoka2018}, [18] \cite{Hosseinzadeh2018}, [19] \cite{ONeill2019}, [20] \cite{Tsvetkov2019}, [21] \cite{Jager2020}, [22] \cite{Muller2020}, [23] \cite{Tsvetkov2021}, [24] \cite{Reguitti2021}, [25] \cite{Valerin2022}, [26] \cite{Bostroem2023}, [27] \cite{Pearson2023}, [28] \cite{Murai2024}, [29] \cite{Meza2024} }
\end{table*}
\begin{table*}
\centering
\caption{Physical parameters of SN II comparison sample.}
\renewcommand{\arraystretch}{1.1}
\setlength{\tabcolsep}{3pt}
\footnotesize
\begin{tabular}{lcccccc}
\hline
SN          & s$_{2}$ (mag/100d) & t$_{\rm start}$  & t$_{\rm end}$ & t$_{\rm PT}$        & M$_V^{\rm 50d}$ & M$_{\rm Ni}$ \\
\hline
            &                    &        &          & \sc{normal luminosity} \\
\hline
1999em      & 0.32 $\pm$ 0.02    & 24.8 &  75.5 & 123.4 $\pm$ 3.6       & $-$16.74$\pm$0.10 & 0.050$^{+0.008}_{-0.009}$ \\
\hline
            &                    &        &       & \sc{intermediate luminosity} \\
\hline
2009ib      & 0.25 $\pm$ 0.04    & 31.0 & 102.0 & 140.2 $\pm$ 3.1       & $-$15.82$\pm$0.34 & 0.046 $\pm$ 0.015\\
2013K       & 0.08 $\pm$ 0.08    & 19.8 &  93.8 & 142.6$^{+5.4}_{-5.3}$ & $-$15.94$\pm$0.75 & 0.012 $\pm$ 0.010\\
2018aoq     & 0.35 $\pm$ 0.04    & 22.4 & 100.4 & 117.0$\pm$1.2         & $-$15.90$\pm$0.18 & 0.01 \\
\hline
            &                    &        &       & \sc{low luminosity} \\
\hline
1999br      & 0.01 $\pm$ 0.02    & 18.5 &  78.5 & --                    & $-$13.61$\pm$0.83 & 0.002 $\pm$ 0.001 \\
2003Z       & 0.39 $\pm$ 0.10    & 27.4 &  94.4 & 131.4$^{+4.3}_{-5.4}$ & $-$14.50$\pm$0.60 & 0.005 $\pm$ 0.003 \\
2005cs      & 0.38 $\pm$ 0.02    &  3.4 & 106.3 & 126.0 $\pm$ 1.0       & $-$14.64$\pm$0.37 & 0.006 $\pm$ 0.003\\
2009md      & 0.63 $\pm$ 0.09    & 18.0 & 100.7 & 117.7 $\pm$ 8.0       & $-$14.88$\pm$0.38 & 0.004 $\pm$ 0.001 \\
ASASSN-14ha & 0.45 $\pm$ 0.01    &  3.1 & 123.8 & 136.8$\pm$1.5         & $-$14.37$\pm$0.50 & 0.0014 $\pm$ 0.0002\\
2016aqf     & $-$0.11 $\pm$ 0.01 &  7.8 &  64.8 & --                    & $-$14.54$\pm$0.28 & 0.008 $\pm$ 0.002\\
2016bkv     & 0.41 $\pm$ 0.05    & 30.7 &  90.6 &   --                  & $-$14.81$\pm$0.06 & 0.0216 $\pm$ 0.0014\\
2018hwm     & 0.43 $\pm$ 0.05    &  5.5 &  92.5 & 143.7$^{+2.0}_{-2.2}$ & --                & 0.0033$^{+0.0026}_{-0.0015}$\\
SN-NGC6412  & $-$0.10 $\pm$ 0.03 & 25.0 &  65.3 & --                    & $-$14.65$\pm$0.16 & 0.0015 $\pm$ 0.0008\\
2018lab     & 0.34 $\pm$ 0.06    &  7.4 &  64.9 & --           & $-$14.96$\pm$0.40 & -- \\
2020cxd     &      --            &   -- &   --  & --                    &  --     & 0.0018 $\pm$ 0.0005\\
2021aai     & 0.02 $\pm$ 0.07    & 10.5 &  73.8 & --                    & $-$14.64 $\pm$ 0.22   & 0.0075 $\pm$ 0.0025\\
2021gmj     & 0.27 $\pm$ 0.02    & 10.6 &  81.7 & -- & $-$15.42$\pm$0.07 & 0.014 $\pm$ 0.001\\
2022acko    & 0.88 $\pm$ 0.02    & 13.4 &  60.3 & --          & $-$15.07$\pm$0.33 & -- \\
\textbf{2018is} & \textbf{1.04 $\pm$ 0.03} & \textbf{24.6} & \textbf{94.5} & \textbf{113.9$\pm$1.1} & \textbf{$-$15.08$\pm$0.20}$^\dagger$ & \textbf{0.0049 $\pm$ 0.0008}$^\dagger$ \\
\hline
\end{tabular}
\newline
{\footnotesize
$^\dagger$ correspond to high extinction scenario
}
\end{table*}

\begin{figure*}
    \includegraphics[scale=0.40]{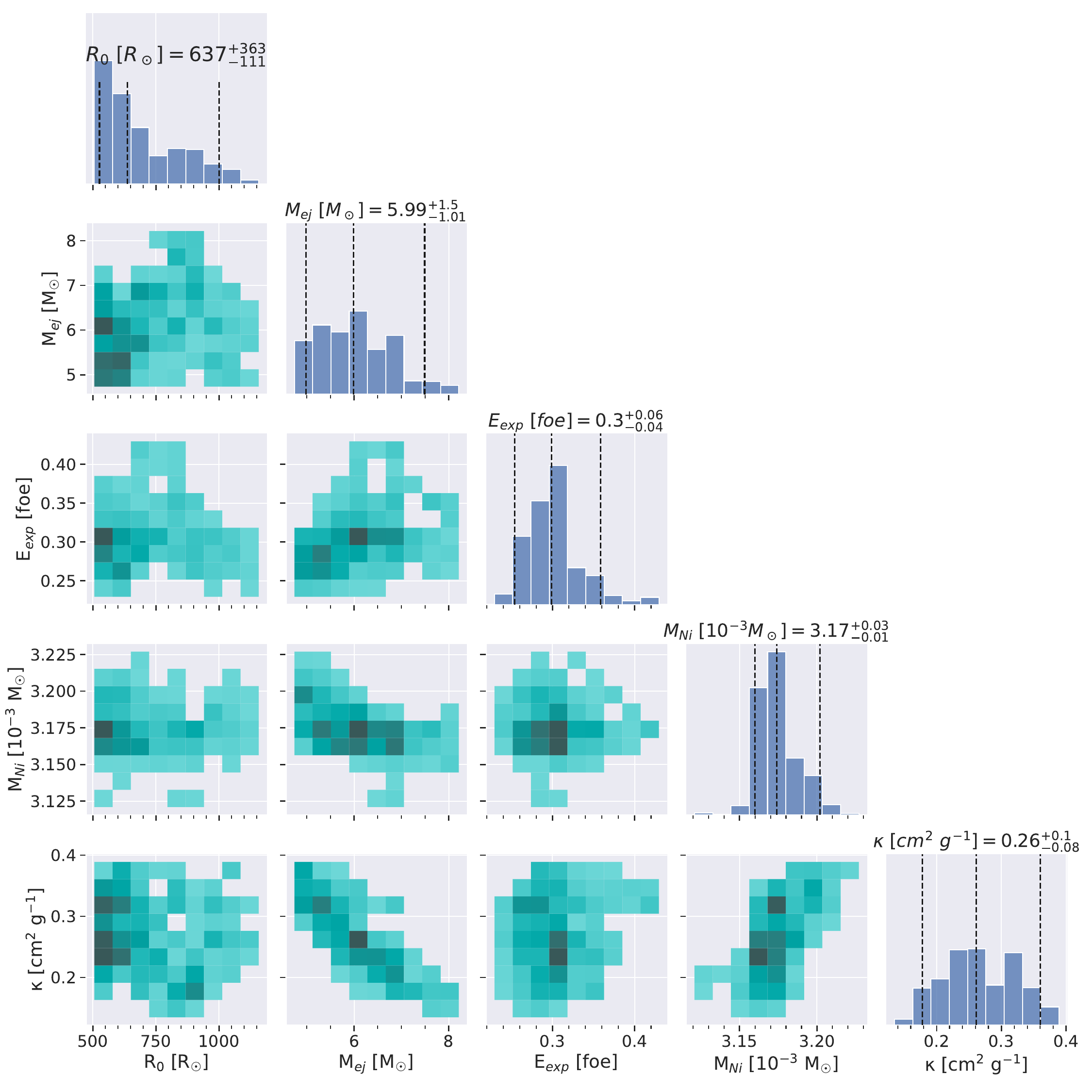}
    \caption{The corner plot displaying the correlation between the modelling parameters of the semi-analytical light curve modelling for the low reddening scenarios, where the diagonal plots show the posterior distribution of the parameters.}
    \label{fig:corner_plot1}
\end{figure*}

\begin{figure*}
    \includegraphics[scale=0.40]{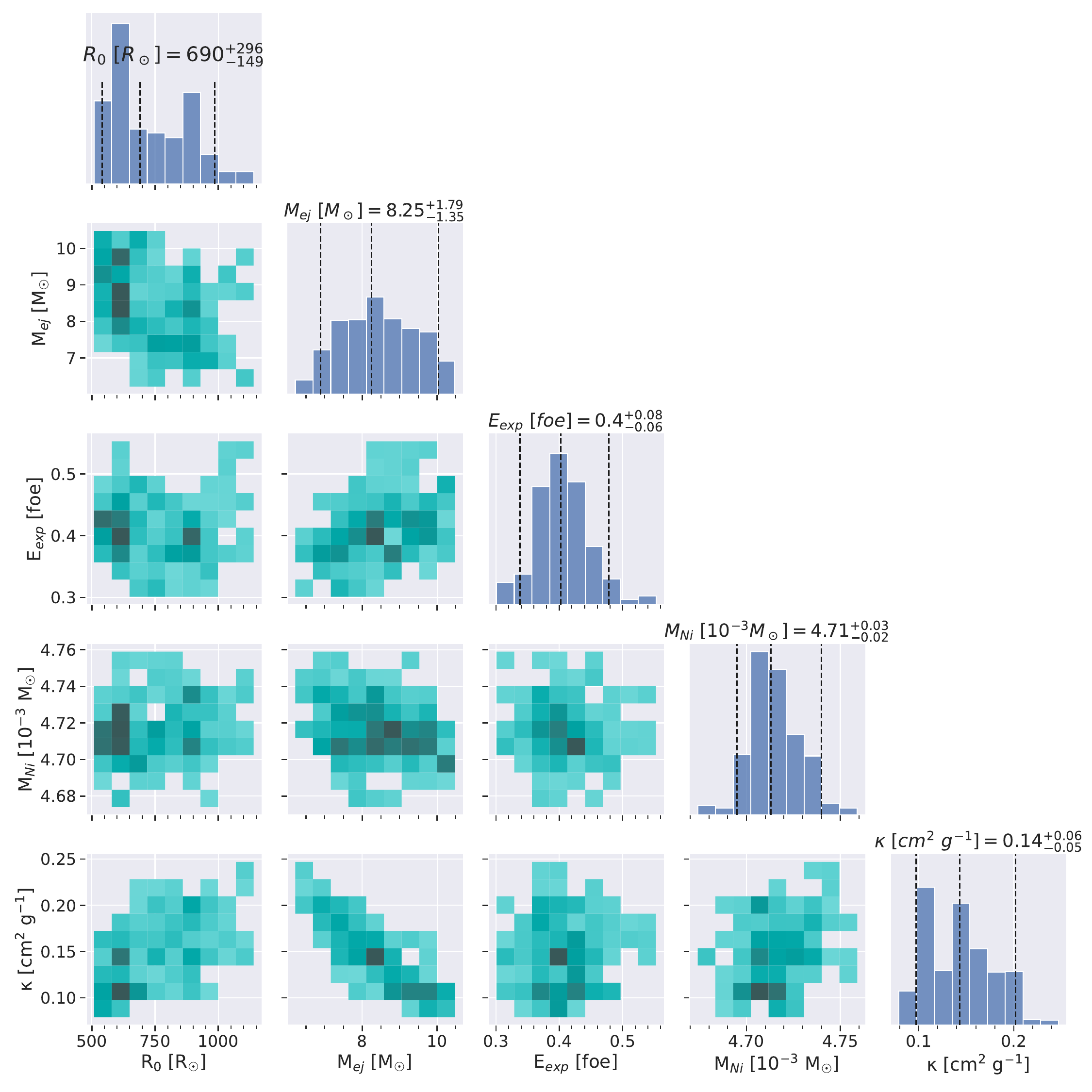}
    \caption{The corner plot displaying the correlation between the modelling parameters of the semi-analytical light curve modelling for the high reddening scenarios, where the diagonal plots show the posterior distribution of the parameters.}
    \label{fig:corner_plot2}
\end{figure*}
%%%%%%%%%%%%%%%%%%%%%%%%%%%%%%%%%%%%%%%%%%%%%%%%%%

% Don't change these lines
\end{document}